\documentclass[aps,preprint,groupedaddress,floatfix]{revtex4-1}
\usepackage{array}
\usepackage{amsmath,amsthm,array}
\usepackage{amsfonts}
\usepackage{CJKutf8} 
\usepackage{multirow}
\usepackage{graphicx}
\usepackage{subfigure}
\usepackage{threeparttable}
\usepackage{array}

\makeatletter
\renewcommand{\thetable}{\@arabic\c@table}
\usepackage[T1]{fontenc}
\usepackage{supertabular}
\usepackage{color}

\usepackage{diagbox}
\usepackage{footmisc}
\usepackage{longtable}
\usepackage{booktabs}
\usepackage{comment}

\begin{document}
\title{Social Media Would Not Lie: Prediction of the 2016 Taiwan Election via Online Heterogeneous Data}

\author{Zheng Xie$^1$, Guannan Liu$^1$, Junjie Wu$^{1,2,~\star}$, and Yong Tan$^3$}

\affiliation{
1. School of Economics and Management, Beihang University, Beijing 100191, China \\
2. Beijing Advanced Innovation Center for Big Data and Brain Computing, Beihang University, Beijing 100191, China\\
3. Foster School of Business, University of Washington, Seattle, WA, 98195, USA\\
$^\star$Corresponding author: wujj@buaa.edu.cn}


\begin{abstract}
The prevalence of online media has attracted researchers from various domains to explore human behavior and make interesting predictions. In this research, we leverage heterogeneous social media data collected from various online platforms to predict Taiwan's 2016 presidential election. 
In contrast to most existing research, we take a ``signal'' view of heterogeneous information and adopt the Kalman filter to fuse multiple signals into daily vote predictions for the candidates. We also consider events that influenced the election in a quantitative manner based on the so-called event study model that originated in the field of financial research. We obtained the following interesting findings. First, public opinions in online media dominate traditional polls in Taiwan election prediction in terms of both predictive power and timeliness. But offline polls can still function on alleviating the sample bias of online opinions. Second, although online signals converge as election day approaches, the simple Facebook “Like” is consistently the strongest indicator of the election result. Third, most influential events have a strong connection to cross-strait relations, and the Chou Tzu-yu flag incident followed by the apology video one day before the election increased the vote share of Tsai Ing-Wen by 3.66\%. This research justifies the predictive power of online media in politics and the advantages of information fusion. The combined use of the Kalman filter and the event study method contributes to the data-driven political analytics paradigm for both prediction and attribution purposes.
\end{abstract}
\keywords{Election Prediction $|$ Heterogeneous Data $|$ Kalman Filter $|$ Event Study Method $|$ Social Media}
\maketitle

\section{Introduction}
Recent years have witnessed the rapid development of social media and their innovative applications in many fields~\cite{asur2010predicting}. For instance, it has been found that the volumes of tweets related to protests on Twitter are associated with real-life protest events~\cite{steinert2015online}. Moreover, film mentions on Twitter can reflect box office revenues~\cite{asur2010predicting}. Additionally, public moods extracted from tweets can predict changes in stock markets~\cite{bollen2011twitter}, and a real-time earthquake reporting system was developed by analyzing only tweets~\cite{sakaki2013tweet}.

The unprecedented prevalence of social media has driven politicians to make use of this channel to propagate their ideas and political views~\cite{metaxas2012social,effing2011social,graham2013between,enli2013personalized} to more directly approach potential voters. It is not unusual to see election candidates post their daily activities and political ideas on social media and even debate on social media before and during the campaign. These behaviors can attract online discussion from massive numbers of netizens and, compared with traditional polls, are an easier way to gather wide-ranging public opinions about the candidates. Some research has shown the predictability of election results based on social media information in various countries and regions, including the United States~\cite{williams2008social,lui2011predictability,digrazia2013more,macwilliams2015forecasting}, the United Kingdom~\cite{franch2013wisdom,burnap2015140}, Germany~\cite{tumasjan2010predicting}, the Netherlands~\cite{sang2012predicting}, and Korea~\cite{song2014analyzing},where netizens' behaviors and posts on social media were analyzed to infer the election results.

The existing research, however, usually exploits a single information source and uses simple descriptive statistics for election predictions, which easily results in hindsight bias and lacks generality. The way to ameliorate these issues is two-fold. On one hand, multiple sources should be included to obtain heterogeneous information for robust predictions. For instance, the keywords searched in Google represent the attention of the public, and the aggregated volumes can be used to predict the trends of influenza~\cite{kang2013using}, stock markets~\cite{preis2013quantifying,curme2014quantifying}, consumer behaviors~\cite{goel2010predicting}, {\it etc}. On the other, massive heterogeneous data obtained in real time are often too chaotic to provide consistent predictions; therefore, a method that can fuse the data and deliver robust predictions is indispensable. Our work in this paper is a novel attempt on this front.

We take Taiwan's 2016 presidential election as a real-life case. Taiwan adopted direct election in 1996, and since then, Kuomintang (KMT) and the Democratic Progressive Party (DPP) have become the two major competing political parties. KMT pursues a ``One China Policy'' and the political legitimacy of the ``Republic of China'', whereas DPP takes ``Taiwan Independence'' as its party program. In 2016, three candidates ran for the presidential election, including Eric Chu from KMT, Tsai Ing-wen from DPP, and James Soong from the People First Party (PFP). The election regulations adopt the ``one man one vote'' principle and execute the majority rule~\cite{fell2005party}.

This research leverages time series data collected from various mainstream online platforms ({\it i.e.}, Facebook, Twitter and Google) and visitation traffic to candidates' campaign pages. These heterogeneous signals represent public opinions and are fed into a \textit{Kalman filter}~\cite{kalman1960new} to estimate the vote shares of each candidate dynamically. The most efficient signals are then identified based on the signal strengths characterized by the \textit{Kalman gain}. In addition to prediction, this research attempts to automatically identify the events that most influenced the election by leveraging the \emph{event study model}~\cite{mackinlay1997event} that originated in the field of financial research.

The results show that the prediction errors for every candidate one day, week, and month before the election are no greater than 2.59\%, 4.58\% and 5.87\%, respectively. The results include some interesting findings. First, online signals appear to be more accurate than traditional polls in election prediction, although the polls can still function on mitigating the sample bias of netizens. In particular, a simple Facebook``Like'' on a candidate's post is the most significant predictor, whereas the seemingly more informative ``Comments'' function is much less important. Second, online signals show clear convergence as the final election day approaches. For example, Google keyword searches fluctuated initially but became a strong indicator in the final stage. Third, bursty events most influential to the campaign have a strong relationship with the cross-strait relation topics. For instance, while the Xi-Ma meeting reduced support of Tsai Ing-wen by 0.55\%, the Chou Tzu-yu flag incident followed by the apology video one day before the election increased her votes by 3.66\%.

\section{Data and Measurements}
To identify the most popular Internet applications in Taiwan, we referred to professional Internet surveys\footnote{Internet Usage in Taiwan: Summary Report of October 2015 Survey.} and web traffic reports from Alexa, comScore and Digital Age (see \emph{SI}, TABLE 1). We selected Facebook, Twitter, Google, and candidates' campaign homepages as the ``online sensors'' of public opinions towards the election and designed various daily updated measurements to characterize the signals during the period from Oct. 31, 2015 to Jan. 16, 2016 consecutively. A 30-day moving average was applied to each measure to avoid excessive fluctuation.

\emph{Facebook}.~Facebook is the most popular social platform in Taiwan and provides an easy way for candidates to reach out to a large audience. For each post by a candidate, users can click the ``Like'' tag to indicate a positive reaction. Hence, we can use the ``daily average number of Likes per post'' to measure a candidate's popularity:
\begin{equation}
s^c_{k, \textit{FAL}}=\frac{1}{m}\sum^{m-1}_{j=0}\frac{\sum_i{like^c_{k-j,i}}/n^c_{k-j,\textit{FA}}}{\sum_c{\sum_i{like^c_{k-j,i}}/n^c_{k-j,\textit{FA}}}},
\end{equation}
where $like^c_{k-j,i}$ is the number of {\it Likes} of post $i$ published by candidate $c$ on day $k-j$, $n^c_{k,\textit{FA}}$ is the total number of the candidate's posts, and $m$ is the window length of the moving average. Analogously, we compute the ``daily average number of Comments per post'' for each candidate as another signal from Facebook:
\begin{equation}
s^c_{k, \textit{FAC}} = \frac{1}{m}\sum^{m-1}_{j=0}\frac{\sum_i{Comment^c_{k-j,i}}/n^c_{k-j,\textit{FA}}}{\sum_c{\sum_i{Comment^c_{k-j,i}}/n^c_{k-j,\textit{FA}}}},
\end{equation}
where $Comment^c_{k-j,i}$ is the number of comments on post $i$ published by candidate $c$ on day $k-j$.

\emph{Twitter}.~We use three candidates' names in both Simplified and Traditional Chinese as keywords(See \emph{SI}, Sect.1.2) to retrieve tweets from Twitter. The measurement ``number of tweets mentioning the candidate'' is calculated as
\begin{equation}
s^c_{k, TW}= \frac{1}{m}\sum^{m-1}_{j=0} \frac{tw^c_{k-j}}{\sum_{c}tw^c_{k-j}},
\end{equation}
where $tw^c_{k-j}$ is the volume of tweets about candidate $c$ on day $k-j$.

\emph{Search Engine}.~We also obtained search data from Google Trends to trace the evolution of a keyword's search volume. We used the three candidates' names in both Simplified and Traditional Chinese as keywords and restricted the search source to Taiwan. The measurement ``search index ratio'' is defined as
\begin{equation}
s^c_{k,\textit{GO}} = \frac{1}{m}\sum^{m-1}_{j=0}\frac{search^c_{k-j}}{\sum_{c}search^c_{k-j}},
\end{equation}
where $search^c_{k-j}$ is the aggregated search indexes of keywords about candidate $c$ on day $k-j$.

\emph{Campaign Homepages.}~We collected the daily traffic to candidates' campaign homepages data from Alexa, and used the ``IP traffic ratio'' as an opinion measure as follows:
\begin{equation}
s^c_{k,\textit{IP}} = \frac{1}{m}\sum^{m-1}_{j=0}\frac{IP^c_{k-j}}{\sum_{c}IP^c_{k-j}},
\end{equation}
where $IP^c_{k-j}$ is the IP traffic volume to candidate $c$'s campaign homepage on day $k-j$.

The above measurements convey different signals for continuous election prediction. We also collected offline election polls published by nineteen authoritative pollsters during the period from Aug. 1, 2015 to Jan. 16, 2016 (see \emph{SI}, Sect.~1.1) for comparison. These polls were published aperiodically and infrequently, so we assume the opinions from a poll remain unchanged until a new poll has been released.

\section{Methods}
\subsection{Vote Prediction Model}
The goal of election prediction is to infer the underlying vote shares of various candidates based on heterogeneous noisy signals. A model that can fuse the signals in such a way to debias the prediction from noise and make dynamic predictions to reflect the evolution of public opinion is desired. We exploit the \emph{Kalman filter}, a linear dynamic model, for this purpose. The filter was adopted in~\cite{jackman2005pooling,fisher2011polls,walther2015picking} for election analysis, but previous studies were mostly based on polls and assumed only two candidates.

In general, a Kalman filter maps hidden states to observed variables with noise, and the current hidden states are assumed to transition from previous states with noise. 
That is,
\begin{equation}\label{equ:kfe}
\begin{array}{lll}
\mathbf{s}^c_k & = &\mathbf{h}_k x^c_k+\mathbf{r}^c_k,~~\mathbf{r}^c_k \sim N(0,\mathbf{R}^c_k),
\\[0.15cm]
x^c_k & = &f_kx^c_{k-1}+q^c_k, ~~q^c_k \sim N(0,\sigma^2_{c,k}),
\\[0.15cm]
x^c_0 & \sim &N(m^c_0,p^c_0),
\end{array}
\end{equation}
where $\mathbf{h}_k$ is a vector that maps the hidden state $x^c_k$ to observed multiple signals in $\mathbf{s}^c_k$, $f_k$ is the state transition coefficient, and $x^c_0$ is the initial value of the hidden state. $\mathbf{r}^c_k$ and $q^c_k$ denote independent Gaussian random noise.

In our case, $x^c_k$ is the genuine vote share of candidate $c$ on day $k$, and $\mathbf{s}^c_k = (s^c_{k,GO},s^c_{k,FAL},s^c_{k,TW},s^c_{k,IP})^{\top}$ contains the observed multiple signals. We set $f_k = 1$ and $\mathbf{h}_k = \mathbf{1}$ for scale equivalence of the variables. The initial vote $m^c_0$ is set as the latest poll result of TISR (see {\em SI}, Sect.~2.1), with $p^c_0=1$ to allow fluctuation. Nevertheless, the final prediction is insensitive to the initial values when the time series is sufficiently long (see \emph{SI}, Sect.~2.2 and Sect.~2.3).

The remaining challenge is to estimate the noise parameters $\mathbf{R}^c_k$ and $\sigma^2_{c,k}$. To reduce the model complexity, we assume $\mathbf{R}^c_k = \mathbf{R}_k$ and $\sigma^2_{c,k}=\sigma^2_k$, $\forall~c$. The maximum a posteriori estimation can then be obtained by maximizing the conditional density function:
\begin{equation}
\begin{array}{c}
\mathcal{J} =  p(x^{tsai}_{1:k},x^{chu}_{1:k},x^{soong}_{1:k},\sigma^2_k,\mathbf{R}_k|\mathbf{s}^{tsai}_{1:k},\mathbf{s}^{chu}_{1:k},\mathbf{s}^{soong}_{1:k})\\[0.15cm]
\propto \prod_c p(x^c_0) \prod^{k}_{j=1}p(\mathbf{s}^c_j|x^c_j,\mathbf{R}_k)p(x^c_j|x^c_{j-1},\sigma^2_k)p(\sigma^2_k,\mathbf{R}_k),
\end{array}
\end{equation}
with $\sum_c x^c_k=1$ and $\sum_c \mathbf{s}^c_k = \mathbf{I}_{4 \times 1}$. Accordingly  (see \emph{SI}, Sect.~2.1),
{\begin{equation}
\label{equ:errors}
\begin{array}{c}
\hat{\sigma}^2_k = \frac{1}{3k}\sum_c \sum^k_{j=1}(\hat{x}^c_{j|j}-f_j \hat{x}^c_{j-1|j-1})^2,
\\[0.15cm]
\hat{\mathbf{R}}_k = \frac{1}{3k}\sum_c \sum^k_{j=1}((\mathbf{s}^c_j-\mathbf{h}_j\hat{x}^c_{j|j-1})(\mathbf{s}^c_j-\mathbf{h}_j\hat{x}^c_{j|j-1})^{\top}-\mathbf{h_j}\hat{p}_{j|j-1}\mathbf{h_j}^{\top}),
\end{array}
\end{equation}}
where $\hat{p}_{j|j-1}$ is the estimated variance of $x^c_{j|j-1}$.

\subsection{Event Detection Method}
Twitter, as an online plaza, aggregates information about different candidates during an election campaign. The volatility of tweets can thus signal influential events. A three-step detection method is designed as follows. Step~\uppercase\expandafter{\romannumeral1} is to perceive events based on massive numbers of tweets. To this end, we watch the statistic $tw_k^c$, {\it i.e.}, the number of tweets about candidate $c$ on day $k$, and trace its volatility in the past $m$ days by comparing it with an upper bound $u^c_{k+1} = \bar{n}+\frac{s}{\sqrt{m}}t_{\alpha/2}(m-1)$, where $\bar{n}$ is the average of $tw_k^c$ on $m$ days and $s$ is the standard deviation. Based on a t-test with significance level $\alpha$, there exists an influential event if $tw^{c}_{k+1}$ surpasses $u^c_{k+1}$. We assume that only one new event is dominant in each burst, which is reasonable for political campaigns.

Step~\uppercase\expandafter{\romannumeral2} is to estimate the event time window. The daily tweets about each candidate are first integrated into a single document; then, the terms in the document are weighted by the \emph{tf-idf} method(see \emph{SI}, Sect.~3.1). The 30 terms with the highest weights in the burst are selected as the typical words for that event. We then proceed to check the overlaps of typical words on the burst day plus or minus five days. The first day with non-zero overlap is deemed to be the start day of the event, and the last day with non-zero overlap is the closing day, which defines the event time window (see \emph{SI}, TABLE 9, TABLE 10, and TABLE 11). We remove suspicious events with a time window of one day.

Step~\uppercase\expandafter{\romannumeral3} is to measure the impact of events on public opinion. We denote the estimated $x_k^c$ initially transited from the previous day as $\hat{x}_{k|k-1}^c$ (see transition function in~\eqref{equ:kfe}) and the final $x_k^c$ calibrated with multiple signals as $\hat{x}_{k|k}^c$ (see mapping function in~\eqref{equ:kfe}). Intuitively, $\hat{x}_{k|k}^c$ has absorbed the information about all pertinent events on day $k$; hence, the change from $\hat{x}_{k|k-1}^c$ (equaling $\hat{x}_{k-1|k-1}^c$ for $f_k=1$ and $\mathbb{E}(q^c_k)=0$) to $\hat{x}_{k|k}^c$ indicates the impact of an event. To measure the significance of the impact, we apply the {\it event study model}~\cite{Binder1998The} from the field of finance as follows:
\vspace{-0.5cm}
\begin{equation}\label{equ:reg}
\hat{x}^c_{k|k}=a+\hat{x}^c_{k-1|k-1}+\sum^{J^c}_{j=1}\gamma^c_jD^c_{j,k}+\varepsilon,
\end{equation}
where $D^c_{j,k}$ is a dummy variable equal to 1 if day $k$ is within the time window of event $j$ for candidate $c$ and is equal to 0 otherwise. $J^c$ is the total number of events detected for candidate $c$, and $a$ is a regression constant. $\gamma^c_{j}$ is the estimator of the effect of event $j$ on candidate $c$, which passes the t-test if event $j$ has a significant effect on public opinion (see \emph{SI}, TABLE 12, TABLE 13, and TABLE 14). In this way, we can identify the events that actually influence the election.

\section{Results}
\subsection{Prediction Performance}
\begin{figure*}[htbp]
	\centering
	\subfigure{
		\begin{minipage}[t]{.6\linewidth}
			\label{fig:raw_ratio_tsai}
			\includegraphics[width=\linewidth]{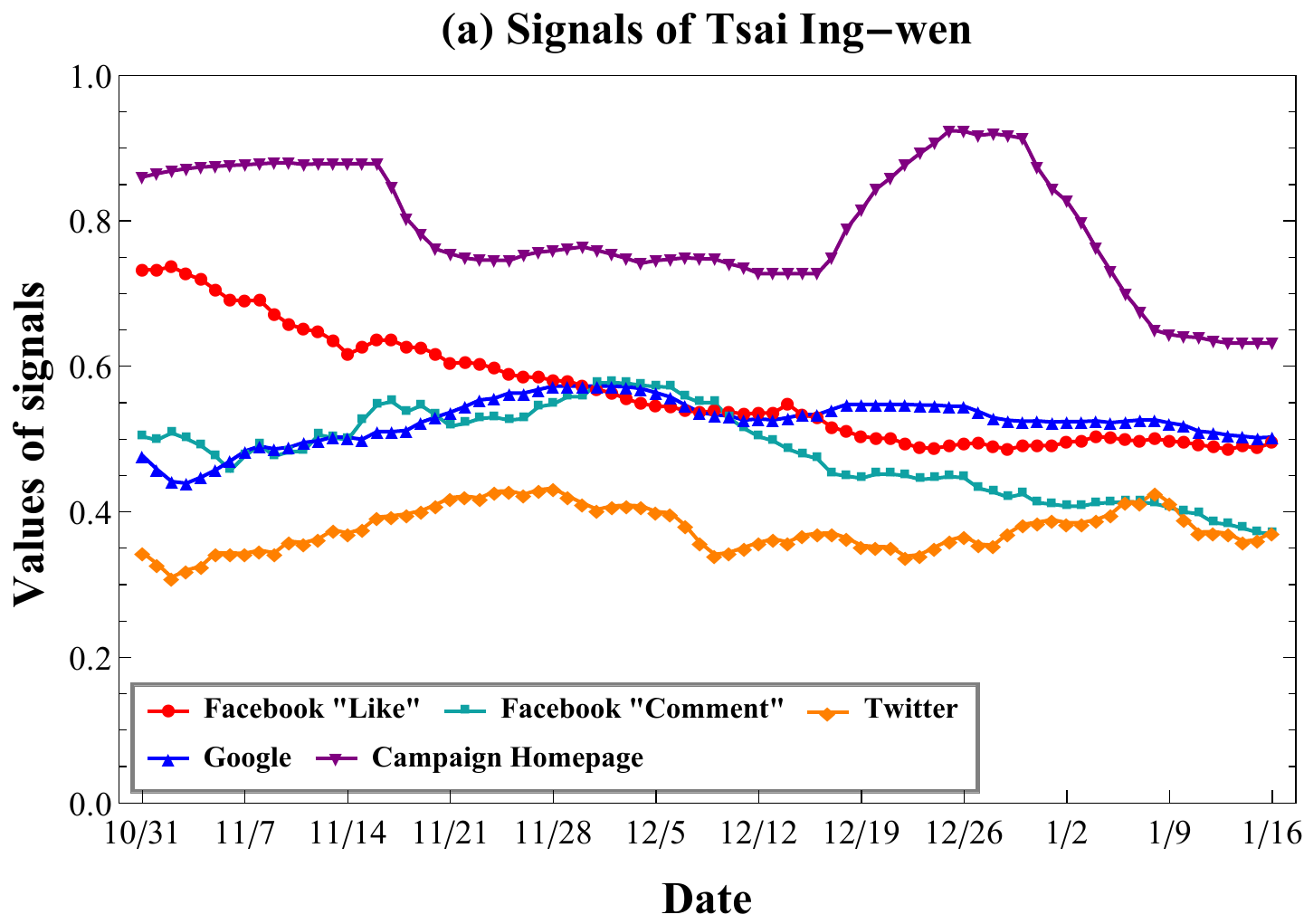}
		\end{minipage}}
		\subfigure{
			\begin{minipage}[t]{.6\linewidth}
                \label{fig:raw_ratio_llchu}
				\includegraphics[width=\linewidth]{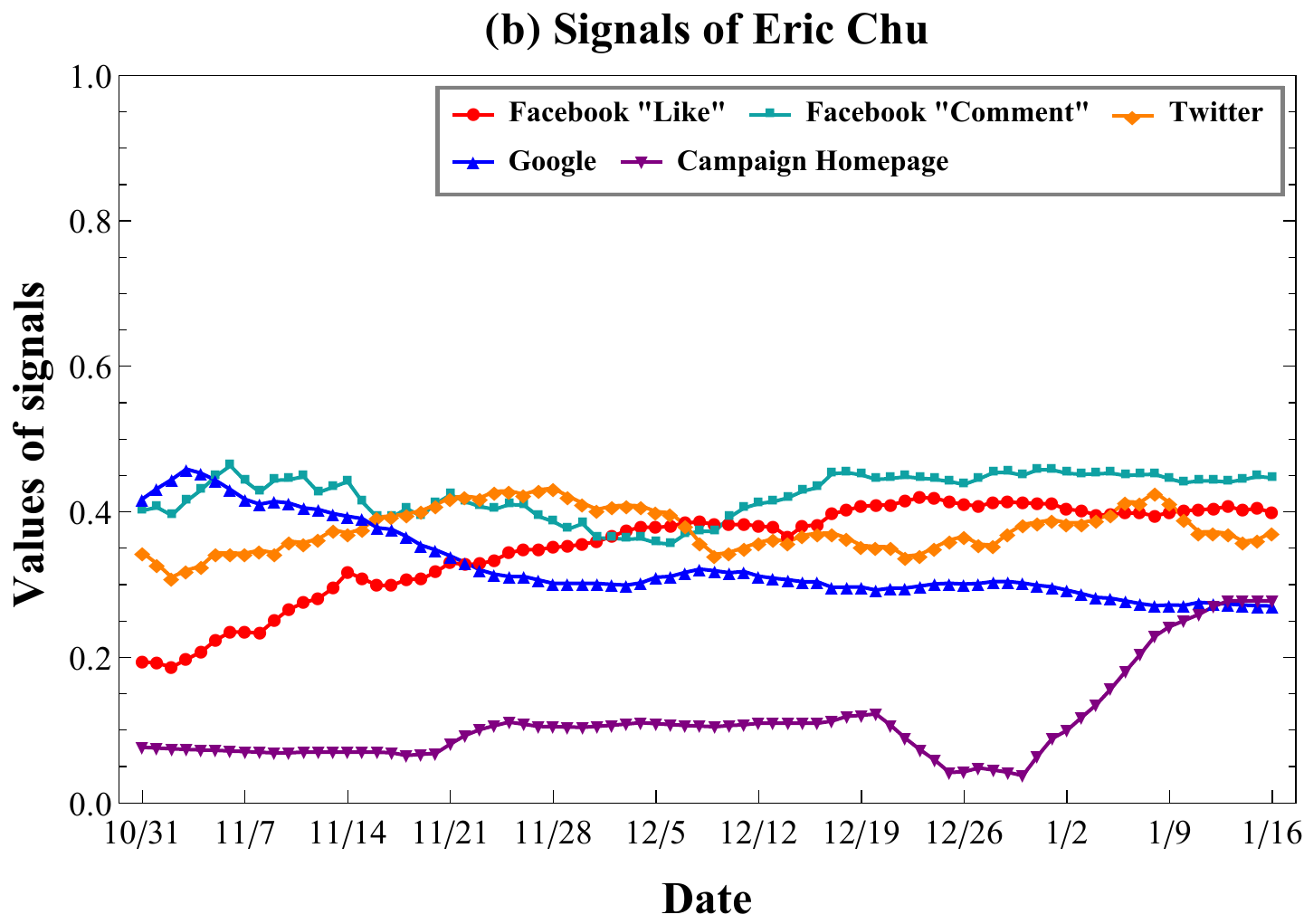}
			\end{minipage}}
			\subfigure{
				\begin{minipage}[t]{.6\linewidth}
                    \label{fig:raw_ratio_soong}
					\includegraphics[width=\linewidth]{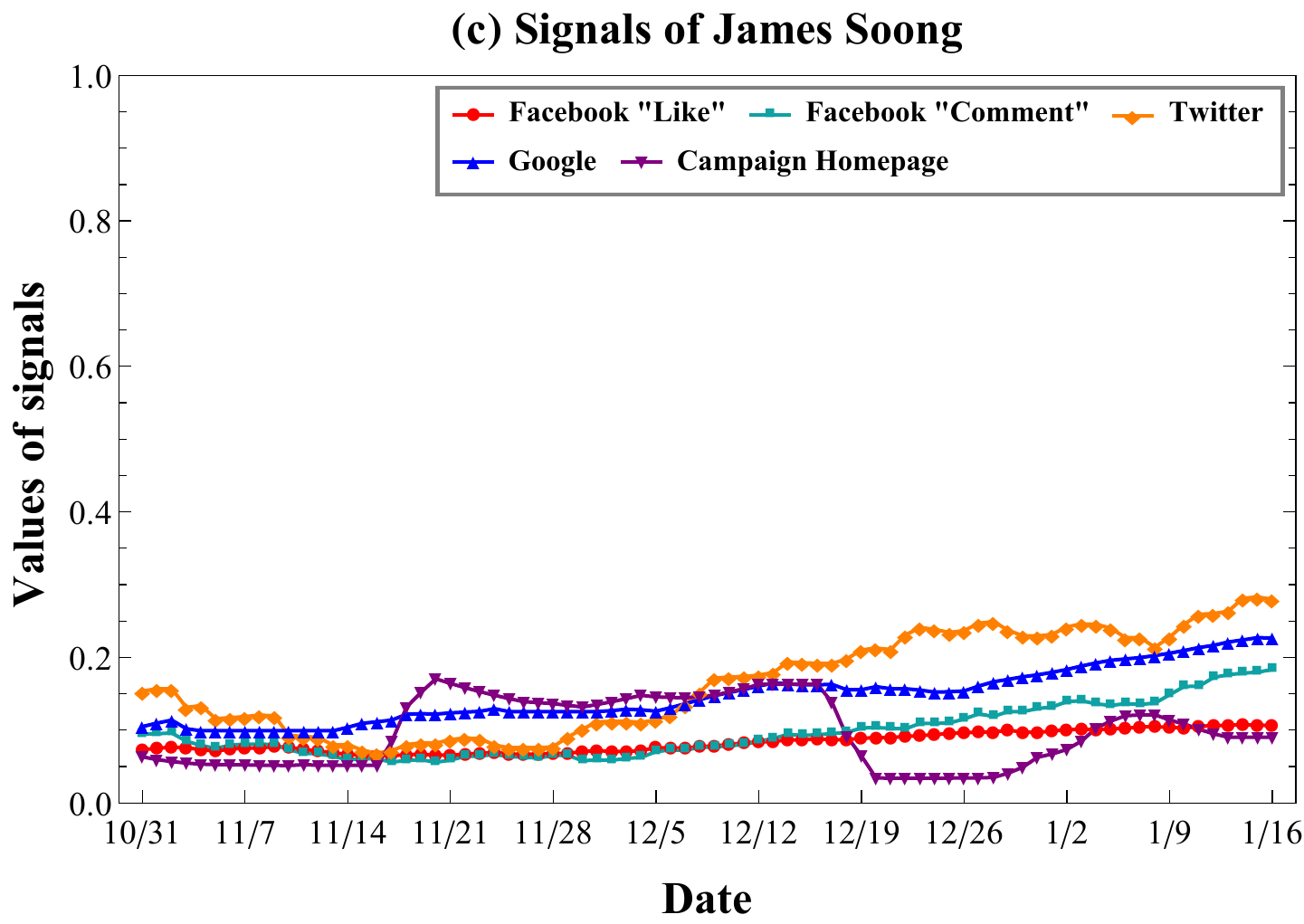}
				\end{minipage}}
				\caption{Signals of public opinions from all the online channels for the three candidates.}
				\label{fig:raw_ratio}
				\vspace{-0.2cm}
			\end{figure*}

Figs.~\ref{fig:raw_ratio}(a)-(c) show various online signals two months before election day. Intuitively, the user behavior in different channels is related to the public opinion towards a candidate, but the signals have vastly different volatilities. This justifies the value of information fusion for election prediction.

\begin{figure*}[t!]
	\centering
	\includegraphics[width=.6\linewidth]{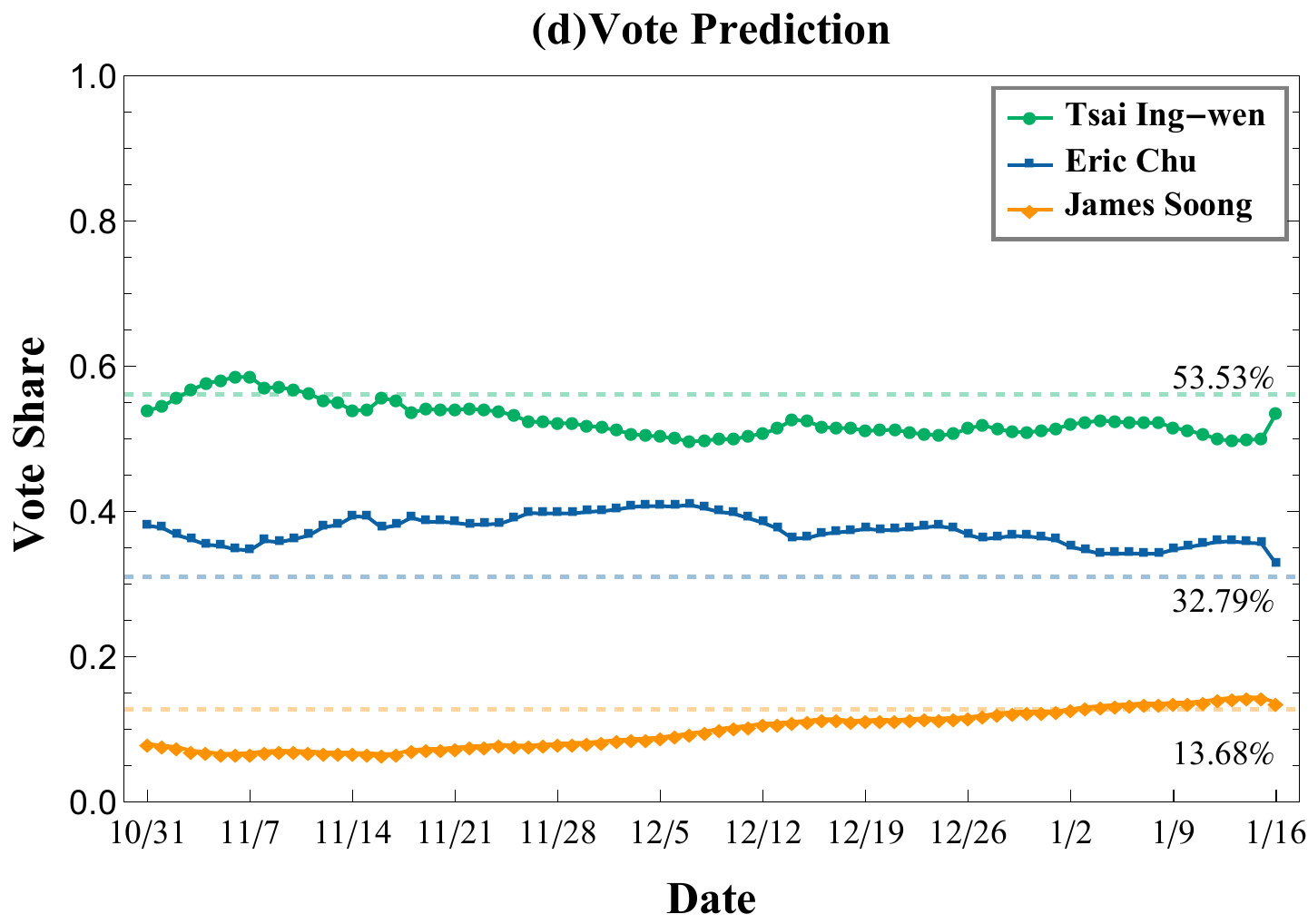}
	\vspace{-0.7cm}
	\caption{Time series of the vote share predictions. The dashed lines are the actual election outcomes. On election day, the errors are less than 2.59\%.}
	\label{fig:prediction}
\end{figure*}
Fig.~\ref{fig:prediction} depicts the dynamic vote predictions after fusing the four types of online signals, {\it i.e.}, $s^c_{k,FAL}$, $s^c_{k,TW}$, $s^c_{k,GO}$ and $s^c_{k,IP}$, by the Kalman filter. Although the four signals behave differently, the fused signal representing the predicted vote share for each candidate is relatively stable and exhibits a clear tendency, confirming the effectiveness of the prediction system for information aggregation. The final result is impressive --- while Tsai's win is easy to predict even in October, the prediction errors for every candidate one day, week, and month before the election day are no greater than 2.59\%, 4.58\% and 5.87\%, respectively.

\begin{figure}[t!]
  \centering
  \includegraphics[width=0.9\linewidth]{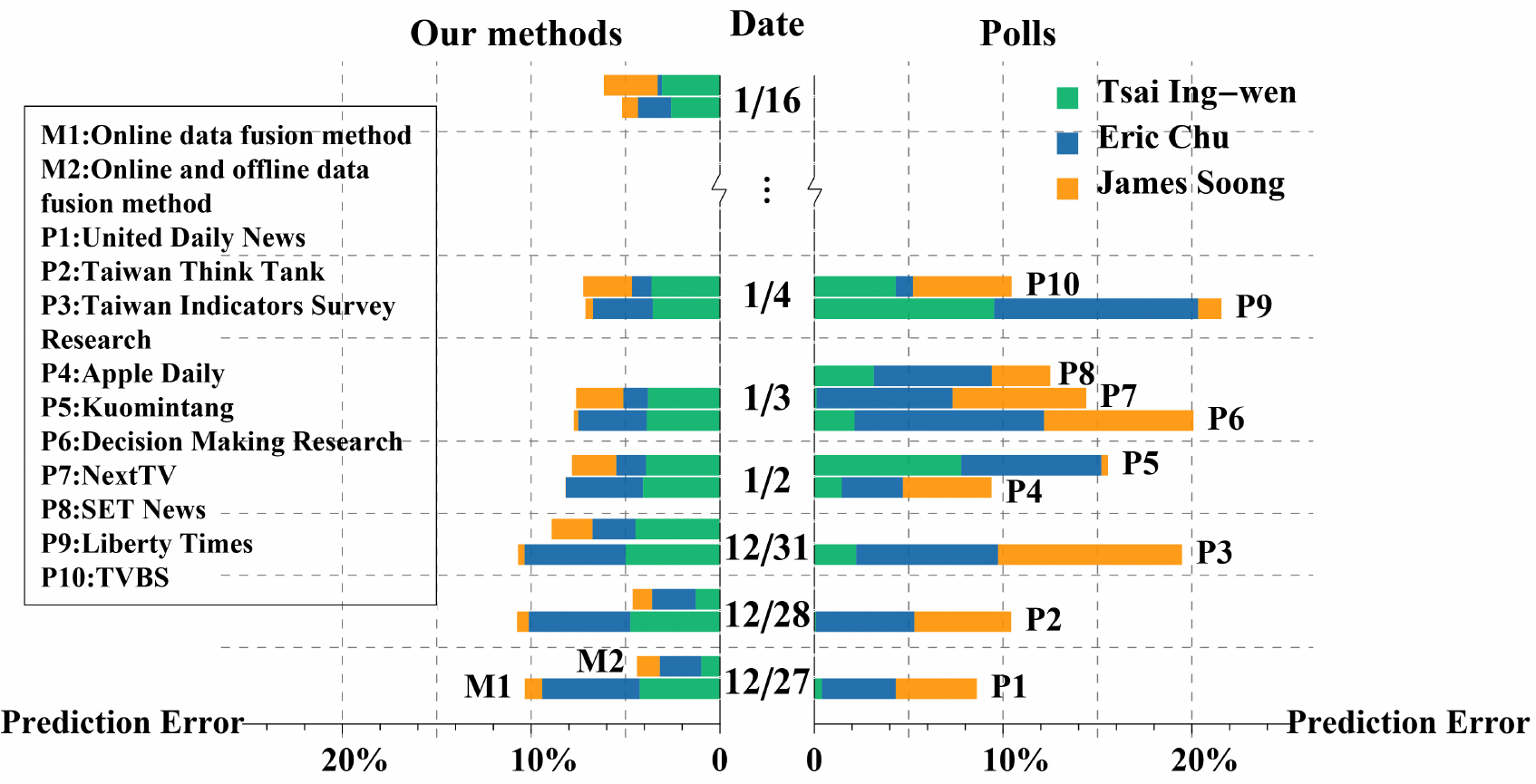}
  \caption{Timeline of the absolute prediction errors of final polls and data fusion methods. The bars on the left side of the timeline represent the prediction errors of the data fusion methods. In each interval between two gray dashed lines, there are two bars. The lower bar represents the absolute error of the online data fusion method, and the upper bar represents the absolute error of the online-offline data fusion method. The interval between two gray horizontal dashed lines indicates one day. The bars on the right side of the timeline show the prediction errors of the final polls from ten pollsters.  Comparison of the bars on both sides shows that the absolute prediction errors of the signal fusion methods are smaller than those of the polls.}
  \label{fig:prederr}
  \vspace{-0.5cm}
\end{figure}

To further justify the predictive power of online signals, we also compare our results with offline polls. As shown in Fig.~\ref{fig:prederr}, during the last two weeks of the election, our predictions (M1) outperform most of the pollsters (P1-P10) greatly, and can improve continuously by absorbing up-to-date information. This is possibly due to the fact that the anonymity of the Internet enables individuals to express their opinions freely and voluntarily, which could reduce the bias relative to that in the tele-interview setting of a traditional poll. Furthermore, currently, news usually breaks online first and then spreads at a tremendously fast pace from online to offline via physical social networks. Therefore, online information can also influence offline voting blocs during campaigns, which mitigates the bias effect of using only the netizen population in our method.

We also try to reduce the sample bias by mixing the prediction results from online signals with those from offline pollsters in older groups (see \emph{SI}, Sect.~2.4). As shown in Fig.~\ref{fig:prederr}, the online-offline data fusion method (M2) indeed outperforms the online data fusion method (M1) in the early stage of the final two weeks, which indicates the power of sample bias correction. But the advantage disappears gradually as the final election day approaches, which again exposes the drawback of offline polls in responding to newly emerging information.

\subsection{Signal Evaluation}
\begin{figure}[t!]
	\centering
	\includegraphics[width=0.8\linewidth]{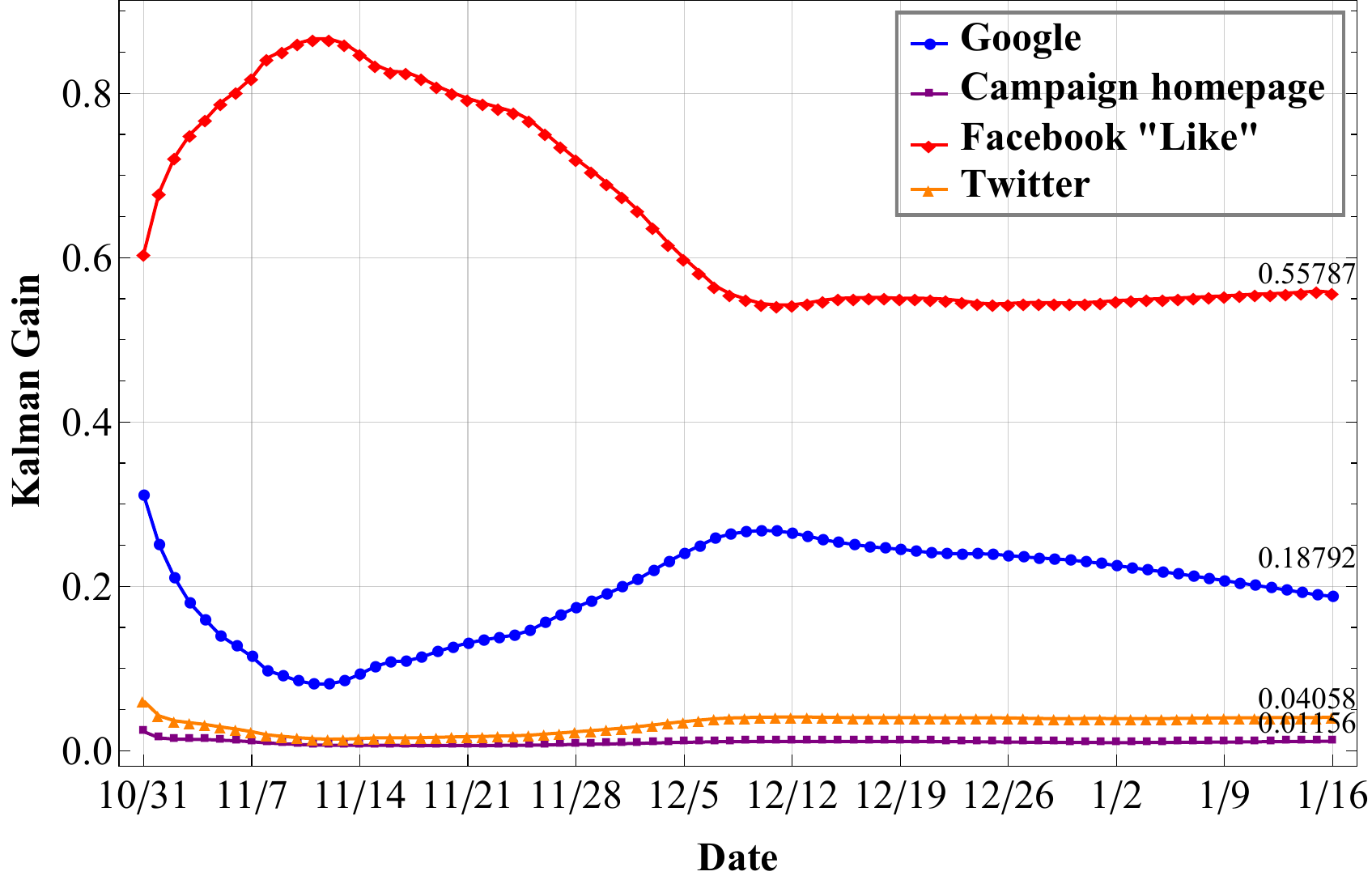}
	\caption{Kalman gain for different online signals. The Kalman gain of the Facebook "Like" constitutes the highest proportion but gradually decreases while the Google signal continuously increases. Twitter and the campaign homepages are not indicative signals for the election. The Kalman gains of all the signals converge to steady states approximately one month before the election.}
	\label{fig:gain}
	\vspace{-0.3cm}
\end{figure}

We also explore the predictive power of various online signals via their daily Kalman gains. As shown in Fig.~\ref{fig:gain}, Facebook ``Likes'' are consistently the strongest indicator among all the signals. This demonstrates the power of social media in collecting public opinions via a simple mechanism, although it is vulnerable to shilling attacks. The predictive power of the Google index appears to be time-sensitive, contributing less initially and becoming the second best indicator one month before the election. One possible explanation is that the election might not be a focal topic in the early stage of the campaign, making Google searches rather random. However, as the election day approaches, the campaign becomes the central topic and drives the public to search for information about the candidates. The two remaining signals, {\it i.e.}, tweet volumes and homepage traffic, appear to be of much weaker predictive value, which may be due to their lack of popularity in Taiwan (see \emph{SI}, TABLE 1) and diverse attitudes about candidates.

\begin{figure*}[htbp]
	\centering
    \includegraphics[width=\linewidth]{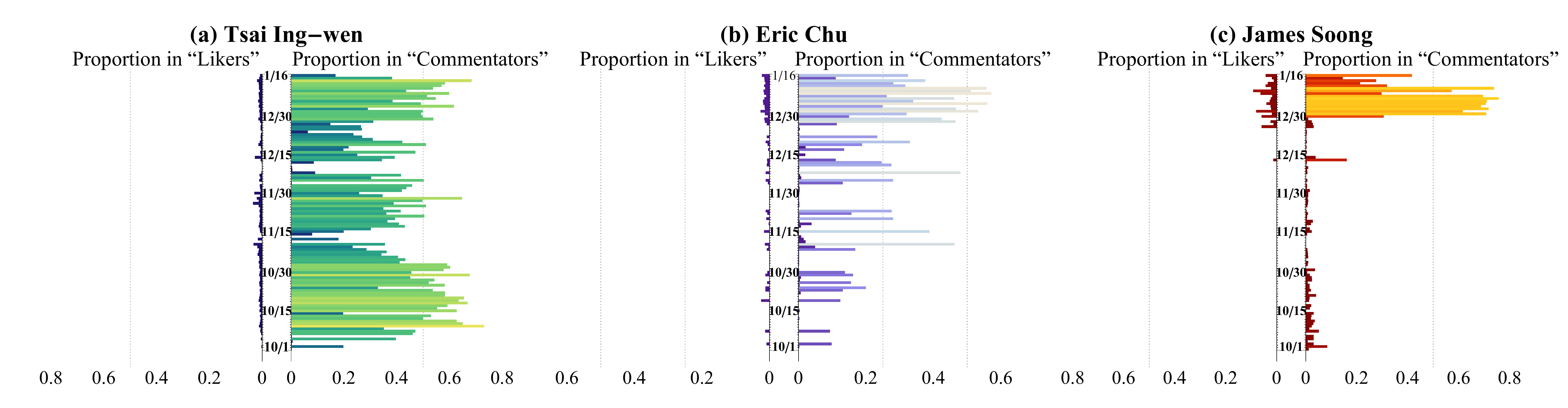}
    \vspace{-0.4cm}
				\caption{Proportions of overlapping users who have both liked and commented on the candidates' posts among "Likers" and "Commentators".The intermediate vertical axis is a timeline covering the whole period of the election. The bar on the left side of the timeline represents the daily proportion of overlapping users to users who have ever liked. The bar on the right side of the timeline represents the daily proportion of overlapping users to users who have ever commented. The number of overlapping users accounts for less than 1\% of all the users who have ``liked'' on average, with the maximum proportions being 3.51\%, 3.74\%, and 9.25\% for the three candidates. By contrast, the overlapping users constitute more than 37.16\%, 14.90\%, and 12.03\% of all users who have commented, on average, for the three candidates, and the maximum ratios are 73.05\%, 59.75\%, and 83.01\%. }
				\label{fig:deu}
				\vspace{-0.2cm}
			\end{figure*}

We further explore the distinct value of the ``Like'' function on Facebook. We compare it with the ``Comment'' function by substituting $s^c_{k,FAL}$ with $s^c_{k,FAC}$ in the Kalman filter. The results indicate that the prediction outcomes become significantly worse ---  the one-day-earlier prediction errors for Tsai and  Chu increase to 5.42\% and 4.86\%, respectively(see \emph{SI},Sect.~2.5). These results indicate the superiority of ``Like''  over ``Comment''. To understand this result, we search for the population of Facebook users who have ever liked or commented on the candidates and obtain the overlapping users who have both liked and commented on a candidate. Fig.~\ref{fig:deu} shows that these users constitute only a small proportion of the ``Like'' users but a much larger proportion of the ``Comment'' ones. Therefore, a considerable proportion of users who have commented on a post may also choose to like the post but not vice versa. In other words, the ``Like'' signal represents the positive attitude of a much larger population than that of the ``Comment'' signal, which may be attributed to the fact that a ``Like'' is a more direct and widely engaged in behavior for online users to express their positive opinions without great effort. Another disadvantage of ``Comment'' lies in its diversity of expression, which can be a blend of contradictory attitudes, including support, praise, opposition and even insult (see {\em SI}, Sect.~2.6).


The overlapping users indeed constitute a group of firm supporters for each candidate who show their support by not only clicking ``Like'' but also going through the effort to publish comments. By further tracking the changes in the overlap ratios during the election, as shown in Fig.~\ref{fig:deu}, we find that the ratio for Tsai is relatively stable, indicating that Tsai has a firm group of supporters regardless of her behavior during the campaign. By contrast, for Chu and Soong, the overlap ratios remain small until election day approaches, suggesting Tsai should partially attribute her success to her firm supporters rather than swing voters. This also explains why we can predict the victory of Tsai two months before election day.

\begin{figure*}[t!]
    \includegraphics[width=1\linewidth]{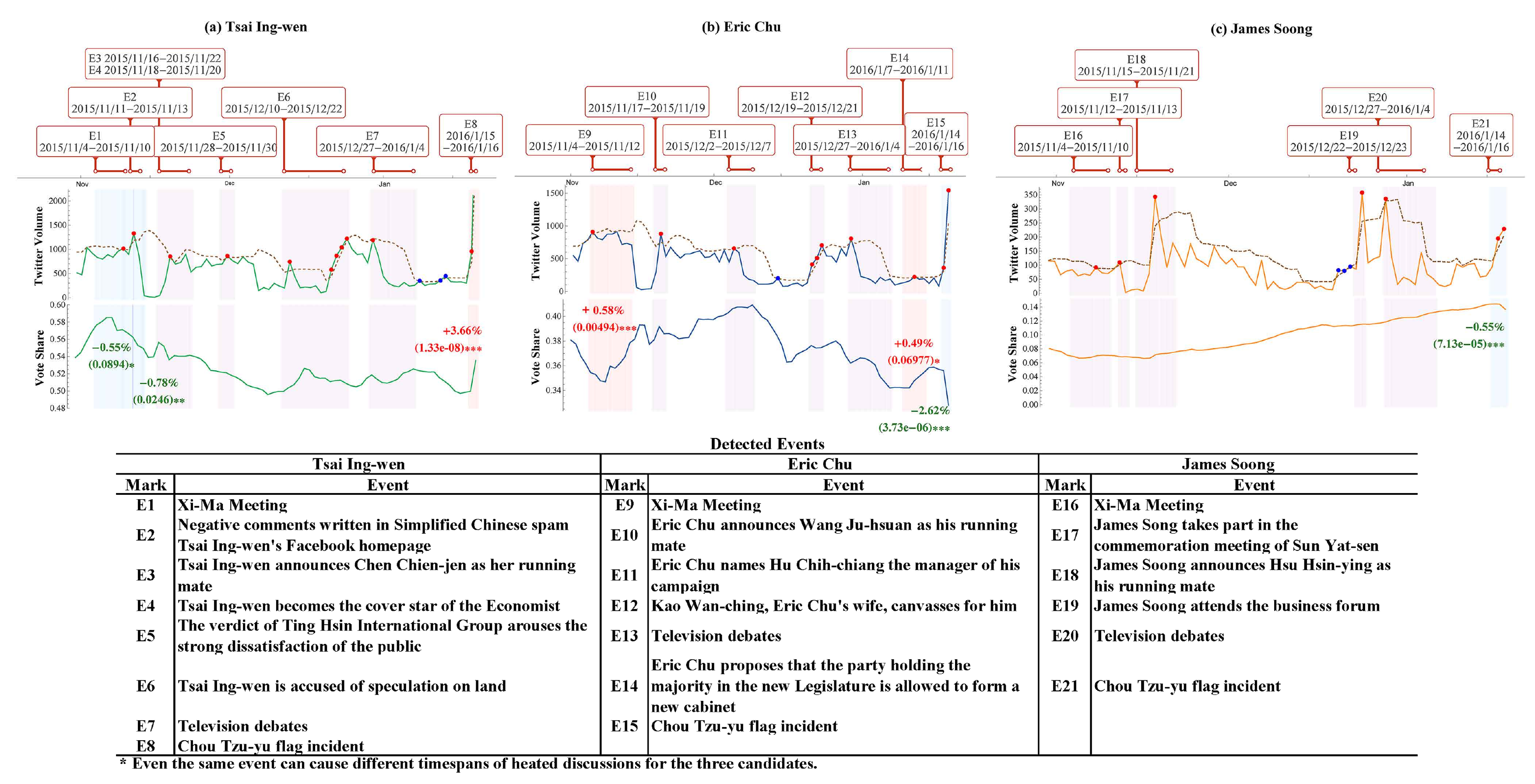}
    \vspace{-1.5cm}
	\caption{Detected events and their influence on potential vote shares. The numbers of detected events during the elections are 8, 7, and 6, respectively, for the Tsai, Chu, and Soong. Each event spans a time window and influences the potential voting rates differently, as denoted by different colors. A light purple bar indicates that the detected event does not have a significant influence on the vote shares. A red bar indicates a positive effect of the event on vote shares, and a blue bar represents a negative effect of the event. The influence of each significant event is marked on the curve, and the number below it in brackets is the p-value of the t-test. The typical words used to determine the event timespan are detailed in \emph{SI},TABLE 9, TABLE 10, and TABLE 11. In addition, the Twitter bursty days detected in Step~\uppercase\expandafter{\romannumeral1} are noted in the Twitter volume time series. The red points represent events with a timespan longer than one day, which are fed into Step~\uppercase\expandafter{\romannumeral2} for further analysis. The blue points are removed.}
	\label{fig:events}
	\vspace{-0.5cm}
\end{figure*}

\subsection{Influential Events}
We apply the event detection method to each candidate's Twitter data to identify influential events. Fig.~\ref{fig:events} shows the results and event descriptions. The most influential events detected with $p$-values less than 0.05 include the meeting between Xi Jinping and Ma Ying-jeou (Xi-Ma Meeting), the emergence of negative comments on Tsai Ing-wen's Facebook homepage possibly by users from mainland China, and the Chou Tzu-yu flag incident. All these events share a common feature; that is, they all belong to the category of \emph{cross-strait relation}, which is always subtle and controversial in Taiwan's political circle. Other seemingly important events from the perspective of the election campaign, such as the TV broadcast of the candidates' debates and various types of electioneering activities in local areas, have insignificant influences on public opinion.

We further assess the influence level of the events, which is measured by the coefficient $\gamma^c_j$ in \eqref{equ:reg}. The Xi-Ma Meeting resulted in a 0.55\% decrease in the vote share of Tsai Ing-wen. This result is not surprising because Tsai was believed to favor Taiwan independence over the ``One China Policy'', and the meeting thus prompted the public to doubt Tsai's ability to handle cross-strait relations. This same event increased Eric Chu's vote share by 0.58\% because he was thought to be more able to develop cross-strait peace after the meeting. 

Despite the abundance of events during the campaign, the \emph{Chou Tzu-yu flag incident} from the entertainment domain is the most influential. Chou Tzu-yu, a 16-year-old Taiwan singer, sparked huge controversy in social media for showing the Taiwan flag as the national flag of China. As the uproar intensified online, Chou's company released a video in which Chou apologized for her behavior by stating that ``there is only one China'' and identifying herself as Chinese. The most subtle point is that the video was released the day before the election, which was described as a humiliation to Taiwan and spread quickly in Taiwan's online social media. As a consequence, this incident increase the vote share of Tsai Ing-wen by approximately 3.66\% and lowered the vote share of Eric Chu by approximately 2.62\%.

\section{Discussion}
The accurate prediction of Taiwan's 2016 Presidential Election suggests an interesting viewpoint that public opinions towards political campaigns can be determined via online user-generated content. This indeed coincides with some recent studies reporting that social media such as Facebook~\cite{effing2011social,enli2013personalized,williams2008social,macwilliams2015forecasting,franch2013wisdom}, Twitter~\cite{steinert2015online,metaxas2012social,effing2011social,enli2013personalized,franch2013wisdom,graham2013between,digrazia2013more,burnap2015140,tumasjan2010predicting,sang2012predicting,song2014analyzing} and Youtube~\cite{effing2011social} are able to aggregate public opinions about political matters. Donald Trump winning the 2016 US Presidential Election was also considered to be a victory for the heavy use of social media such as Twitter~\cite{Yaqub2017Analysis}. Nevertheless, this finding remains controversial in academia, and the above studies have often been criticized for the unreliability of single-source information~\cite{You2015A} and/or the unrepresentativeness of online user populations~\cite{Gayo2011Limits}. Our study attempts to address these concerns.

First, we introduce multiple online channels as different types of signals to produce more robust predictions. These signals, while reflecting more or less latent public opinions, have varied fluctuations due to their different sensitivities to campaign dynamics and possible fake responses from the Internet ``water army'' (see Fig.~\ref{fig:raw_ratio}). The fusion of these signals can help to filter out some noise by consensus learning to highlight the tendencies. Moreover, although one signal might contribute more to some specific election prediction, such as the Facebook ``Like'' for the Taiwan election, it is unlikely to find it omnipotent for different elections. The fusion of these signals could help to mitigate the risk of selection bias. This information fusion scheme gives our study some important extensibility --- the four channels, namely, Facebook, Twitter, Google Trends and campaign homepages, could be considered to be the fundamental and preemptive online information sources for different elections.

We also find that although selection bias of the online voting population exists, its influence on the prediction results is limited. Prediction based on pure online information is much more accurate than the polls released by Taiwan's mainstream pollsters (see Fig.~\ref{fig:prederr}).
The reason behind this may be two-fold.  On one hand, online users who pay close attention to election campaigns likely become active voters and constitute a large voting population on election day~\cite{Gopoian1994Late,Henderson2016Changing}. On the other hand, we should not underestimate the information exchange between online social networks and offline physical networks~\cite{Bond2012A,Kramer2014Experimental}. Older people who seldom interact with the Internet still have access to online information via ordinary family communications or traditional media's reports on Internet opinions. This communication contributes to the opinion conformance across online and offline networks and further improves the representativeness of the online voting population. In fact, compared with traditional polls, which are susceptible to questionnaire wording~\cite{Bryan2011Motivating}, reporting error~\cite{Rogers2016Unacquainted}, ballot order~\cite{Wang2014Context}, and social desirability bias~\cite{Rogers2016Unacquainted,Rand2009Dynamic}, online big data enables a much larger sample and thus can improve the sample resistance to human manipulation. The real-time availability of online data, which enables the timely update of predictions based on continuously incoming information, is another major advantage relative to polls.

Our study also suggests that the Kalman filter with the event detection model (see \emph{Materials and Methods}) could be packaged as a fundamental kit for political vote analytics. Specifically, the Kalman filter is responsible for the dynamic prediction of vote shares given multi-source time-varying signals and multiple candidates. Meanwhile, the event detection model is responsible for the automatic identification of influential events during the campaign, which provides a causal explanation for the predictions. In other words, the two models together could provide {\em interpretable predictions} to political vote analytics, which is deemed particularly valuable for a big-data-driven research paradigm~\cite{Hofman2017Prediction}.

The Kalman filter has been adopted in previous studies but either for backward review given the final result or for forward prediction given multiple historical elections data. Our study shows that while we cannot obtain the true vote shares until election day, we can still fine-tune the model parameters by using up-to-date time series signal data for the current election, which solves the problems in leveraging the Kalman filter for election prediction. Moreover, given the sum-to-one constraint in a statistical learning framework (see \eqref{equ:errors}), the Kalman filter is capable of building models for more than two election candidates. One may consider the inclusion of some other relatively stable factors, such as the globalization trend, economic status, the technology environment, {\it etc.}, in the prediction model, which can be achieved by setting appropriate initial values of the Kalman filter. Nevertheless, our study shows that the Kalman filter is insensitive to the initial values as long as the prediction is based on a sufficiently long time series (see {\em SI},Sect~2.2). In this case, the signals should have fully ``absorbed'' the influences of the macro factors.

Our study provides some political insight into the Taiwan presidential election. It is interesting that the simple ``Like'' function on Facebook collects the public opinions about candidates (see {\em Signals Evaluation} in {\em Results}), although it has been reported to be vulnerable to shilling attacks in electronic commerce~\cite{De2014Paying}. The ``Like'' function is more beneficial than the ``Comment'' function, although the latter actually expresses more complex sentiments and richer opinions. This difference is attributes to the widespread use of Facebook in Taiwan (see {\em SI}, TABLE 1) and the easy-to-use characteristic and emotional unambiguity of the ``Like'' function. Another interesting finding is that the most influential events during the Taiwan election campaign are all closely related to cross-strait relations (see {\em Influential Events} in {\em Results}). In particular, in line with the findings in~\cite{Gerber2011How}, the events more closely associated with public sentiment (such as the Chou Tzu-yu flag incident) appear to have a greater impact than those with merely political meaning (such as the Xi-Ma Meeting).

We provide accurate prediction and automatic causal analysis of the 2016 Taiwan Presidential Election, which illustrates the feasibility of applying a data-driven paradigm for political vote analytics. Although our focus is on Taiwan, the proposed signal fusion approach and the event detection model can be applied to other elections or referendums, especially those using majority rule. Considering the different Internet applications used across countries and areas, we may need to adjust the input online information sources and design new measurements for the new signals. Furthermore, we should consider how the election systems of particular countries or areas differ and require adjustment of the prediction model. For example, the US election system is not a direct election but relies on the Electoral College system with 538 electoral votes. Hence, we have to incorporate information about the states and locations of online users into the prediction. However, this information is often unavailable. Nevertheless, we can still consider online users as the voters for a ``virtual'' direct election and obtain the predictive results as the popular votes for the candidates, which could still indicate the winner if there is a large difference in vote share among candidates. The recent 2016 US Presidential Election demonstrates the power of voices on social media.

\section*{References}
\bibliographystyle{naturemag}

\clearpage
\section*{Supplemental Information}
\setcounter{section}{0}
\section{Data}
\label{sub:data}
To measure the public opinions of the three candidates, we collected offline data from pollsters and online data from social media, search engines, and campaign homepages.
\subsection{Offline Data}
We collected a set of polls published by nineteen pollsters from 1 August 2015 to 16 January 2016. The pollsters include Decision Making Research, Kuomintang, Apple Daily, TVBS, Taiwan Indicators Survey Research (TISR), Liberty Times, SET News, Want Want China Times, Taiwan Think Tank, ETtoday, Shih Hsin University, Apollo Survey \& Research, Now News, Taiwan Competitiveness Forum, My-formosa, Business Weekly, United Daily News, NextTV, and Now News. Among the polls from October 2015, the results of TISR are closest to the average of all the polls not done by political orientations. 
In addition, TISR, which is conducted every 15 days using random-digit telephone surveys, is the most frequently updated poll during the election. Thus, we select TISR as the representative of the polls. In each survey, in addition to the overall vote preference of the candidates, the polls investigate the opinions with regards to six population age groups  (\emph{i.e.}, the people aged between 20 and 30, 30 and 40, 40 and 50, 50 and 60, 60 and 70, above 70 years). In this study, we collect the support rate in each age group from the polls published by TISR\footnote[1]{Taiwan Indicators Survey Research(TISR), \protect\url{http://www.tisr.com.tw/?page_id=700}}.
 \subsection{Online Data}
Generally, public opinions are concentrated on popular social media sites. To identify the most popular websites in Taiwan, we referred to professional Internet surveys (\emph{i.e.}, Internet Usage in Taiwan: Summary Report of October 2015 Survey~\footnote[2]{Internet Usage in Taiwan: Summary Report of October 2015 Survey, \protect\url{http://www.twnic.net.tw/download/200307/20150901e.pdf}\label{fn:twsurvey}})
and web traffic reports (\emph{i.e.}, Alexa, comScore and Digital Age). The top websites are listed in TABLE~\ref{tab:popnet}. We selected several sites primarily composed of user-generated content and classified them into three categories: social network (\emph{i.e.}, Facebook and Twitter), search engine (\emph{i.e.}, Google), and campaign homepages. We extracted the daily measurements from the above platforms as signals of collective opinions during the period of October 1, 2015 to January 16, 2016. To weaken the effects of measurement loss and violent fluctuations on prediction and to capture the trends, we apply a moving average to the measurements.
 The number of moving days used in the following analysis is 30.

\emph{Social Network.}
 Facebook, which provides an easy way for candidates to reach out to a large audience, is the most popular social service platform in Taiwan. For each published post, users can "Like" or "Comment" on it to indicate their attitudes towards the candidate. In our study, we use Facebook API to retrieve all the posts published by the three candidates during the election period, the corresponding timestamp, the number of \emph{Likes}, the id list of users that pressed the "Like" button, the number of "Comments", the "Comment" content and user id, etc.

\begin{CJK}{UTF8}{gkai}
\begin{center}
\tablecaption{Top 10 websites in Taiwan\label{tab:popnet}}
\tablefirsthead{
\hline
Rank & Alexa & comScore & Digital Age \\}
\tabletail{
\hline}
\begin{supertabular}{p{.25\textwidth}<{\centering}|p{.25\textwidth}<{\centering}|p{.25\textwidth}<{\centering}|p{.25\textwidth}<{\centering}}
 \hline
 1 & Yahoo & Google & Facebook\\
 2 & Facebook & Yahoo & YouTube\\
 3 & Google & Facebook & Yahoo\\
 4 & Pixnet & YouTube & Google\\
 5 & ETtoday & Pixnet & Chinatimes\\
 6 & YouTube & UDN & Ruten\\
 7 & Giga Circle & Wikipedia & UDN\\
 8 & UDN & Xuite & Gamer \\
 9 & Xuite & Apple Daily & Mobile01\\
 10 & LIFE & Yam & Apple Daily\\
 \hline
 \end{supertabular}
 \begin{tablenotes}
\scriptsize
\item[1]{Source:

1. Alexa, \url{http://www.alexa.com/topsites/countries/TW};

2. comScore, \url{http://www.comscore.com/Products/Audience-Analytics/Media-Metrix-Multi-Platform};

3. Digital Age, \url{https://www.bnext.com.tw/article/35475/BN-ARTICLE-35475}.}
\end{tablenotes}
\end{center}
\end{CJK}

To reflect the candidates' popularity in terms of Facebook "Like", we calculate the daily average number of "Like" per post for each candidate $s^c_{k, FAL}$ as follows,
\begin{equation}
s^c_{k, FAL}=\frac{1}{m}\sum^{m-1}_{j=0}\frac{\sum_i{like^c_{k-j,i}}/n^c_{k-j,FA}}{\sum_c{\sum_i{like^c_{k-j,i}}/n^c_{k-j,FA}}},
\end{equation}
where $like^c_{k-j,i}$ is the number of \emph{Likes} of the $i-th$ post published by candidate $c$ on day $k-j$, and $n^c_{k-j,FA}$ is the number of posts published by candidate $c$ on day $k-j$.
To compare "Like" with "Comment"
 on Facebook, we also extract daily average number of "Comments" per post for each candidate as an indicator.
\begin{equation}
s^c_{k, FAC}=\frac{1}{m}\sum^{m-1}_{j=0}\frac{\sum_i{Comment^c_{k-j,i}}/n^c_{k-j,FA}}{\sum_c{\sum_i{Comment^c_{k-j,i}}/n^c_{k-j,FA}}},
\end{equation}
where $Comment^c_{k-j,i}$ is the number of comments on the $i-th$ post published by candidate $c$ on day $k-j$.

Some recent reports~\cite{tumasjan2010predicting, digrazia2013more, o2010tweets, mejova2013gop} have suggested that there is a statistically significant correlation between Twitter and election outcomes in terms of volume and sentiment. Thus, although Twitter is not as popular as Facebook in Taiwan, we select it as a measurement of voting preference. By querying the Twitter API, we obtained 283412 candidate-related Twitter messages posted from 1 October 2015 to 16 January 2016 in real time. To focus on Taiwan public opinions, we only use party names, the three candidates' names and their morphs in Simplified and Traditional Chinese as keywords to retrieve tweets. The keyword list is presented in TABLE~\ref{tab:kwt}. By analyzing Twitter sentiment, we find that more than 80\% of the retrieved tweets are news and do not represent public opinions.
Thus, we use $s^c_{k, TW}$ to represent the number of tweets mentioning a specific candidate,
\begin{equation}
s^c_{k, TW}=\frac{1}{m}\sum^{m-1}_{j=0}\frac{tw^c_{k-j}}{\sum_{c}tw^c_{k-j}},
\end{equation}
where $tw^c_{k-j}$ is the Twitter volume about candidate $c$ on day $k-j$.

\emph{Search Engine.} Search queries usually reflect information that users hope to obtain, which can be regarded as an indicator of their opinions and preferences. Thus, we also investigate Google, the most used search engine in Taiwan. We obtained the search data from Google Trends~\footnote[3]{Google Trends, \protect\url{https://trends.google.com/trends/}}, which provides the time series of the search index for given words. Google Trends not only allows users to view the search index in a specific region for given words but also to compare the search frequencies of multiple keywords. In our study, we take the candidates' names in Simplified and Traditional Chinese as keywords and further restrict the search source to Taiwan. Then, we calculate the ratio of search indexes $s^c_{k-j, GO}$ as a signal reflecting the vote share for candidate $c$,
\begin{equation}
s^c_{k,GO} = \frac{1}{m}\sum^{m-1}_{j=0} \frac{search\_volume^c_{k-j}}{\sum_{c}search\_volume^c_{k-j}},
\end{equation}
where $search\_volume^c_{k-j}$ is the search index of keywords about candidate $c$ on day $k-j$.

\emph{Campaign Homepages.} The three candidates set up campaign websites\footnote[4]{Tsai Ing-wen,\protect\url{http://iing.tw/};\\ Eric Chu, \protect\url{http://www.doer.tw/};\\ James Soong, \protect\url{http://www.pfp.org.tw/}} to promote themselves and raise funds. The IP traffic of candidates' homepages can be used to determine their popularity. We collect the daily traffic data and take the proportion $s^c_{k-j, IP}$ as a measurement of opinions about candidate $c$: 
\begin{equation}
s^c_{k,IP} = \frac{1}{m}\sum^{m-1}_{j=0}\frac{IP^c_{k-j}}{\sum_{c}IP^c_{k-j}},
\end{equation}
where $IP^c_{k-j}$ is the IP traffic of candidate $c$'s campaign homepage on day $k-j$. The traffic data are from Alexa~\footnote[5]{Alexa, \protect\url{http://www.alexa.com/}}, which is a subsidiary of Amazon that provides commercial web traffic data for given sites.

\begin{CJK}{UTF8}{gkai}
\begin{center}
\tablecaption{Keyword list used to retrieve tweets in Twitter\label{tab:kwt}}
\tablefirsthead{
\hline
Candidate & Keywords \\}
\tabletail{
\hline}
\begin{supertabular}{p{.15\textwidth}<{\centering}|p{.85\textwidth}<{\centering}}
\hline
   Tsai Ing-wen& 蔡英文 (Tsai Ing-wen, candidate's name in Simplified Chinese), 空心菜 (a
nickname of Tsai Ing-wen), 小英 (a nickname of Tsai Ing-wen), 陈建仁 (Tsai Ing-
wen's vice presidential candidate), 民进党 (Democratic Progressive Party, the party
name in Simplified Chinese), \begin{CJK}{UTF8}{bsmi} 民進黨 (the party name in
Traditional Chinese)\end{CJK}, 绿营 (a nickname of the party), \begin{CJK}{UTF8}
{bsmi}綠營 (a nickname of the party)\end{CJK}\\
  \hline
   Eric Chu & 朱立伦(Eric Chu, candidate's name in Simplified Chinese), \begin{CJK}
{UTF8}{bsmi}朱立倫 (candidate's name in Traditional Chinese)\end{CJK}, 阿伦 (a nickname of
Eric Chu), 朱市长 (a nickname of Eric Chu), 王如玄 (Eric Chu's vice presidential
candidate), 国民党 (Kuomintang, the party name in Simplified Chinese), \begin
{CJK}{UTF8}{bsmi}國民黨 (the party name in Traditional Chinese)\end{CJK}, 蓝营 (a
nickname of the party), \begin{CJK}{UTF8}{bsmi} 藍營 (a nickname of the
party)\end{CJK}\\
   \hline
   James Soong & 宋楚瑜 (James Soong, candidate's name in Simplified Chinese), 宋爷爷 (a
nickname of James Soong), 宋省长 (a nickname of James Soong), 徐欣莹 (James Soong's
vice presidential candidate), 亲民党 (People First Party, the party name in
Simplified Chinese)，\begin{CJK}{UTF8}{bsmi}親民黨 (the party name in Traditional
Chinese)\end{CJK}\\
   \hline
\end{supertabular}
\end{center}
\end{CJK}

\clearpage
\section{vote Prediction Model and Results}
\subsection{Kalman Filter}
\label{sub:kf}
The core of the Kalman filter is defined by the following set of equations:
\begin{equation}\label{equ:obs}
\mathbf{s}^c_k = \mathbf{h_k}x^c_k+\mathbf{r}^c_k,\mathbf{r}^c_k \sim N(0,\mathbf{R}^c_k),
\end{equation}
\begin{equation}\label{equ:trans}
x^c_k = f_kx^c_{k-1}+q^c_k, q^c_k \sim N(0,\sigma^2_{k,c}),
\end{equation}
\begin{equation}\label{equ:init}
x^c_0 \sim N(m^c_0,p^c_0).
\end{equation}

Equation \ref{equ:obs} defines the time trends of the online measures, $\mathbf{s}^c_k=(s^c_{k,GO}, s^c_{k,FAL}, s^c_{k,TW}, s^c_{k,IP})^T$, where $\mathbf{h}_k$ is a vector that maps the state to the observed signals. Equation \ref{equ:trans} defines the vote shares trend, where $f_k$ is the state transition coefficient applied to the previous state. $\mathbf{r}^c_k$ and $q^c_k$ are independent Gaussian random noise with mean zero, observation covariance $\mathbf{R}^c_k$ and transition variance $\sigma^2_{k,c}$. Equation \ref{equ:init} is the starting value that sets the dynamic system in motion. The logic behind the set of equations is that the online measures are flawed signals with the true vote states represented by the mean with mixing noise.
 The goal of the model is to fuse the flawed signals to estimate the daily state and to further transfer the estimate to next day to make a prediction.

To recursively estimate the daily vote state at time $k$, the prediction of vote shares $\hat{x}^c_{k|k-1}$ is first derived by the state transition equation, a variation of equation \ref{equ:trans}:
\begin{equation}\label{equ:stransi}
\hat{x}^c_{k|k-1}=f_k\hat{x}^c_{k-1|k-1},
\end{equation}
\begin{equation}
p^c_{k|k-1}=f^2_k p^c_{k-1|k-1}+\sigma^2_{k,c},
\end{equation}
where $\hat{x}^c_{k|k-1}$ is the vote state prediction for candidate $c$ at time $k$ given the signals up to $k-1$,
and $\hat{x}^c_{k-1|k-1}$ is the updated estimation of the vote state at time $k-1$ given the signals up to $k-1$. $p^c_{k|k-1}$ and $p^c_{k-1|k-1}$ are the prediction covariance and updated estimation covariance, respectively.
Meanwhile, online measure $\mathbf{s^c_k}$ is observed. Then, it is feasible to update the state estimation $\hat{x}^c_{k|k}$ by absorbing the new signals $\mathbf{s^c_k}$ into the prediction $\hat{x}^c_{k|k-1}$.
We use a weighting function to express the combination of the state prediction and signals as follows:
\begin{equation}\label{equ:supdate}
\hat{x}^c_{k|k}=f_k\hat{x}^c_{k|k-1}+\mathbf{k^c_k}(\mathbf{s^c_k}-\mathbf{h_k}\hat{x}^c_{k|k-1}),
\end{equation}
\begin{equation}
p^c_{k|k}=p^c_{k|k-1}-\mathbf{k}^c_k\mathbf{h}_kp^c_{k|k-1},
\end{equation}
where $\mathbf{k^c_k}$ is the Kalman gain, which is used to weight the state prediction and various signals in the fusion. $p^c_{k|k}$ is the updated estimate covariance.
By minimizing the updated state estimation error $x^c_k-\hat{x}^c_{k|k}$, we can derive the Kalman Gain
\begin{equation}\label{equ:kgcal}
\mathbf{k^c_k}=p^c_{k|k-1}\mathbf{h^T_k}(\mathbf{h_k}p^c_{k|k-1}\mathbf{h^T_k}+\mathbf{R^c_k})^{-1}.
\end{equation}
When the updated estimate is obtained, we can use equation \ref{equ:stransi} to predict the next-day vote share.

As shown in equation~\ref{equ:stransi} to equation~\ref{equ:kgcal}, daily vote state estimation depends on the linear dynamic system parameters, such as $\mathbf{R}_k$, $\sigma^2_{k,c}$, $f_k$,$\mathbf{h}_k$, $x^0_c$, and $p^0_c$. In particular, the two noise parameters $\mathbf{R}^c_k$ and $\sigma^2_{k,c}$ determine the impact of the signals on the state update. To estimate the two parameters, we adopt maximum posterior estimations, which can be obtained by maximizing the conditional density function. As $\sum_c x^c_k=1$ and $\sum_c \mathbf{s}^c_k = \mathbf{I}_{4 \times 1}$, there is a trade-off between the vote states and measurements of the three candidates, which implies that a change in one candidate's vote state is equal to the sum of the changes in that of the other two candidates'. To characterize this constraint, we assume that the noise of measurements and state transitions for the three candidates are the same, that is, $\mathbf{R}^c_k = \mathbf{R}_k$ and $\sigma^2_{k,c}=\sigma^2_k$. Then, we express the conditional density function as follows,
\begin{equation}\label{equ:likelihood}
\begin{split}
J&=p(x^{tsai}_{1:k},x^{llchu}_{1:k},x^{soong}_{1:k},\sigma^2_k,\mathbf{R}_k|\mathbf{s}^{tsai}_{1:k},\mathbf{s}^{llchu}_{1:k},\mathbf{s}^{soong}_{1:k})\\
&\propto p(x^{tsai}_{1:k},x^{llchu}_{1:k},x^{soong}_{1:k},\sigma^2_k,\mathbf{R}_k,\mathbf{s}^{tsai}_{1:k},\mathbf{s}^{llchu}_{1:k},\mathbf{s}^{soong}_{1:k})\\
&=p(\mathbf{s}^{tsai}_{1:k},\mathbf{s}^{llchu}_{1:k},\mathbf{s}^{soong}_{1:k}|x^{tsai}_{1:k},x^{llchu}_{1:k},x^{soong}_{1:k},\sigma^2_k,\mathbf{R}_k)\\
&p(x^{tsai}_{1:k},x^{llchu}_{1:k},x^{soong}_{1:k}|\sigma^2_k,\mathbf{R}_k)p(\sigma^2_k,\mathbf{R}_k)\\
&=\prod_c p(x^c_0)\prod^k_{j=1}p(\mathbf{s}^c_{j}|x^c_{j},\mathbf{R}_k)p(x^c_j|x^c_{j-1},\sigma^2_k)p(\sigma^2_k,\mathbf{R}_k).\\
\end{split}
\end{equation}

According to the multiplication rule of probability, we have
\begin{small}
\begin{equation}\label{equ:distristate}
\begin{array}{ll}
p(\mathbf{s}^c_j|x^c_j,\mathbf{R}_k)=\frac{1}{(2\pi)^{\frac{n}{2}}|\mathbf{R}_k|^{\frac{1}{2}}}exp\{-\frac{1}{2}(\mathbf{s}^c_j-\mathbf{h}_jx^c_{j})^T\mathbf{R}^{-1}_k(\mathbf{s}^c_j-\mathbf{h}_jx^c_j)\},
\end{array}
\end{equation}
\end{small}
where $n$ is the dimension of the measurement vector.

Similar to equation~\ref{equ:distristate}, we can readily obtain:
\begin{equation}\label{equ:distriobs}
\begin{array}{ll}
p(x^c_j|x^c_{j-1},\sigma^2_k)=\frac{1}{\sqrt{2\pi\sigma^2_k}}exp\{-\frac{1}{2\sigma^2_k}(x^c_j-f_j x^c_{j-1})^2\}.
\end{array}
\end{equation}

Inserting equation~\ref{equ:distristate} and equation~\ref{equ:distriobs} into equation~\ref{equ:likelihood} yields
\begin{small}
\begin{equation}\label{equ:likelihood_den}
\begin{split}
&J=C_1 \prod_c\prod^k_{j=1}\frac{1}{|\mathbf{R}_k|^{\frac{1}{2}}}\frac{1}{\sqrt{\sigma^2_k}}exp\{-\frac{1}{2}(\mathbf{s}^c_j-\mathbf{h}_jx^c_j)^T\mathbf{R}^{-1}_k(\mathbf{s}^c_j-\mathbf{h}_jx^c_j)\\
&-\frac{1}{2\sigma^2_k}(x^c_j-f_jx^c_{j-1})^2\},\\
\end{split}
\end{equation}
\end{small}
where
\begin{equation}
\begin{array}{ll}
C_1=(2\pi)^{-\frac{k(n+1)}{2}}p(\sigma^2_k,\mathbf{R}_k)\prod_cp(x^c_0).
\end{array}
\end{equation}
Because $J$ and ln$J$ have the same maximum value, we transfer equation ~\ref{equ:likelihood_den} into its logarithmic form,
\begin{equation}
\begin{split}
&lnJ= lnC_1-\frac{1}{2}\sum_c\sum^k_{j=1}(ln|\mathbf{R}_k|+ln\sigma^2_k+(\mathbf{s}^c_j-\mathbf{h}_jx^c_j)^T\mathbf{R}^{-1}_k\\
&(\mathbf{s}^c_j-\mathbf{h}_jx^c_j)+\frac{1}{\sigma^2_k}(x^c_j-f_jx^c_{j-1})^2).\\
\end{split}
\end{equation}

Taking the partial derivatives of $ln J$ with respect to $\sigma^2_k$ and $\mathbf{R}_k$ gives
\begin{small}
\begin{equation}
\begin{array}{ll}
\frac{\partial ln J}{\partial \sigma^2_k}=-\frac{1}{2}\sum_c\sum^k_{j=1}(\frac{1}{\sigma^2_k}-\frac{1}{(\sigma^2_k)^2}(x^c_j-f_jx^c_{j-1})^2)=0
\end{array}
\end{equation}
\begin{equation}
\begin{split}
&\frac{\partial ln J}{\partial \mathbf{R}_k}=\frac{1}{2}\sum_c\sum^k_{j=1}(\mathbf{R}^{-T}_k(\mathbf{s}^c_j-\mathbf{h}_jx^c_{j})(\mathbf{s}^c_j-\mathbf{h}_jx^c_{j})^T\mathbf{R}^{-T}_{k}-\mathbf{R}^{-T}_k)=0,\\
\end{split}
\end{equation}
\end{small}
 and the maximum posterior estimations of $\sigma^2_k$ and $\mathbf{R}_k$ can be written as
 \begin{equation}
 \begin{array}{ll}
 \sigma^2_k = \frac{1}{3}\sum_c\frac{1}{k}\sum^k_{j=1}(x^c_{j}-f_jx^c_{j-1})^2,
 \end{array}
 \end{equation}
 \begin{equation}
 \begin{array}{ll}
 \mathbf{R}_k=\frac{1}{3}\sum_c\frac{1}{k}\sum^k_{j=1}(\mathbf{s}^c_j-\mathbf{h}_jx^c_{j})(\mathbf{s}^c_j-\mathbf{h}_jx^c_{j})^T.
 \end{array}
 \end{equation}
 In the following, we construct the unbiased estimations for the variance $\sigma^2_k$ and covariance matrix $\mathbf{R}_k$. Since the daily vote states $x^c_{j}$ and $x^c_{j-1}$ are normally unavailable, a suboptimal estimator can be obtained by replacing them with the filtering estimations $\hat{x}_{j|j}$ and $\hat{x}_{j-1|j-1}$ and the prediction $\hat{x}_{j|j-1}$,
 \begin{equation}
 \begin{array}{ll}
 \hat{\sigma}^2_k = \frac{1}{3}\sum_c\frac{1}{k}\sum^k_{j=1}(\hat{x}^c_{j|j}-f_j\hat{x}^c_{j-1|j-1})^2,
 \end{array}
 \end{equation}
 \begin{equation}
 \begin{array}{ll}
 \hat{\mathbf{R}}_k=\frac{1}{3}\sum_c\frac{1}{k}\sum^k_{j=1}(\mathbf{s}^c_j-\mathbf{h}_j\hat{x}^c_{j|j-1})(\mathbf{s}^c_j-\mathbf{h}_j\hat{x}^c_{j|j-1})^T.
 \end{array}
 \end{equation}
 Denoting the innovation vector by
 \begin{equation}\label{equ:innov}
 \begin{array}{ll}
 \mathbf{\varepsilon}^c_j = \mathbf{s}^c_j-\hat{\mathbf{s}}^c_{j|j-1},j=1,...,k,
 \end{array}
 \end{equation}
 we have
 \begin{equation}
 \begin{array}{ll}
 E[\mathbf{\varepsilon}^c_j]=E[\mathbf{s}^c_j-\hat{\mathbf{s}}^c_{j|j-1}]=0,\\
 E[\mathbf{\varepsilon}^c_j(\mathbf{\varepsilon}^{c}_j)^T]=E[(\mathbf{s}^c_j-\hat{\mathbf{s}}^c_{j|j-1})(\mathbf{s}^c_j-\hat{\mathbf{s}}^c_{j|j-1})^T]=\mathbf{P}_{\hat{\mathbf{s}}^c_{j|j-1}}.
 \end{array}
 \end{equation}

 In view of equation~\ref{equ:supdate} and equation ~\ref{equ:innov}, the following equation holds.
 \begin{equation}
 \hat{x}^c_{j|j}-f_j\hat{x}^c_{j-1|j-1}=\mathbf{k}_j(\mathbf{s}^c_j-\hat{\mathbf{s}}^c_{j|j-1})=\mathbf{k}_j\mathbf{\varepsilon}^c_j.
 \end{equation}
Thus, we have
\begin{equation}
\begin{split}
E[\hat{\sigma}^2_k]&=\frac{1}{3}\sum_c\frac{1}{k}\sum^k_{j=1}E[(\hat{x}^c_{j|j}-f_j\hat{x}^c_{j-1|j-1})^2]\\
&=\frac{1}{3}\sum_c\frac{1}{k}\sum^k_{j=1}E[\mathbf{k}_j\mathbf{\varepsilon}^c_j(\mathbf{\varepsilon}^c_j)^T\mathbf{k}^T_j]\\
&=\frac{1}{3}\sum_c\frac{1}{k}\sum^k_{j=1}\mathbf{k}_j\mathbf{P}_{\hat{s}^c_{j|j-1}}\mathbf{k}^T_j\\
&=\frac{1}{3}\sum_c\frac{1}{k}\sum^k_{j=1}(p^c_{j|j-1}-p^c_{j|j})\\
&=\frac{1}{3}\sum_c\frac{1}{k}\sum^k_{j=1}(f^2_kp^c_{j-1|j-1}+\sigma^2_k-p^c_{j|j}),
\end{split}
\end{equation}
and
\begin{equation}
\begin{split}
E[\hat{\mathbf{R}}_k]&=\frac{1}{3}\sum_c\frac{1}{k}\sum^k_{j=1}(\mathbf{s}^c_j-\mathbf{h}_j\hat{x}^c_{j|j-1})(\mathbf{s}^c_j-\mathbf{h}_j\hat{x}^c_{j|j-1})^T\\
&=\frac{1}{3}\sum_c\frac{1}{k}\sum^k_{j=1}\mathbf{\varepsilon}^c_j(\mathbf{\varepsilon}^c_j)^T\\
&=\frac{1}{3}\sum_c\frac{1}{k}\sum^k_{j=1}\mathbf{P}_{\hat{\mathbf{s}}^c_{j|j-1}}\\
&=\frac{1}{3}\sum_c\frac{1}{k}\sum^k_{j=1}(\mathbf{h}_jp^c_{j|j-1}\mathbf{h}^T_j+\mathbf{R}_k).\\
\end{split}
\end{equation}

Therefore, the unbiased estimate of $\sigma^2_k$ is
\begin{equation}
\begin{array}{ll}
\hat{\sigma}^2_k=\frac{1}{3}\sum_c\frac{1}{k}\sum^k_{j=1}(\mathbf{k}_j\mathbf{\varepsilon}_j\mathbf{\varepsilon}^T_j\mathbf{k}_j+p^c_{j|j}-f^2_jp^c_{j-1|j-1}).
\end{array}
\end{equation}
To ensure that $\hat{\sigma}^2_k>0$, we take a suboptimal estimate,
\begin{equation}\label{equ:qesti}
\begin{split}
\hat{\sigma}^2_k&=\frac{1}{3}\sum_c\frac{1}{k}\sum^k_{j=1}(\mathbf{k}_j\mathbf{\varepsilon}_j\mathbf{\varepsilon}^T_j\mathbf{k}_j)\\
&=\frac{1}{3}\sum_c\frac{1}{k}\sum^k_{j=1}(\hat{x}^c_{j|j}-f_j\hat{x}^c_{j-1|j-1})^2.
\end{split}
\end{equation}

The unbiased estimate of $\mathbf{R}_k$ is
\begin{equation}\label{equ:resti}
\begin{split}
\hat{\mathbf{R}}_k&=\frac{1}{3}\sum_c\frac{1}{k}\sum^k_{j=1}(\mathbf{\varepsilon}^c_j(\mathbf{\varepsilon}^c_j)^T-\mathbf{h}_jp^c_{j|j-1}\mathbf{h}^T_j)\\
&=\frac{1}{3}\sum_c\frac{1}{k}\sum^k_{j=1}((\mathbf{s}^c_j-\mathbf{h}_j\hat{x}^c_{j|j-1})(\mathbf{s}^c_j-\mathbf{h}_j\hat{x}^c_{j|j-1})^T-\mathbf{h}_jp^c_{j|j-1}\mathbf{h}^T_j)\\
\end{split}
\end{equation}

At time $k$, we use equation~\ref{equ:qesti}, equation~\ref{equ:resti} and the signals up to time $k$ to update $\sigma^2_k$ and $\mathbf{R}_k$ and then apply the updated estimates to the next-day state calculations.
In addition to the covariance of observation noise $\mathbf{R}_k$ and the variance of state transition noise $\sigma^2_k$, the mapping vector $\mathbf{h}_k$, the state transition coefficient $f_k$, and the initial state values $x^c_0$ and $p^c_0$ affect the state estimation. However, as shown by equation~\ref{equ:qesti} and equation~\ref{equ:resti}, those parameters determine the update of $\mathbf{R}_k$ and $\sigma^2_k$, which indicates that it is not feasible to estimate them simultaneously via the maximum posterior method. Instead, we set the parameters as follows based on some assumptions.

\emph{Mapping vector $\mathbf{h_k}$. }Since the signals are normalized, the scale of the signals and that of the state estimations are the same. Thus, $\mathbf{h_k}$ is set to be a unit vector.

\emph{State transition coefficient $f_k$.} We introduce signals and update prediction at a daily frequency, which makes it possible to absorb the latest information into the prediction. Therefore, we assume that daily prediction is equal to the previous updated state estimation. The state transition coefficient $f_k$ is set to be 1.

\emph{Initial vote share $m^c_0$.} At the beginning of the prediction, there is little information about the three candidates' vote ratios. We use the latest poll result of the pollster TISR as a rough estimation of the vote share. We also change the value to the mean of each candidate's signal $x^c_0=(s^c_{k,GO}+s^c_{k,FAL}+s^c_{k,TW}+s^c_{k,IP})/4$, where $c\in\{ Tsai\ Ing-wen, Eric\ Chu, James\ Soong\}$, and an equal value $m^c_0=1/3$. However, the changes have almost no effects on the prediction results. Details are provided in ~\ref{sub:rts}.

\emph{Initial state variance $p^c_0$.} The initial state variance reflects our belief of the state value $x^c_0$, which is a rough estimate. Thus, we make the variance large, $p^c_0=1$. We test different settings and find that the fusion result is not sensitive to changes. The test results are shown in ~\ref{sub:rtpn}.

\subsection{Robustness Tests for Initial Candidate Vote Ratio}
\label{sub:rts}
\begin{figure*}[htbp]
\centering
\includegraphics[width=.8\linewidth]{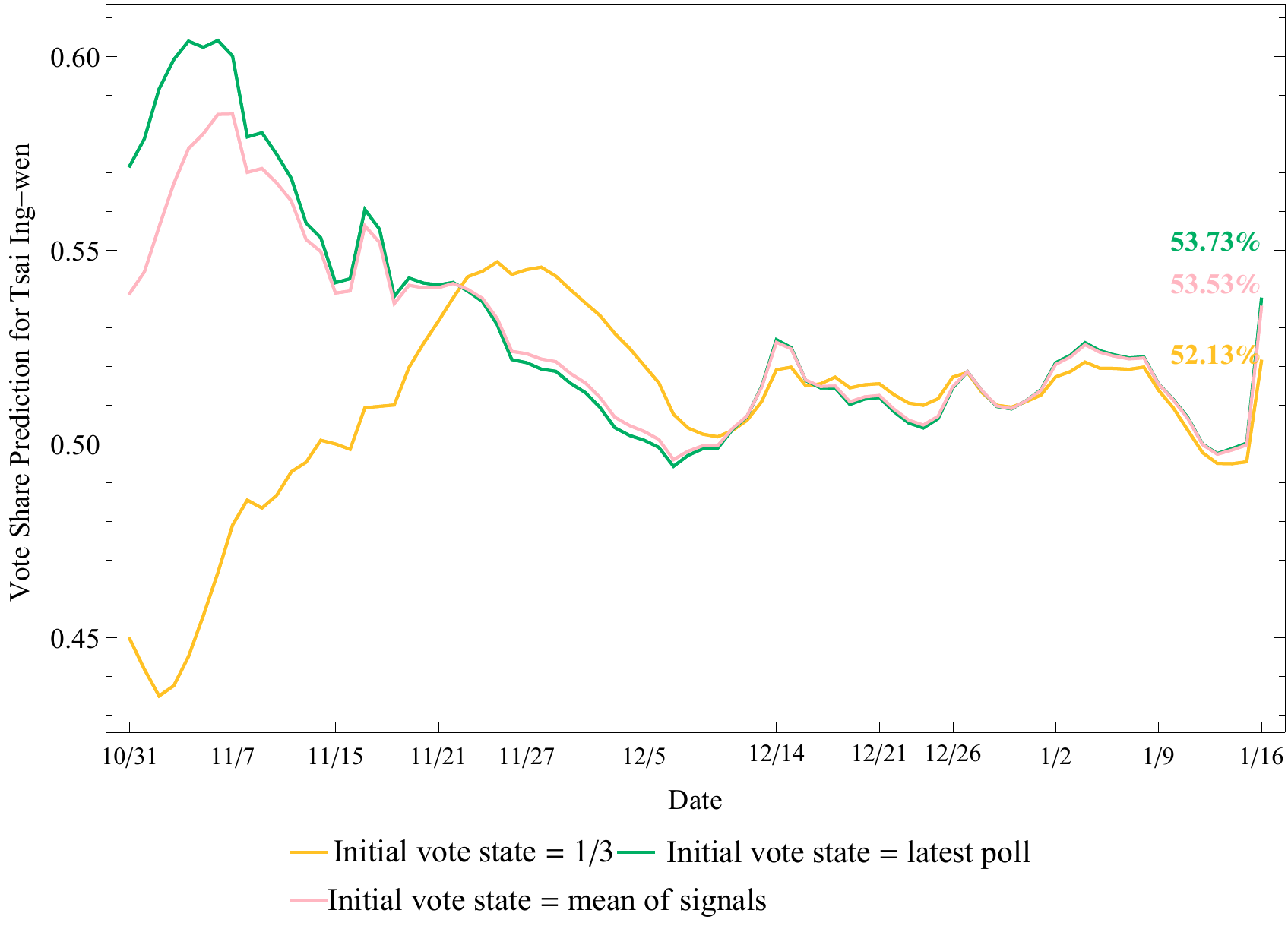}
\caption{The predicted vote share of Tsai Ing-wen based on three sets of initial vote ratios. After November 15, 2016, the differences between the three predictions are no greater than 5\%. On election day, the differences are less than 1.60\%.}
\label{fig:rtstsai}
\end{figure*}

\begin{figure*}[htbp]
\centering
\includegraphics[width=.8\linewidth]{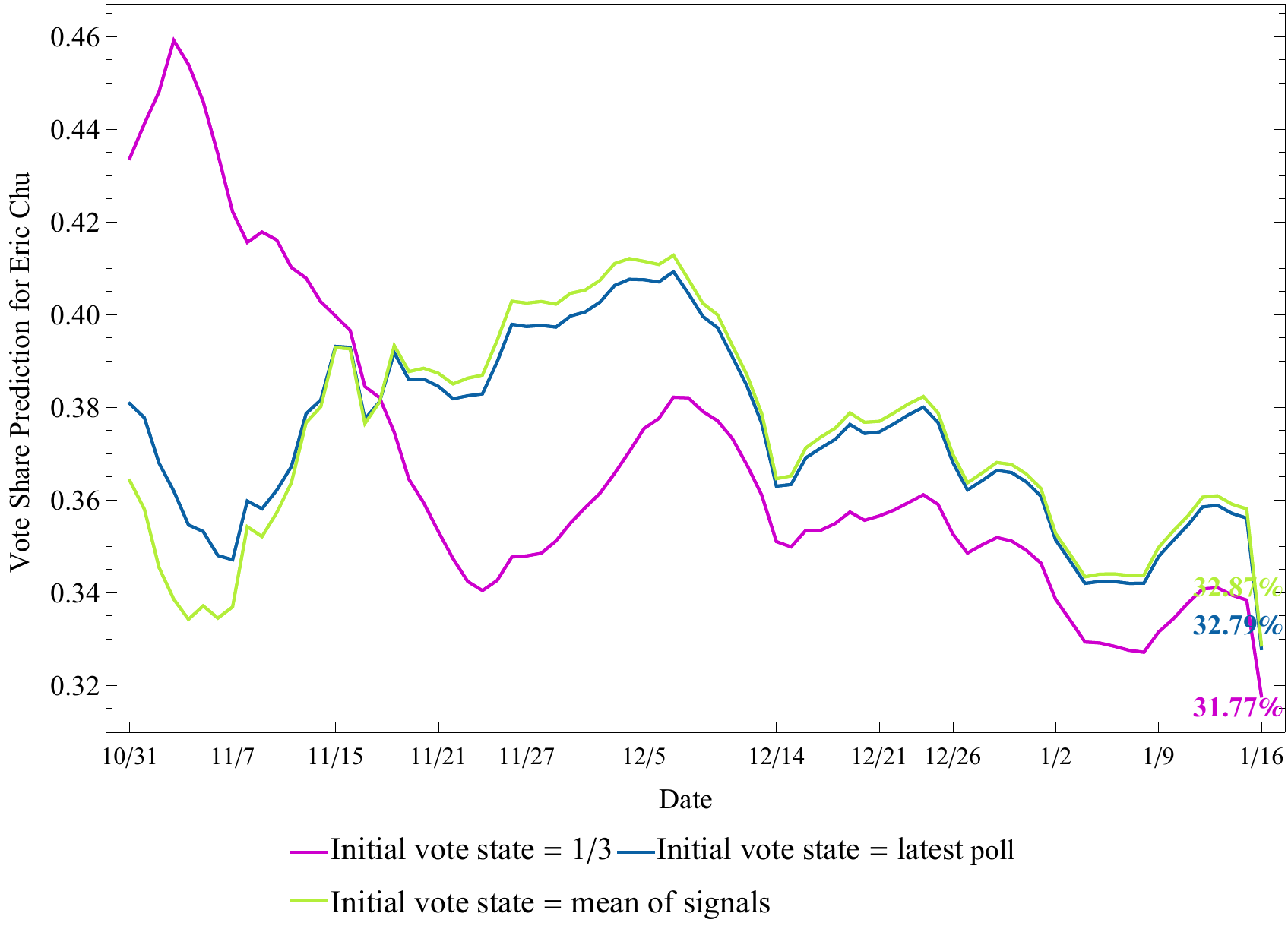}
\caption{The predicted vote share of Eric Chu based on three sets of initial vote ratios. After November 13, 2016, the differences between the three predictions are no greater than 5\%. On election day, the differences are less than 1.10\%.}
\label{fig:rtsllchu}
\end{figure*}

\begin{figure*}[htbp]
\centering
\includegraphics[width=.8\linewidth]{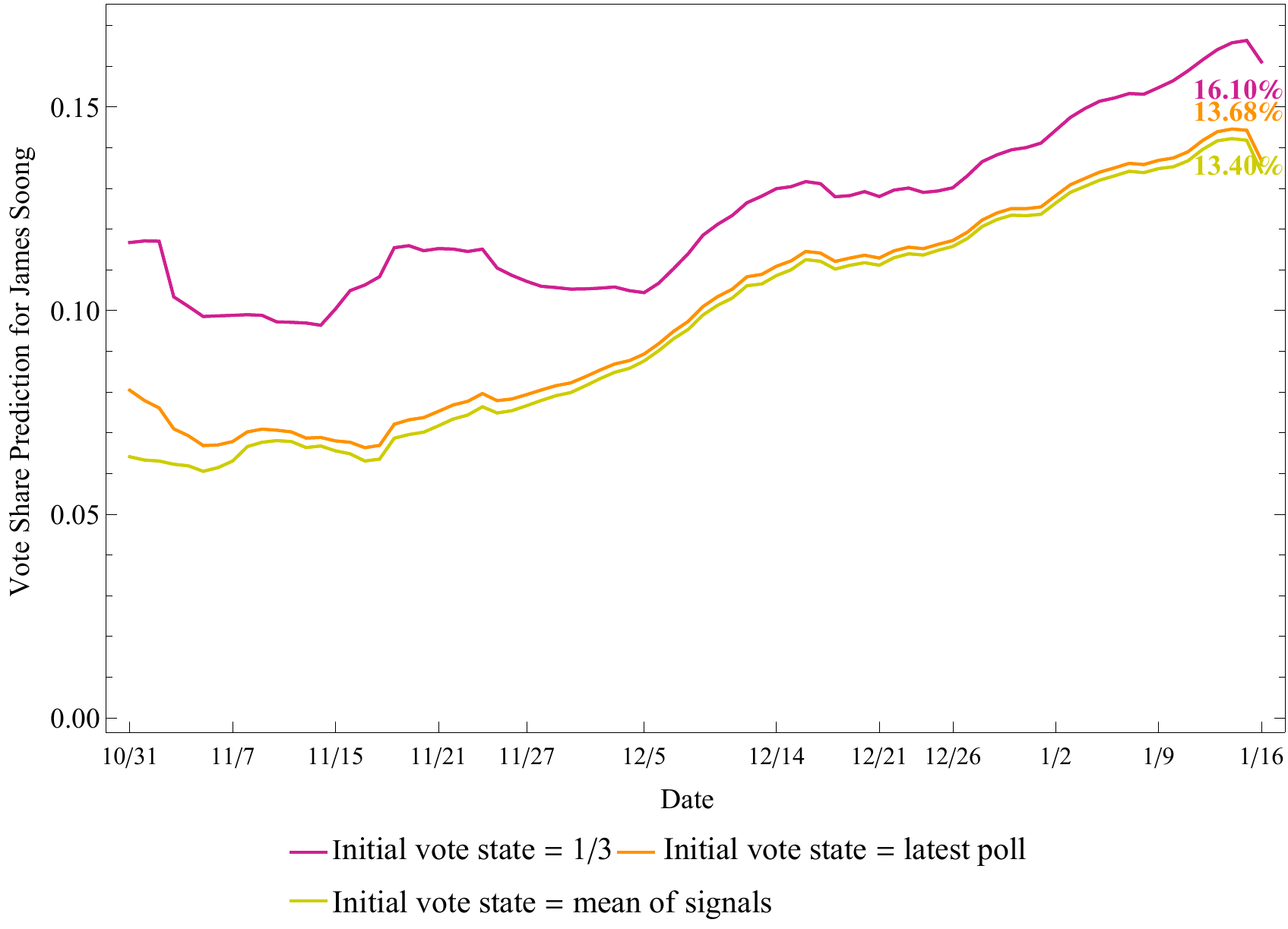}
\caption{The predicted vote share of James Soong based on three sets of initial vote ratios. After November 4, 2016, the differences between the three predictions are no greater than 5\%. On election day, the differences are less than 2.70\%.}
\label{fig:rtssoong}
\end{figure*}

The initial candidate vote ratio $m^c_0$ starts the prediction process. Since the candidate vote preference is unknown at the beginning of the prediction, we make a rough estimate. To test the influence of the initial estimate on the prediction, we examine three different sets of parameters. First, we set $m^c_0=1/3$. The logic of this setting is that when little information about the candidates' vote ratios is available, we assume that the three candidates receive equal vote shares. However, in contrast to the unknown vote ratio of the first setting, we believe that polls and social media signals provide clues about candidates' popularity. Therefore, in the second and third settings, we take the latest poll results of the pollster TISR and the mean of social media signals as the initial state values, respectively. The results are shown in FIG.~\ref{fig:rtstsai}, FIG.~\ref{fig:rtsllchu}, and FIG.~\ref{fig:rtssoong}.

All the figures show that the predictions under the three settings are consistent with respect to both trend and numerical value. Since mid-November 2016, the differences between the three cases are no greater than 5\%. On election day, the differences are no greater than 3\%. The small differences between the three cases indicate that the prediction is insensitive to the initial state value, especially more than one month after the start of the prediction.

\clearpage
\subsection{Robustness Tests for Process Noise}
\label{sub:rtpn}
As described in Sec.~\ref{sub:kf}, the daily state prediction $\hat{x}^c_{k|k-1}$ depends on the updated previous state estimate.
Then, the prediction is fused with the signals $\mathbf{s}^c_k$ to update the current state estimation $\hat{x}^c_{k|k}$. As shown in equation~\ref{equ:supdate} and equation~\ref{equ:kgcal}~\cite{kalman1960new}, the fusion weight primarily relies on the process noise $p^c_{k|k-1}$ and the measurement noise $\mathbf{R}^c_k$.

The process noise $p^c_{k|k-1}$ can be iteratively calculated as follows,
\begin{equation}\label{equ:pn}
\begin{split}
&p^c_{k|k-1}=f_kp^c_{k-1|k-1}+\sigma^2_{k,c},\\
&p^c_{k|k}=(I-\mathbf{k^c_{k}}\mathbf{h_{k}})p^c_{k|k-1},\\
&p^c_{0|0}=p^c_0.
\end{split}
\end{equation}
As $\sigma^2_{k,c}$ and $\mathbf{R}^c_k$ are derived by maximum posterior estimation, the fusion weight can be obtained as long as $p^c_0$ is determined. We consider two parameter sets for $p^c_0$:

First, we set $p^c_0=0$. This setting assumes that the variance of the initial state distribution is 0, which implies that we are relatively sure of the mean value of the initial state.

Second, we set $p^c_0=1$. This setting assumes that the variance of the initial state distribution is 1. Because the state value is greater than 0 but less than 1, the setting implies that we are relatively uncertain about the mean value of the initial state.

\begin{figure}
\centering
\includegraphics[width=1\linewidth]{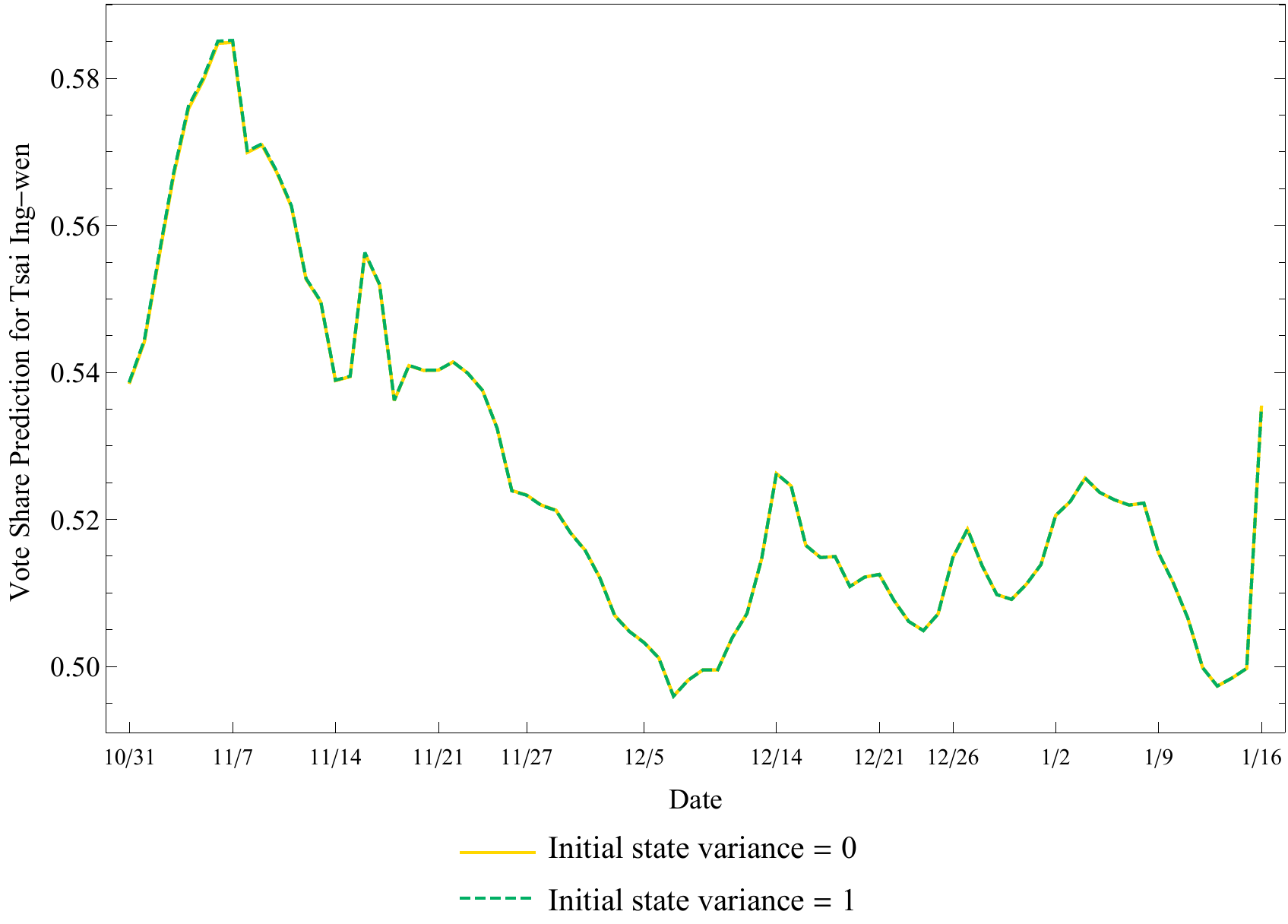}
\caption{The predicted vote share of Tsai Ing-wen based on two initial state variance settings.}
\label{fig:rttsai}
\end{figure}

The time series of the prediction for the three candidates based on the above two parameter sets are shown in FIG.~\ref{fig:rttsai}-FIG.~\ref{fig:rtsoong}. In the three figures, the curve of the first setting is almost coincident with that of the second setting. This coincidence indicates that the daily prediction is not sensitive to the setting of $p^c_0$.

\begin{figure}
\centering
\includegraphics[width=1\linewidth]{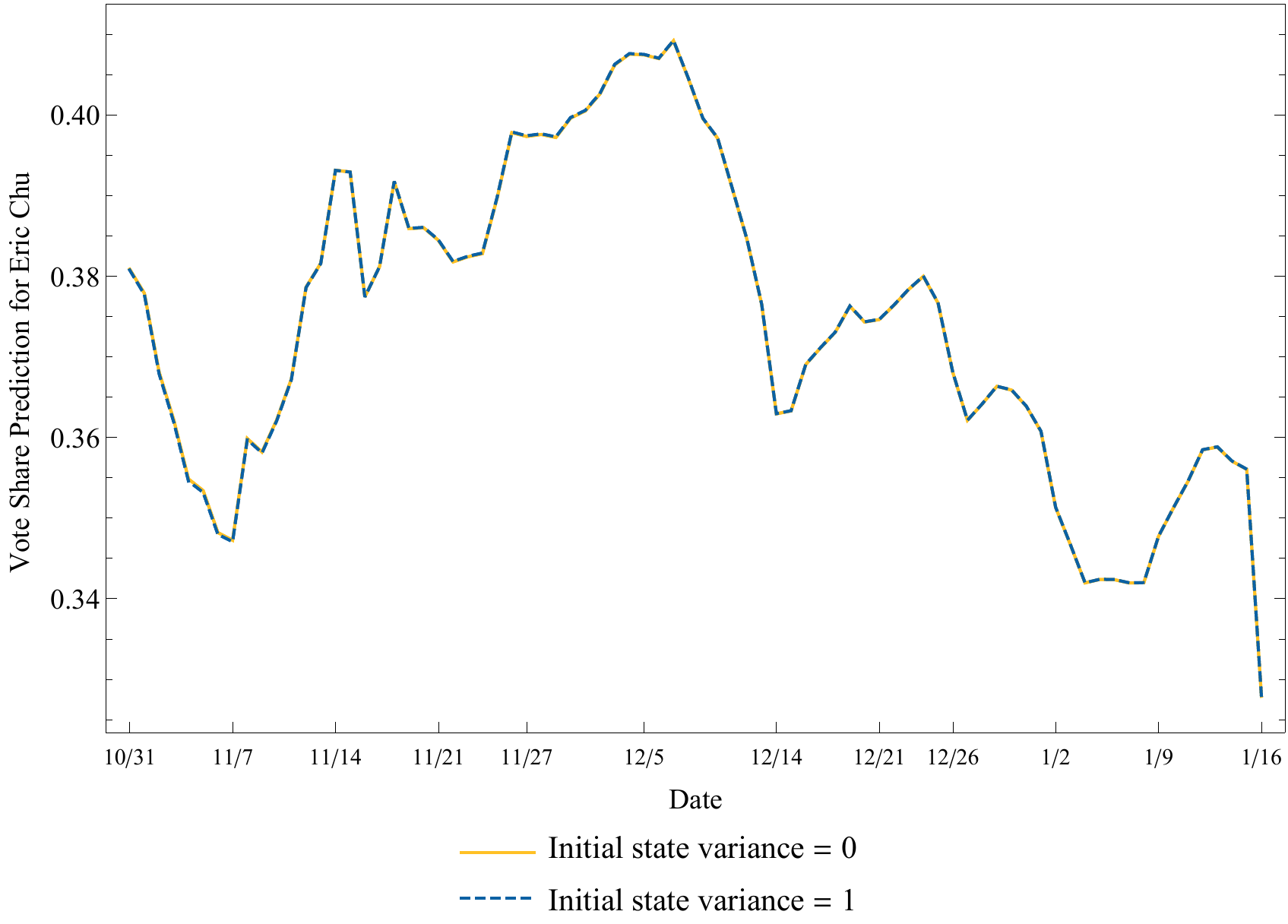}
\caption{The predicted vote share of Eric Chu based on two initial state variance settings.}
\label{fig:rtllchu}
\end{figure}

\begin{figure}
\centering
\includegraphics[width=1\linewidth]{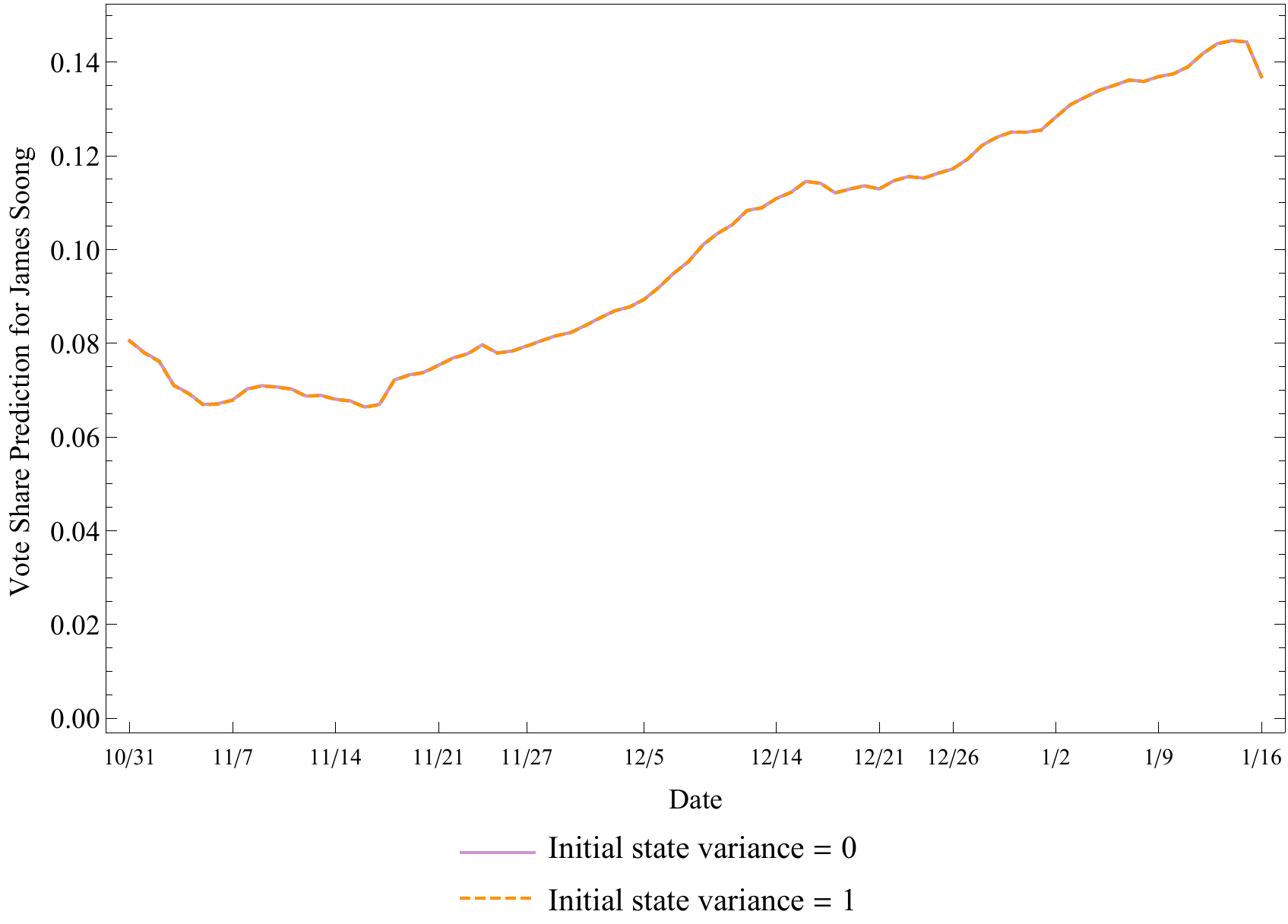}
\caption{The predicted vote share of James Soong based on two initial state variance settings.}
\label{fig:rtsoong}
\end{figure}

\clearpage
\subsection{Online and Offline Data Fusion Method} 
\label{sub:pred_online_offline}
In the results part of the main body, we compare our predictions with those of the polls and the outcomes of the online and offline data fusion method. The comparison is shown in Fig. 2 in the main body, which illustrates that our online data fusion method is superior in terms of prediction accuracy. However, due to the space constraints, we do not elaborate on the comparative method, the online and offline data fusion method. We now present the detailed results.

According to the Internet survey report of Taiwan\footnotemark[1], more than 90\% of Taiwan residents aged between 20 and 45 years have accessed the Internet since May 2015. This figure is over 80\% in the population aged between 45 and 55 years. By contrast, only 49.5\% of residents aged over 55 years have used the Internet during the same time period. Thus, we take the online data fusion result merely as a representation for the group aged between 20 and 50 years. With respect to the age-adjusted sampling methods adopted by pollsters, we take the poll results for the 50 to 60 year-old, 60 to 70 year-old and over 70 year-old groups as the vote share estimations of the corresponding age groups. Therefore, the final daily vote share prediction $y^c_{k}$ for candidate $c$ at time $k$ is weighted as follows,
\begin{equation}
\begin{split}
y^c_k=w_{20-50}\hat{x}^c_{k|k-1}+w_{50-60}z^c_{50-60,k}\\+w_{60-70}z^c_{60-70,k}+w_{70}z^c_{>70,k},
\end{split}
\end{equation}
where $w_{i}$ is the population proportion of age group $i$, which is obtained from the Ministry of the Interior of Taiwan\footnote[6]{Taiwan demographics, \protect\url{http://statis.moi.gov.tw/micst/stmain.jsp?sys=100}}. $z^c_{i,j}$ is the most recent TISR poll result of age group $i$ for candidate $c$ on day $k$.

\begin{figure}
   \centering
   \includegraphics[width=1\linewidth]{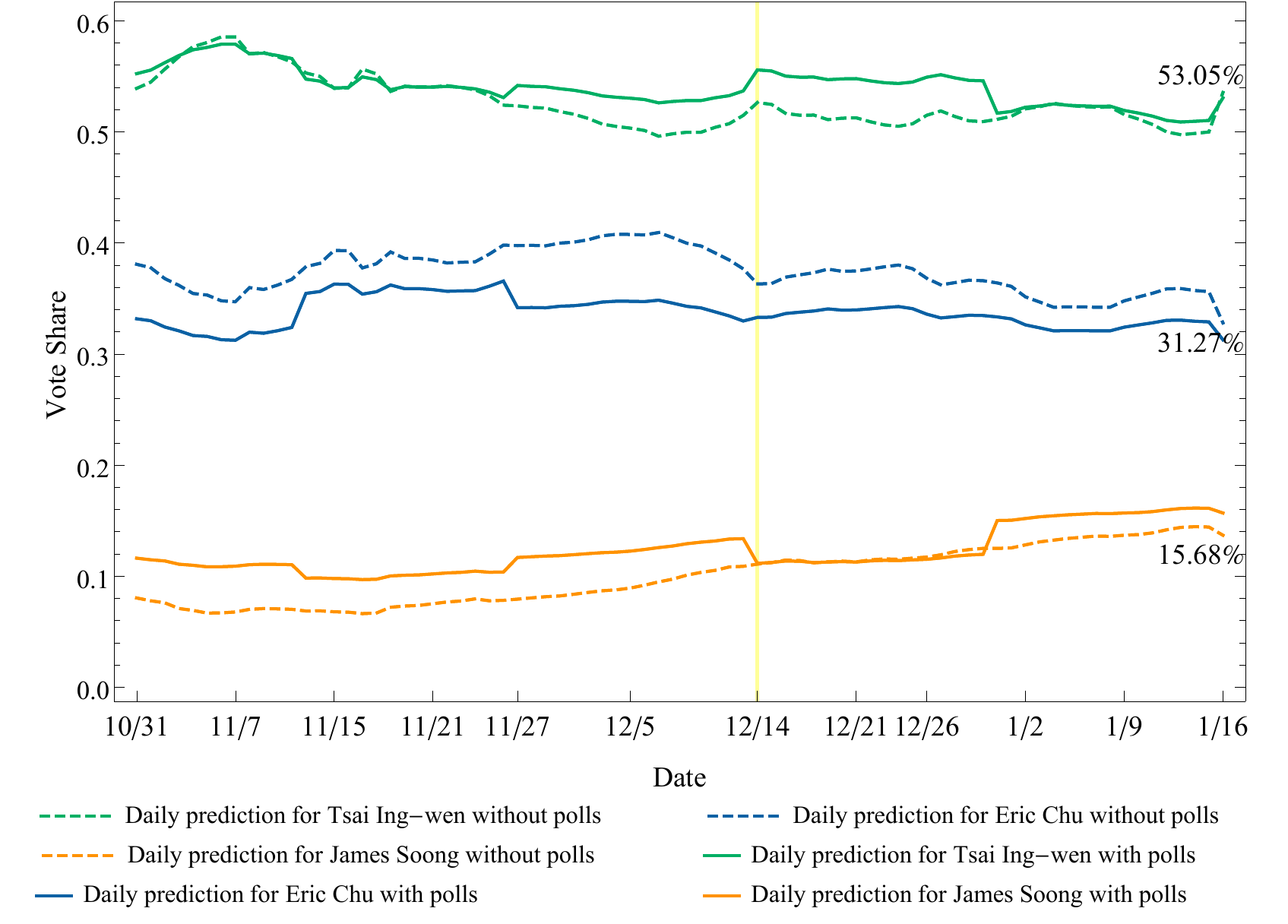}
   \caption{Prediction results of the online and offline data fusion method.}
 \label{fig:pcompare_online_offline}
 \end{figure}

The results are shown in FIG.~\ref{fig:pcompare_online_offline}. The errors of the online-offline fusion method are greater than those of the online fusion method, which range from 0.23\% to 3.07\%.

\clearpage
\subsection{Prediction Based on Comment Volume}
\begin{figure}[htbp]
   \centering
   \includegraphics[width=1\linewidth]{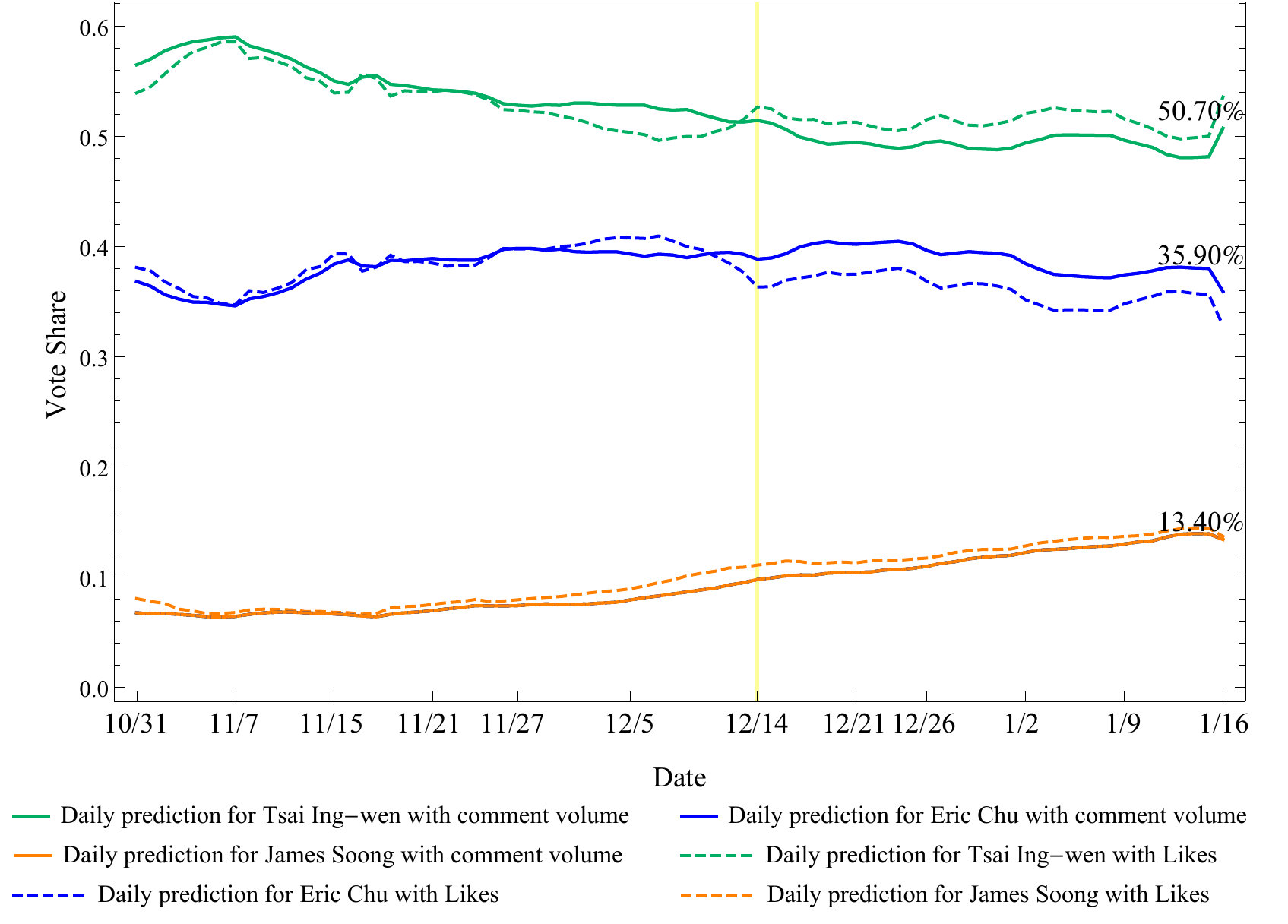}
   \caption{Prediction based on comment volume.}
 \label{fig:pcompare}
 \end{figure}

To compare the use of "Like" and ``Comments'' in election prediction, we substitute $s^c_{k,FAL}$ with $s^c_{k,FAC}$ in the Kalman filter. The results are plotted in FIG.~\ref{fig:pcompare}. The results indicate that the prediction outcomes become worse. The prediction errors for the three candidates are 5.42\%, 4.86\%, and 0.56\%. The figure also indicates that there is a negative correlation between the number of comments and the number of ``Likes" before December 14, 2015. However, after that point, they are consistent in time. This change also provides evidence that the various types of signals become consistent as election day approaches. 

\clearpage
\subsection{Topics in Comments}
To compare the behavior patterns of overlapping users and users who only commented on the candidates, we apply Latent Dirichlet Allocation (LDA) model to extract topics from their corresponding comments. The results are presented in TABLE~\ref{tab:lc_tsai} to TABLE~\ref{tab:c_soong}. The representative words from the topics of overlapping users are mainly supportive attitudes, while the words from the topics of users who only commented on candidates are mixed, with both positive and negative words. The negative words are marked in red.

\clearpage
\begin{CJK}{UTF8}{gkai}
\begin{center}
\begin{longtable}{|p{.1\textwidth}|p{.2\textwidth}|p{.7\textwidth}|}
\caption[Topics of overlapping users of Tsai Ing-wen]{Topics of overlapping users of Tsai Ing-wen}\label{tab:lc_tsai} \\

\hline \multicolumn{1}{|c|}{\textbf{Topic Number}} & \multicolumn{1}{c|}{\textbf{Summary}} & \multicolumn{1}{c|}{\textbf{Keywords}} \\ \hline
\endfirsthead

\multicolumn{3}{c}%
{{\bfseries \tablename\ \thetable{} -- Topics of overlapping users of Tsai Ing-wen}} \\
\hline \multicolumn{1}{|c|}{\textbf{Topic Number}} &
\multicolumn{1}{c|}{\textbf{Summary}} &
\multicolumn{1}{c|}{\textbf{Keywords}} \\ \hline
\endhead

\hline \multicolumn{3}{|r|}{{Continued on next page}} \\ \hline
\endfoot

\hline \hline
\endlastfoot
  1 & Commentators express their excitement about Tsai Ing-wen's successful election. & 总统 (president)，期待 (look forward)，小英 (a nickname of Tsai Ing-wen)，感动 (touched)，终于 (finally)，四年 (four years)，一刻 (the moment)，明天 (tomorrow)，迎接 (welcome)，倒数 (final stage of the election)，命运 (destiny)，公道 (justice)，真好 (it's good)，到来 (arrival)，为荣 (feel proud)\\
  \hline
  2 & Taiwan commentators criticize mainland netizens of negative comments on Tsai Ing-wen's Facebook homepage.& 大陆 (mainland)，留言 (messages on Facebook)，对岸 (Mainland China)，自由 (freedom of speech)，网友 (netizens)，民主 (democratic)，言论 (remarks)，水准 (level)，独立 (independence)，开放 (open to)，同胞 (compatriot)，尊重 (respect)，攻击 (attack)，简体字 (Simplified Chinese)，可怜 (poor) \\
  \hline
  3 & Commentators encouraged young people to return home to vote for Tsai Ing-wen.& 投票 (vote)，年轻人 (young people)，监票 (scrutinize balloting)，选票 (votes)，返乡 (return home to vote)，回家 (go back home)，记得 (remember)，全民 (all the people)，明天 (tommorow)，胜选 (win the election)，青年 (youth)，务必 (must)，告诉 (tell)，神圣 (sacred)，踊跃 (enthusiastically) \\
  \hline
  4 & Commentators cheer for Tsai Ing-wen& 总统 (president)，加油 (come on)，小英 (a nickname of Tsai Ing-wen)，台湾 (Taiwan)，万岁 (long live)，辛苦 (hard work)，未来 (in the future)，骄傲 (proud)，带领 (lead)，交给 (give)，恭喜 (Congratulations)，同心 (with one heart)，狗民 (the dog people)，人选 (candidates)，民族 (nation)，福气 (good fortune) \\
  \hline
  5 & Commentators expect that Tsai Ing-wen will bring a new look to Taiwan after being elected.& 台 湾(Taiwan)，未来 (in the future)，改变 (change)，改革 (reform)，美好 (wonderful)，期待 (look forward)，创造历史 (make history)，幸福 (happiness)，来临 (coming)，下一代 (the next generation)，执政 (be in power)，初衷 (original intention)，世代 (generation)，年轻人 (young people)\\
  \hline
  6 & Commentators remind Tsai Ing-wen to look after herself.& 加油 (cheer up)，主席 (chairman)，辛苦 (work hard)，身体 (body)，保重 (take care of)，总统 (president)，好好 (well)，健康 (health)，休息 (have a rest)，平安 (peace)，天气 (weather)，小心 (look after)，保暖 (keep warm)，团结 (hold together)，感冒 (have a cold)\\
  \hline
  7 & Commentators expect Tsai Ing-wen to reform the judicial system.& 改革 (reform)，司法 (judicial)，政府 (government)，劳工 (labor)，法官 (judge)，年金 (annuity)，黑心 (evil mind)，制度 (system)，无罪 (innocent)，法院 (court)，立法 (legislation)，选举 (election)，保护 (protect)，浪费 (waste)，基层 (grass-roots unit) \\
  \hline
  8 & Commentators wish Tsai Ing-wen a happy new year.& 快乐 (happy)，总统 (president)，新年 (new year)，幸福 (happiness)，祝福 (wish)，当选 (be elected)，平安 (peace)，成功 (successful)，顺利 (all right)，健康 (healthy)，开心 (joyful)，心想事成 (may all your wishes come true)，愉快 (pleasant)，晚安 (good night)，美好 (wonderful)\\
  \hline
  9 & Commentators express support for Tsai Ing-wen and her deputy.& 支持 (support)，总统 (president)，英派 (faction of Tsai Ing-wen)，革新 (reform)，向前 (forward)，期待 (look forward)，力挺 (firmly support)，未来 (in the future)，团队 (team)，最佳 (best)，英仁双全 (a good campaign team of Tsai Ing-wen and Chen Chien-jen)，团结 (unity)，自信 (confidence)，参选 (participate in the election)，翻转 (reverse) \\
  \hline
  10 & Commentators from Taoyuan, Taichung, Kaohsiung, Taipei and Tainan express their support for Tsai Ing-wen.& 市长 (mayor)，桃园 (Taoyuan)，台中 (Taichung)，候选人 (candidate)，政见 (political view)，直播 (broadcast)，高雄 (Kaohsiung)，倒数 (in the final stage)，选举 (election)，参加 (participate)，台北 (Taipei)，网路 (Internet)，辩论 (debate)，台南 (Tainan)，发表 (make a statement)\\
  \hline
  11 &Commentators accuse the KMT of black-heartedness and judicial injustice. &正义 (justice)，公平 (fair)，国民党 (the KMT)，改革 (reform)，社会 (social)，转型 (transition)，司法 (judicial)，不公 (unfair)，落实(implement)，执政 (in power)，政党 (political party)，政策 (policy)，黑心 (evil mind)，利益 (benefit)，诈骗 (fraud) \\
  \hline
  12 & Commentators praise Tsai Ing-wen for her hard work.& 谢谢 (thanks)，努力 (try hard)，辛苦 (work hard)，感谢 (appreciate)，主席 (chairman)，感恩 (grateful)，感动 (touched)，用心 (with heart)，团队 (team)，付出 (make great efforts)，团结 (unity)，勇敢 (brave)，承担 (responsibility)，英雄 (hero)，智慧 (intelligence) \\
  \hline
  13 & Commentators expect Tsai Ing-wen to seek justice for Chou Tzu-yu.& 主席 (chairman)，希望 (hope)，好好 (well)，谢谢 (thanks)，婚姻 (marriage)，支持 (support)，一月 (January)，终于 (finally)，上任 (take office)，周子瑜 (Chou Tzu-yu)，明天 (tomorrow)，欺负 (bully)，黄安 (Huang An)，投票 (vote)\\
  \hline
  14 & Commentators accuse the KMT of framing Tsai Ing-wen.& 国民党 (the KMT)，小心 (pay attention to)，小英 (a nickname of Tsai Ing-wen)，抹黑 (disgrace)，对手 (competitor)，马英九 (Ma Ying-Jeou)，历史 (history)，手段 (means)，时刻 (the moment)，理会 (take notice of)，不用 (no need to)，清算 (expose and criticize)，面对 (facing)，一路 (all the way)，制造 (make trouble) \\
  \hline
  15 & Commentators praise Tsai Ing-wen for her ability and lofty vision.& 国家 (nation)，政治 (political)，诚信 (credibility)，领导人 (leader)，高度 (lofty vision)，未来 (in the future)，格局 (landscape)，气度 (tolerance)，展现 (show)，政客 (politico)，利益 (benefit)，智慧 (intelligence)，能力 (competence)，领袖 (leader)，民主 (democracy)\\
  \hline
  16 & Commentators express expectations of Tsai Ing-wen's policies on economics, education, and cross-strait relations.& 经济 (economic)，政策 (policy)，发展 (development)，产业 (industry)，教育 (education)，社会 (social)，文化 (cultural)，生活 (life)，两岸 (Taiwan and mainland)，和平 (peaceful)，希望 (hopeful)，人才 (talented persons)，合作 (cooperate)，资源 (resource)，民主 (democracy)\\
  \hline
  17 & Commentators hope the DPP could occupy more than half of the seats in the congress.& 国会 (congress)，过半 (occupy more than half of the seats)，加油 (cheer up)，改革 (reform)，小英 (a nickname of Tsai Ing-wen)，执政 (take office)，英仁 (Tsai Ing-wen and Chen Chien-jen)，希望 (hope)，投票 (vote)，带领 (lead)，革新 (renovation)，力拼 (fight fiercely)，成功 (successfully)，改造 (transform)，总统府 (presidential palace)\\
  \hline
  18 & Commentators hope that Tsai Ing-wen will bring a stable life with dignity for Taiwan.& 希望 (hope)，改变 (change)，带来 (bring)，努力 (make efforts)，经济 (economic)，安居乐业 (live and work in peace)，尊严 (dignity)，政府 (government)，幸福 (happiness)，冲刺 (sprint)，安定 (stable)，期待 (expect)，环境 (environment)，日子 (life)\\
  \hline
  19 & Commentators debate the 1992 Consensus and cross-strait relations.& 台湾 (Taiwan)，中华民国 (Republic of China)，中国 (China)，国民党 (the KMT)，共识 (consensus)，两岸 (Taiwan and mainland)，独立 (independence)，国旗 (flag)，国家 (nation)，马英九 (Ma Ying-Jeou)，中共 (the Communist Party of China)，关系 (relations)，台独 (Taiwan Independence)，九二共识 (the 1992 Consensus)，和平 (peaceful) \\
  \hline
  20 & Hakka commentators support Tsai Ing-wen.& 客家 (Hakka)，心中 (in my heart)，努力 (make efforts)，精神 (spirit)，值得 (deserve)，文化 (cultural)，好好 (well)，感觉 (feel)，学习 (study)，朋友 (friends)，基隆 (Keelung)，故乡 (hometown)，感到 (feel)，温暖 (warm)，孩子 (children)\\
  \hline
  21 &Commentators discuss the relations between Taiwan, mainland China, Japan and the United States. & 大陆 (mainland)，台湾 (Taiwan)，日本 (Japan)，美国 (the United States)，支持 (support)，独立 (independence)，民主 (democracy)，经济 (economic)，中华民国 (Republic of China)，马英九 (Ma Ying-Jeou)，同胞 (compatriot)，土地 (land)，中共 (the Communist Party of China)，总统 (president)，开放 (open to)\\
  \hline
  22 & Commentators accuse the KMT of incompetence.& 国民党 (the KMT)，选票 (vote)，教训 (lesson)，无能 (incompetence)，打倒 (down with)，马英九 (Ma Ying-Jeou)，反省 (introspection)，国家 (nation)，终结 (end)，民进 (the DPP)，政权 (regime)，抹黑 (discredit)，下台 (deserve to lose power)，时代力量 (the power of the moment)\\
  \hline
  23 & Commentators scold Ma Ying-Jeou and support Tsai Ing-wen.&总统 (president)，支持 (support)，陈建仁 (Chen Chien-jen)，第一 (first)，百姓 (common people)，合作 (cooperate)，哈哈哈 (Hahaha)，国庆 (National Day)，恭贺 (congratulations)，放心 (at ease)，成功 (succeed)，精英 (elite)，心愿 (wish)，实实在在 (down-to-earth)，马英狗 (Ma Ying-Dog, the irony of Ma Ying-Jeou) \\
  \hline
  24 & Commentators hope that Tsai Ing-wen will make reforms for young people.& 希望 (hope)，台湾 (Taiwan)，未来 (in the future)，机会 (opportunity)，好好 (well)，下一代 (next generation)，努力 (make efforts)，年轻人 (young people)，解决 (solve)，改变 (change)，越来越 (more and more)，珍惜 (cherish)，子孙 (descendants)，着想 (for the sake of)，把握 (grasp) \\
   \hline
\end{longtable}
\end{center}
\end{CJK}

\clearpage
\begin{CJK}{UTF8}{gkai}
\begin{center}
\begin{longtable}{|p{.1\textwidth}|p{.2\textwidth}|p{.7\textwidth}|}
\caption[Topics of users who commented on Tsai Ing-wen]{Topics of users who commented on Tsai Ing-wen}\label{tab:c_tsai} \\

\hline \multicolumn{1}{|c|}{\textbf{Topic Number}} & \multicolumn{1}{c|}{\textbf{Summary}} & \multicolumn{1}{c|}{\textbf{Keywords}} \\ \hline
\endfirsthead

\multicolumn{3}{c}%
{{\bfseries \tablename\ \thetable{} -- Topics of users who commented on Tsai Ing-wen}} \\
\hline \multicolumn{1}{|c|}{\textbf{Topic Number}} &
\multicolumn{1}{c|}{\textbf{Summary}} &
\multicolumn{1}{c|}{\textbf{Keywords}} \\ \hline
\endhead

\hline \multicolumn{3}{|r|}{{Continued on next page}} \\ \hline
\endfoot

\hline \hline
\endlastfoot
1 & Commentators expect Tsai Ing-wen to reform the judicial system.& 改革 (reform)，正义 (justice)，司法 (judicial)，公平 (fair)，希望 (hope)，当选 (being successfully elected)，法官 (Judge)，年金 (annuity)，黑心 (evil mind)，不公 (unfair)，转型 (transition)，无罪 (innocent)，期待 (expect)，推动 (promote)，落实 (implement)\\
  \hline
  2 & Commentators express expectations of Tsai Ing-wen's policies on economics, education, and cross-strait relations.& 经济 (economic)，发展 (development)，产业 (industry)，国家 (nation)，政治 (politics)，执政 (take office)，解决 (solve)，教育 (education)，人才 (talented persons)，政策 (policy)，外交 (foreign affairs)，倒退 (go backwards)，停滞 (stagnation)，科技 (technology)，世界 (world)  \\
  \hline
  3 & Commentators remind Tsai Ing-wen to look after herself.& 加油 (cheer up)，总统 (president)，辛苦 (work hard)，小英 (a nickname of Tsai Ing-wen)，身体 (health)，保重 (take care of)，休息 (have a rest)，责任 (responsibility)，谢谢 (thanks)，照顾 (look after)，努力 (make efforts)，健康 (healthy)，烂摊子 (an awful mess)，未来 (in the future)，婚姻 (marriage)  \\
  \hline
  {\color{red}{4}} & {\color{red}{Commentators scold Tsai Ing-wen.}}&{\color{red}{贱人 (bitch)，台独 (Taiwan independence)，陈水扁 (Chen Shuibian)，中华民国 (Republic of China)，不好 (bad)，上台 (take office)，空心 (hollow)，对不起 (sorry)，李登辉 (Li Denghui)，支持 (support)，当选 (being elected)，陈建仁 (Chen Chien-jen)，马英九 (Ma Ying-Jeou)，选择 (select)，全家 (the whole family)}}\\
  \hline
  {\color{red}{5}} & {\color{red}{Commentators call Tsai Ing-wen the governor of Taiwan and take Taiwan as one of the provinces of China.}} & {\color{red}{省长 (governor of the province, Taiwan)，台湾省 (Taiwan province)，恭喜 (congratulations)，中华人民共和国 (the People's Republic of China)，主席 (chairman)，贺电 (message of congratulation)，同胞 (compatriot)，女士 (female)，当选 (being elected)，保护 (protect)，祝贺 (celebrate)，领导人 (leader)，法国 (France)，大使馆 (embassy)，协助 (assist)}}\\
  \hline
  6 & Commentators hope that Tsai Ing-wen will bring a stable life for Taiwan.& 希望 (hope)，百姓 (the common people)，幸福 (happiness)，生活 (life)，快乐 (happy)，总统 (president)，谦卑 (modest)，平安 (peace)，祝福 (wish)，安定 (stable)，未来 (in the future)，期待 (expect)，国家 (nation)，带来 (bring)，全民 (all the people)\\
  \hline
  7 & Commentators say that they do not want to be victims of the Communist Party's tyranny.& 支那 (Shina)，万岁 (long live)，共产党 (the Communist Party)，恐怖 (terrorist)，天安门 (Tian'anmen)，六四事件 (Tian'anmen Square Protests)，法国 (France)，中共 (the Communist Party)，崩溃 (collapse)，攻击 (attack)，打倒 (down with)，暴力 (violent)，推翻 (overthrowing), 谢谢 (thanks) \\
  \hline
  8 & Commentators expect Tsai Ing-wen to be the first female president of Taiwan.& 总统 (president)，第一 (first)，加油 (cheer up)，小英 (a nickname of Tsai Ing-wen)，希望 (hope)，期待 (expect)，明天 (tomorrow)，女性 (female)，未来 (in the future)，恭喜 (congratulations)，骄傲 (proud)，女人 (female)，诞生 (come into the world)，民选 (elected by people)，绿地 (fans of Tsai Ing-wen's faction)\\
  \hline
   {\color{red}{9}} & {\color{red}{Commentators criticize that Tsai Ing-wen is the running dog of Japan.}} &{\color{red}{日本 (Japan)，台湾 (Taiwan)，美国 (the United States)，慰安妇 (comfort woman)，台独 (Taiwan independence)，自愿 (voluntarily)，爸爸 (father)，李登辉 (Li Denghui)，杂种 (son of a bitch)，走狗 (stooge)，日杂 (bastard of Japan)，汉奸 (traitor)，南海 (the South China Sea)，骗子 (liar)，陈水扁 (Chen Shuibian)}} \\
  \hline
  10 & Commentators say that the KMT discredit Tsai Ing-wen.& 国民党 (the KMT)，朱立伦 (Eric Chu)，马英九 (Ma Ying-Jeou)，选票 (vote)，教训 (lesson)，小心 (take care of)，选举 (election)，抹黑 (discredit)，执政 (take office)，对手 (competitor)，反省 (introspection)，诚信 (credibility)，选民 (voter)，政党 (Party)，下台 (step down) \\
  \hline
  11 & Commentators hope that the DPP could occupy more than half of the seats in the congress.& 国会 (congress)，过半 (occupy more than half of the seats)，民进党 (the DPP)，政党 (Party)，改革 (reform)，时代 (the force of the moment)，支持 (support)，执政 (take office)，当选 (being elected)，英仁 (Tsai Ing-wen and Chen Chien-jen)，立法 (legislation)，法院 (court)，选票 (vote)，席次 (seats)，总统府 (presidential palace)\\
  \hline
  {\color{red}{12}} &{\color{red}{ Commentators say that Taiwan has been an indivisible part of the Chinese territory since ancient times.}} & {\color{red}{中国 (China)，台湾 (Taiwan)，不可分割 (inalienable)，永远 (always)，领土 (territory)，自古 (from ancient times)，统一 (unified)，中国共产党 (the Communist Party)，讨厌 (hate)，统治 (rule)，主权 (sovereignty)，国家 (nation)，分割 (separate)，文化 (cultural)，世界 (world)}} \\
  \hline
   {\color{red}{13}} & {\color{red}{Commentators support the use of military force to liberate Taiwan.}} & {\color{red}{独立 (independence)，台湾 (Taiwan)，台独 (Taiwan independence)，大陆 (mainland)，上台 (take office)，赶紧 (quickly)，支持 (support)，武力 (force against)，统一 (unified)，当选 (being elected)，解放军 (liberation army)，理由 (reason)，收复 (resume)，解放 (liberate)，导弹 (missile)}}\\
  \hline
   {\color{red}{14}} & {\color{red}{Commentators accuse Tsai Ing-wen of land speculation.}}&{\color{red}{地皮 (land)，蔡英文 (Tsai Ing-wen)，土地 (land)，正义 (justice)，家族 (family)，辩论 (debate)，民进党 (the DPP)，用心 (with heart)，抹黑 (discredit)，贪污 (corruption)，回报 (rewards)，空心 (hollow)，居住 (live)，投资 (investment)，不要脸 (have no sense of shame)}} \\
  \hline
  15 & Commentators congratulate Tsai Ing-wen on a successful election.& 总统 (president)，当选 (being elected)，恭喜 (congratulations)，小英 (a nickname of Tsai Ing-wen)，希望 (hope)，加油 (cheer up)，第一 (first)，开心 (happy)，终于 (finally)，青天白日旗 (Blue Sky with a White Sun)，责任 (responsibility)，初衷 (initial intention)，恭贺 (celebrate)，福气 (good luck)，任重道远 (take a heavy burden and embark on a long road)\\
  \hline
  16 & Commentators reply to and scold netizens possibly from the mainland who post a large number of negative comments on Tsai Ing-wen's Facebook homepage.& 自由 (freedom of speech)，民主 (democracy)，言论 (remarks)，对岸 (mainland)，留言 (messages on Facebook)，网友 (netizens)，大陆 (mainland)，评论 (comment)，享受 (enjoy)，发言 (speech)，开放世界 (open world)，光临 (welcome)，体验 (experience)，网路 (Internet)，感受 (feel) \\
  \hline
  17 & Commentators encourage netizens to return home to vote for Tsai Ing-wen.& 投票(vote)，选票(votes)，明天(tomorrow)，选举(election)，回家(go home)，回来(go back)，回去(be back)，年轻人(young people)，监票(scrutineer)，神圣(holy)，返乡(return home)，改变(change)，记得(remember)，朋友(friends)，担心(worry)\\
  \hline
   18& Commentators hope that Tsai Ing-wen will not disappoint the public after being in power.& 希望 (hope)，别让选民失望 (don't make voters disappointed), 期待 (look forward)，辜负 (let down)，期望 (expect)，承诺 (promise)，当选 (be elected)，初衷 (initial intention)，教育 (education)，说到做到 (suit the actions to the words)，未来 (in the future)，上任 (take office)，谢谢 (thanks)\\
  \hline
   19& Commentators hope that Tsai Ing-wen will make reforms for young people.& 希望 (hope)，未来 (in the future)，改变 (change)，加油 (cheer up)，期待 (expect)，年轻人 (young people)，努力 (make efforts)，下一代 (next generation)，年轻 (young)，孩子 (children)，创造 (create)，真心 (sincerely)，美好 (wonderful)，机会 (opportunity)，翻转 (reverse)\\
  \hline
  20 & Commentators say that the Chou Tzu-yu flag incident made people wake up and try to live with dignity. & 感谢 (thanks)，黄安 (Huang An)，国籍 (nationality)，全世界 (all over the world)，团结 (unite)，大声 (loudly)，尊严 (dignity)，骄傲 (proud)，告诉 (tell)，放弃 (give up)，宝岛 (Taiwan)，全家 (all the family)，自信 (confident)，觉醒 (awake)，找回 (get back)\\
  \hline
  21 & Commentators praise Tsai Ing-wen for her ability and lofty vision.& 领导人 (leader)，高度 (lofty vision)，小英 (a nickname of Tsai Ing-wen)，智慧 (intelligence)，格局 (setup)，主席 (chairman)，气度 (generous)，总统 (president)，展现 (exhibit)，团结 (unite)，包容 (tolerance)，态度 (attitude)，领袖 (leader)，加倍奉还 (multiply)，勇者 (a brave man)\\
  \hline
  22 & Commentators support the 1992 Consensus and oppose Taiwan independence.& 共识 (consensus)，两岸 (mainland and Taiwan)，现状 (status)，九二共识 (the 1992 Consensus)，关系 (relations)，承认 (admit)，和平 (peaceful)，台独 (Taiwan independence)，马习会 (Ma-Xi meeting)，大陆 (mainland)，反对 (oppose)，马英九 (Ma Ying-Jeou)，公开 (make public)，原则 (principle)，政策 (policy)\\
  \hline
  23 & Commentators criticize mainland netizens for a large number of negative comments on Tsai Ing-wen's Facebook homepage.& 大陆 (mainland)，台湾 (Taiwan)，网友 (netizen)，同胞 (compatriot)，朋友 (friends)，素质 (qualification)，反攻 (counterattack)，希望 (hope)，民众 (the common people)，开放 (open)，支持 (support)，解救 (rescue)，水深火热 (be filled with untold suffering)，中共 (the Communist Party)，统一 (unite)\\
  \hline
   24 & Commentators hope that Tsai Ing-wen will propose more specific political views in the television debates.&政策 (policy)，主席 (chairman)，辩论 (debate)，政见 (political view)，提出 (propose)，发表 (publish)，口号 (slogan)，解决 (solve)，看法 (opinion)，意义 (significance)，重点 (focal point)，宣传 (propagate)，内容 (content)，候选人 (candidate)，参加 (participate)  \\
  \hline
   {\color{red}{25}}  &  {\color{red}{Commentators sneer at Tsai Ing-wen that she dare not claim Taiwan Independence.}}&  {\color{red}{独立 (independence)，台湾 (Taiwan)，支持 (support)，建国 (found a state)，当选 (be elected)，主席 (chairman)，万岁 (long live)，大陆 (mainland)，早日 (early)，大妈 (aunty)，成立 (establish)，建立 (found)，废除 (abolish)，共和国 (republic)，带领 (lead)}}\\
  \hline
  26 & Commentators claim that the mainland is ruled with absolute power.&民主 (democracy)，自由 (freedom)，进步党 (progressive party)，独裁 (dictatorship), 万岁 (long live)，华人(Chinese)，全世界 (all over the world)，笑话 (joke)，人权 (human rights)，专制 (autocratic)，捍卫 (defend)，全球 (global)，国家 (nation)，社会 (social)\\
  \hline
 {\color{red}{27}} & {\color{red}{Commentators scold Tsai Ing-wen as a shameless politico.}}& {\color{red}{悲哀 (sad)，可怜 (poor)，小姐 (Ms)，女人 (female)，无耻 (shameless)，天下 (the world)，选票 (vote)，政客 (politico)，不要脸 (have no sense of shame)，百姓 (common people)，眼睛 (eyes)，世界 (world)，可笑 (joke)，说话 (remarks)，实在 (really)}}\\
  \hline
 {\color{red}{28}} & {\color{red}{Commentators criticize that Tsai Ing-wen proposes policies for the Hakka but ignores other ethnic groups. }}& {\color{red}{民意 (public opinion)，选择 (select)，代表 (represent)，选举 (election)，客家人 (Hakka)，尊重 (respect)，声音 (sound)，感觉 (feel)，族群 (ethnic groups)，团结 (unite)，文化 (cultural)，接受 (accept)，倾听 (listen attentively to)，政府 (government)，运动 (movement)}} \\
  \hline
   29& Commentators support Taiwan independence.&中华民国 (Republic of China)，国家 (nation)，国旗 (flag)，总统 (president)，国际 (International)，认同 (identification)，承认 (admit)，中华 (China)，宪法 (constitution)，代表 (represent)，联合国 (the United Nations)，国歌 (national anthem)，国号 (national title)，主权 (sovereignty)，台独 (Taiwan independence)\\
  \hline
  30& Commentators accuse Ma Ying-Jeou of incompetence.&政府 (government)，总统 (president)，国民党 (the KMT)，马英九 (Ma Ying-Jeou)，中华民国 (Republic of China)，无能 (incompetence)，政党 (party)，下台 (step down)，流亡 (exile)，政权 (political power)，不好 (bad)，责任 (responsibility)，诈骗 (cheat)，中共 (the Communist Party)，灭亡 (die out)\\
  \hline
  {\color{red}{31}}& {\color{red}{Commentators criticize that Tsai Ing-wen's policies are devoid of substance.}}& {\color{red}{政治 (politics)，选举 (election)，政客 (politico)，空心 (hollow)，民众 (common people)，利益 (benefit)，口号 (slogan)，选票 (vote)，政党 (party)，欺骗 (cheat)，上台 (take office)，抹黑 (discredit)，对立 (oppose)，无耻 (shameless)，利用 (make use of)}}\\
  \hline
  32 & Commentators feel touched because of Tsai Ing-wen's successful election.& 小英 (a nickname of Tsai Ing-wen)，感动 (touched)，谢谢 (thanks)，终于 (finally)，四年 (four years)，明天 (tomorrow)，胜利 (victory)，总统 (president)，期待 (expect)，感谢 (thanks)，辛苦 (work hard)，加油 (cheer up)，迎接 (welcome)，感言 (reflections)，感恩 (feel grateful)\\
  \hline
  {\color{red}{33}}&{\color{red}{Commentators post the content of "eight honors and eight shames" of the mainland.}}& {\color{red}{小英 (a nickname of Tsai Ing-wen)，民主 (democracy)，贱人 (bitch)，建仁 (Chen Chien-jen)，价值 (value)，自由 (free, one of the keywords of eight honors and eight shames proposed by mainland China)，核心 (core)，平等 (equal, one of the keywords of the eight honors and eight shames proposed by mainland China)，文明 (civil, one of the keywords of the eight honors and eight shames proposed by mainland China)，爱国 (patriotic, one of the keywords of the eight honors and eight shames proposed by mainland China)，诚信 (sincere, one of the keywords of the eight honors and eight shames proposed by mainland China)，和谐 (harmonious, one of the keywords of the eight honors and eight shames proposed by mainland China)，友善 (friendly, one of the keywords of the eight honors and eight shames proposed by mainland China)，公正 (judicial, one of the keywords of the eight honors and eight shames proposed by mainland China)，阿姨 (aunty)}}\\
  \hline
  34 & Commentators hope Tsai Ing-wen will focus on the issue of wages.& 劳工 (labor)，希望 (hope)，工作 (job)，年轻人 (young people)，生活 (life)，社会 (social)，照顾 (take care of)，政策 (policy)，改善 (improve)，弱势 (disadvantaged)，政府 (government)，福利 (welfare)，小孩 (children)，薪资 (salary)，经济 (economic)\\
  \hline
  35 & Commentators express their anger over the Chou Tzu-yu flag incident.& 周子瑜 (Chou Tzu-yu)，道歉 (apology)，选票 (vote)，事件 (incident)，国旗 (national flag)，关心 (care for)，公道 (justice)，欺负 (bully)，希望 (hope)，孩子 (children)，女孩 (girl)，伤害 (hurt)，可怜 (poor)，黄安 (Huang An)，发声 (state)\\
  \hline
  36 & Commentators possibly from mainland scold Taiwan netizens.& 天涯 (one of the famous Chinese)，发电 (send a message)，观光团 (tourist party)，大陆 (mainland)，评论 (comment)，前来 (come here)，围观 (onlookers)，弯弯 (Taiwanese)，茶叶蛋 (tea eggs)，听说 (hear about)，台巴子 (Taiwanese)，独立 (independence)，民主 (democracy)，出征 (go to battle)，吱吱 (Taiwanese)\\
  \hline
\end{longtable}
\end{center}
\end{CJK}

\clearpage
\begin{CJK}{UTF8}{gkai}
\begin{center}
\begin{longtable}{|p{.1\textwidth}|p{.2\textwidth}|p{.7\textwidth}|}
\caption[Topics of overlapping users of Eric Chu]{Topics of overlapping users of Eric Chu}\label{tab:lc_llchu} \\

\hline \multicolumn{1}{|c|}{\textbf{Topic Number}} & \multicolumn{1}{c|}{\textbf{Summary}} & \multicolumn{1}{c|}{\textbf{Keywords}} \\ \hline
\endfirsthead

\multicolumn{3}{c}%
{{\bfseries \tablename\ \thetable{} -- Topics of overlapping users of Eric Chu}} \\
\hline \multicolumn{1}{|c|}{\textbf{Topic Number}} &
\multicolumn{1}{c|}{\textbf{Summary}} &
\multicolumn{1}{c|}{\textbf{Keywords}} \\ \hline
\endhead

\hline \multicolumn{3}{|r|}{{Continued on next page}} \\ \hline
\endfoot

\hline \hline
\endlastfoot
1 & Commentators expect Eric Chu to be a leader of Taiwan. & 国民党 (the KMT)，希望 (hope)，未来 (in the future)，朱玄 (Eric Chu and Wang Ju-hsuan)，辛苦 (work hard)，政党 (political party)，期待 (look forward)，带领 (lead)，振作 (cheer up)，神圣 (holy)，表现 (performance)，佩服 (admire)，迈向 (march toward)，投票 (vote)，全家人 (the whole family)\\
  \hline
  2 & Commentators express their support for Eric Chu's campaign team.& 支持 (support)，力挺 (fervently support)，投票 (vote)，乡亲 (folks)，热情 (enthusiasm)，第一 (first)，游行 (march)，黄志雄 (Huang Chih-hsiung)，高雄 (Kaohsiung)，阿婆 (grandma)，客家 (Hakka)，温暖 (warm)，翻转 (reverse)，服务 (serve)，当选 (be elected)\\
  \hline
  3 & Commentators express expectations of Eric Chu's policies on economics, education, and cross-strait relations.& 发展 (develop)，未来 (in the future)，希望 (hope)，政策 (policy)，稳定 (stable)，创业 (start a business)，和平 (peaceful)，青年 (youth)，产业 (industry)，两岸 (mainland and Taiwan)，年轻 (young)，务实 (practical)，开放 (open)，支持 (support)，财经 (finance) \\
  \hline
  4 & Commentators encourage Eric Chu not to give up and stage a comeback after four years.& 加油 (cheer up)，市长 (mayor)，永不放弃 (never give up)，修女 (nun)，道歉 (apology)，阿伦 (Eric Chu)，选战 (campaign)，当选 (be elected)，痛定思痛 (learn from the painful experience)，轻易 (easily)，忍心 (have the heart to)，厉害 (excellent)，大众 (public)，再来 (again)，感谢 (appreciate) \\
  \hline
  5 & Commentators expect that Eric Chu could promote economic development.& 加油 (cheer up)，支持 (support)，经济 (economic)，安定 (stable)，力量 (power)，带领 (lead)，逆转 (reverse)，王如玄 (Wang Ju-hsuan)，温暖 (warm)，专业 (professional)，安心 (peace of mind)，团结 (unite)，领导人 (leader)，生活 (life)，正面 (positive)\\
  \hline
  6 & Commentators hope that the KMT could occupy more than half of the seats in the congress.& 支持 (support)，国民党 (the KMT)，加油 (cheer up)，朱立伦 (Eric Chu)，唯一 (only)，力量 (power)，主席 (chairman)，总统 (president)，国会 (congress)，过半 (occupy more than half of the seats)，团结 (unite)，当选 (be elected)，长期 (long-term)，胜选 (win the election)，王如玄 (Wang Ju-hsuan)\\
  \hline
  7 & Commentators express their anger over the Chou Tzu-yu flag incident.& 中华民国 (Republic of China)，台独 (Taiwan independence)，国旗 (national flag)，国家 (nation)，共识 (consensus)，政治 (politics)，周子瑜 (Chou Tzu-yu)，九二共识 (the 1992 Consensus)，团队 (team)，心中 (in the heart)，勇敢 (brave)，接受 (accept)，国歌 (national anthem)，逆转 (reverse)，清醒 (awake) \\
  \hline
  8 & Commentators hope Eric Chu would provide more opportunities for young people.& 检讨 (self-criticism)，反省 (self-reflection)，改革 (reform)，年轻 (young)，培养 (cultivate)，智慧 (intelligence)，年轻人 (youth)，出发 (set out)，承诺 (promise)，民心 (popular support)，基层 (the basic level)，族群 (ethnic groups)，声音 (sound)，用心 (attentively)，倾听 (listen attentively to)\\
  \hline
  9 & Commentators praise Eric Chu for his hard work.& 辛苦 (work hard)，谢谢 (thanks)，努力 (make efforts)，感谢 (appreciate)，付出 (efforts)，警察 (police)，市长 (mayor)，人员 (staff)，工作 (job)，能力 (capability)，感恩 (feel grateful)，医护 (doctor and nurse)，下一代 (next generation)，前途 (prospect)，有目共睹 (be there for all to see)\\
  \hline
  10 & Commentators from southern Taiwan support Eric Chu.& 好好 (well)，选举 (election)，南部 (south)，支持者 (supporter)，感动 (touched)，骄傲 (proud)，一刻 (the moment)，失望 (disappointed)，嘉义 (Ciayi)，参加 (participate)，祝福 (wish)，信心 (confidence)，工作 (job)，好多 (a lot of)，真的 (really)\\
  \hline
  11 & Commentators hope Eric Chu will protect the flag of the Republic of China.& 国旗 (national flag)，投票 (vote)，青天白日旗 (Blue Sky with White Sun)，国民 (countrymen)，淡水阿嬷 (grandma)，感动 (touched)，党员 (party member)，理性 (reasonable)，帮忙 (help)，感恩 (appreciate)，飘扬 (waving)，觉醒 (awake)，尊严 (dignity)，踊跃 (active)，政绩 (achievements in official career) \\
  \hline
  12 & Commentators wish Eric Chu a happy new year.& 加油 (cheer up)，快乐 (happy)，平安 (safe)，新年 (new year)，祝福 (wish)，当选 (be elected)，健康 (healthy)，身体 (health)，幸福 (happiness)，美好 (wonderful)，早安 (good morning)，保重 (take care)，如意 (the best)，喜乐 (gladness)，顺利 (everything goes well)\\
  \hline
  13 & Commentators hope Eric Chu will reform education policies.& 主席 (chairman)，改革 (reform)，希望 (hope)，教育 (education)，人才 (talented persons)，国会 (congress)，反对 (oppose)，建议 (advice)，改变 (change)，办法 (methods)，小孩 (children)，党内 (in the party)，学生 (student)，父母 (parents)，制度 (system)\\
  \hline
  14 & Commentators are against the imports of American pork with lean meat essence, which is possibly supported by the DPP.& 蔡英文 (Tsai Ing-wen)，瘦肉 (lean meat)，总统 (president)，美国 (the United States)，当选 (be elected)，中华民国 (Republic of China)，开放 (open)，贪污 (corruption)，反对 (oppose)，正义 (justice)，进口 (import)，暴力 (violence)，猪肉 (pork)，拒绝 (refuse)，全民 (the whole people) \\
  \hline
  15 & Commentators criticize Tsai Ing-wen and James Soong.& 蔡英文 (Tsai Ing-wen)，民进党 (the DPP)，执政 (take office)，高雄 (Kaohsiung)，地皮 (land)，宋楚瑜 (James Soong)，陈水扁 (Chen Shuibian)，日本 (Japan)，台北 (Taipei)，台南 (Tainan)，绿营 (faction of Tsai Ing-wen)，抹黑 (disgrace)，游行 (march)，流亡 (exile)，贿选 (vote-buying)\\
  \hline
  16 & Commentators hope Eric Chu will bring a stable life for Taiwan.&安定 (stable)，台湾 (Taiwan)，社会 (society)，和平 (peaceful)，国家 (nation)，国民党 (the KMT)，执政 (take office)，生活 (life)，繁荣 (prosperous)，信心 (confidence)，百姓 (the common people)，两岸 (mainland and Taiwan)，感谢 (appreciate)，放弃 (give up)，幸福 (happiness)\\
  \hline
  17 & Commentators expect Eric Chu to propose polices in favor of young people.& 未来 (in the future)，中华民国 (Republic of China)，幸福 (happiness)，总统府 (presidential palace)，经济 (economic)，下一代 (next generation)，努力 (make efforts)，团结 (unite)，朱玄 (Eric Chu and Wang Ju-hsuan)，福祉 (happiness and benefit)，繁荣 (prosperous)，政见 (political view)，加薪 (increase salary )，子孙 (descendants)，最佳 (best)\\
  \hline
\end{longtable}
\end{center}
\end{CJK}

\clearpage
\begin{CJK}{UTF8}{gkai}
\begin{center}
\begin{longtable}{|p{.1\textwidth}|p{.2\textwidth}|p{.7\textwidth}|}
\caption[Topics of users who commented on Eric Chu]{Topics of users who commented on Eric Chu}\label{tab:c_llchu} \\

\hline \multicolumn{1}{|c|}{\textbf{Topic Number}} & \multicolumn{1}{c|}{\textbf{Summary}} & \multicolumn{1}{c|}{\textbf{Keywords}} \\ \hline
\endfirsthead

\multicolumn{3}{c}%
{{\bfseries \tablename\ \thetable{} -- Topics of users who commented on Eric Chu}} \\
\hline \multicolumn{1}{|c|}{\textbf{Topic Number}} &
\multicolumn{1}{c|}{\textbf{Summary}} &
\multicolumn{1}{c|}{\textbf{Keywords}} \\ \hline
\endhead

\hline \multicolumn{3}{|r|}{{Continued on next page}} \\ \hline
\endfoot

\hline \hline
\endlastfoot
 {\color{red}{1}} & {\color{red}{Commentators criticize Eric Chu of being shameless and telling lies.}}&{\color{red}{无耻 (shameless)，不要脸 (have no sense of shame)，政客 (politico)，恶心 (feel nauseated)，骗子 (liar)，说谎 (tell lies)，诚信 (credibility)，诈骗 (cheat)，集团 (group)，垃圾 (shit)，天下 (the world)，谎言 (lies)，无能 (incompetence)，可怜 (poor)，虚伪 (untruthful)}}\\
  \hline
  {\color{red}{2}} &{\color{red}{Commentators criticize the KMT for its failure to protect Chou Tzu-yu.}} &{\color{red}{保护 (protect)，周子瑜 (Chou Tzu-yu)，执政党 (ruling party)，道歉 (apology)，政府 (government)，回家 (go home)，消费 (consume)，中国 (China)，女孩 (girl)，欺负 (bully)，事件 (incident)，国民 (countrymen)，抗议 (protest)，孩子 (children)，公开 (public)}} \\
  \hline
  {\color{red}{3}}& {\color{red}{Commentators scold Eric Chu.}}&{\color{red}{淡水 (a place of Taiwan)，广告 (advertisement)，丢脸 (have no sense of shame)，留言 (leave message)，老婆 (wife)，笑话 (joke)，对不起 (sorry)，佩服 (respect)，厉害 (really bad)，党徽 (emblem of a political party)，不用 (don't have to)，原谅 (forgive)，下辈子 (next generation)，可怜 (poor)，说谎 (tell lies)}}\\
  \hline
  4 & Commentators hope Eric Chu will focus on the issue of wages.& 劳工 (labor)，政府 (government)，薪资 (salary)，经济 (economic)，工资 (wage)，薪水 (salary)，老板 (boss)，工作 (job)，执政 (take office)，加薪 (increase salary)，企业 (entrepreneur)，休假 (have a holiday)，补助 (allowance)，请假 (off work)，政策 (policy)\\
  \hline
  5 & Commentators hope the KMT will occupy more than half of the seats in the congress.& 国民党 (the KMT)，改革 (reform)，国会 (congress)，总统 (president)，王金平 (Wang Jin-pyng)，政党 (party)，第一 (first)，院长 (chief)，立法 (legislation)，法院 (court)，过半 (occupy more than half of the seats)，执政 (take office)，委员 (committee member)，席次 (seat)，代表 (representative)\\
  \hline
  6 & Commentators hope Eric Chu will bring a better life for Taiwan citizens.& 加油 (cheer up)，未来 (in the future)，希望 (hope)，带领 (lead)，经济 (economic)，主席 (chairman)，走向 (head for)，世界 (global)，天佑 (God blesses)，美好 (wonderful)，国际 (international)，总统 (president)，生活 (life)，领导 (leader)，下一代 (next generation) \\
  \hline
  7 & Commentators wish Eric Chu a happy new year.& 加油 (cheer up)，主席 (president)，快乐 (happy)，祝福 (wish)，当选 (be elected)，辛苦 (work hard)，平安 (safe)，顺利 (everything goes well)，努力 (make efforts)，新年 (new year)，成功 (successful)，健康 (healthy)，感恩 (feel grateful)，幸福 (happiness)，谢谢 (thanks)\\
  \hline
  {\color{red}{8}} & {\color{red}{Commentators accuse Wang Ju-hsuan, Eric Chu's campaign partner, of land speculation.}}&{\color{red}{ 王如玄 (Wang Ju-hsuan)，朱立伦 (Eric Chu)，蔡英文 (Tsai Ing-wen)，土地 (land)，副手 (deputy)，地皮 (land)，小英 (a nickname of Tsai Ing-wen)，蔡正元 (Tsai Cheng-yuan)，邱毅 (Chiu Yi)，公布 (publish)，财产 (property)，买卖 (buying and selling)，正义 (justice)，抹黑 (disgrace)，公开 (public)}}\\
  \hline
  {\color{red}{9}} & {\color{red}{Commentators believe that the KMT needs self-criticism.}}&{\color{red}{国民党 (the KMT)，检讨 (self-criticism)，反省 (self-reflection)，希望 (hope)，执政 (take office)，机会 (opportunity)，感觉 (feel)，改革 (reform)，执政党 (ruling party)，在野党 (opposition party)，改变 (change)，期待 (look forward)，下次 (next time)，四年 (four years)}}\\
  \hline
  10 & Commentators encourage Eric Chu not to be discouraged and to participate in the election after four years.& 加油 (cheer up)，努力 (make efforts)，支持 (support)，永远 (forever)，主席 (chairman)，辛苦 (work hard)，失败 (failure)，成功 (success)，希望 (hope)，未来 (in the future)，气馁 (discourage)，四年 (four years)，期待 (look forward)，没关系 (it doesn't matter)，回来 (go back)\\
  \hline
 {\color{red}{11}} & {\color{red}{Commentators accuse the KMT of corruption and criticize that the ruling party did not solve the problems with respect to the welfare of civil servants.}} & {\color{red}{国民党 (the KMT)，执政 (take office)，无罪 (innocent)，黑心 (evil mind)，政府 (government)，改革 (reform)，司法 (judicial)，人员 (officials)，军公教 (civil servants)，百姓 (common people)，警察 (police)，贿选 (vote-buying)，标准 (standard)，退休 (retire)，法官 (judge)}} \\
  \hline
 {\color{red}{12}}& {\color{red}{Commentators are against Eric Chu because the KMT abolished the nomination of Hong Xiuzhu in the 2016 presidential election and replaced her with Eric Chu.}}& {\color{red}{国民党 (the KMT)，投票 (vote)，废票 (invalid ballot)，王金平 (Wang Jin-pyng)，历史 (history)，洪秀柱 (Hong Hsiu-chu)，下台 (step down)，柱姐 (a nickname of Hong Hsiu-chu)，第一 (first)，明年 (next year)，选情 (election prospect)，罪人 (sinner)，含泪 (tearful)，拒绝 (refuse)，辅选 (assist Eric Chu in the election)}}\\
  \hline
  {\color{red}{13}} & {\color{red}{Commentators accuse Ma Ying-Jeou of corruption and incompetence.}}&{\color{red}{国民党 (the KMT)，马英九 (Ma Ying-Jeou)，朱立伦 (Eric Chu)，贪污 (corruption)，剩下 (remaining)，支持者 (supporter)，执政 (take office)，政党 (party)，选票 (vote)，淡水 (a place of Taiwan)，能力 (competence)，黑金 (black money)，政绩 (achievements in one's official career)，候选人 (candidate)，无能 (incompetence)}} \\
  \hline
  {\color{red}{14}} &{\color{red}{Commentators encourage netizens to vote for James Soong to avoid parties' conflicts of the DPP and the KMT.}}&{\color{red}{宋楚瑜 (James Soong)，小英 (a nickname of Tsai Ing-wen)，总统 (president)，支持 (support)，能力 (competence)，投票 (vote)，主席 (chairman)，保宋弃朱 (support James Soong and give up voting for Eric Chu)，时代 (times)，机会 (opportunity)，百姓 (common people)，全民 (the whole people)，建议 (advise)，恶斗 (tough fight)，改变 (change)}}\\
  \hline
  15& Commentators criticize Tsai Ing-wen.&空心菜 (hollow Tsai)，民进党 (the DPP)，空心 (hollow)，暴力 (violence)，对立 (the opposition)，地皮 (land)，蔡英文 (Tsai Ing-wen)，媒体 (media)，水准 (level)，绿营 (faction of Tsai Ing-wen)，制造 (make)，打败 (defeat)，抹黑 (disgrace)，恐怖 (horrible)，落选 (lose the election)\\
  \hline
  16 & Commentators are angry over the Chou Tzu-yu flag incident.& 国旗 (national flag)，台独 (Taiwan independence)，中华民国 (Republic of China)，黄安 (Huang An)，周子瑜 (Chou Tzu-yu)，青天白日旗 (Blue Sky with White Sun)，中国 (China)，大陆 (mainland)，小心 (take care of)，检举 (impeach)，捍卫 (defend)，封杀 (ban)，投票 (vote)，举报 (tip-off)，道歉 (apology)\\
  \hline
  {\color{red}{17}}& {\color{red}{Commentators criticize the KMT because the party abolished the nomination of Hong Xiuzhu through unjust procedures.}} & {\color{red}{国民党 (the KMT)，民主 (democracy)，党员 (party member)，包容 (tolerance)，骗人 (cheat)，团结 (unite)，制度 (system)，柱柱 (a nickname of Hong Hsiu-chu)，反对 (oppose)，党内 (in the party)，开除 (be fired)，尊重 (respect)，党籍 (party membership)，正义 (justice)，程序 (procedure)}}\\
  \hline
  18 & Commentators expect Eric Chu to propose polices in favor of young people.& 年轻人 (young people)，民意 (public opinions)，年轻 (young)，基层 (the basic level)，希望 (hope)，人才 (talented persons)，民众 (the public)，机会 (opportunity)，国家 (country)，百姓 (common people)，声音 (sound)，世代 (generation)，青年 (youth)，培养 (cultivate)，未来 (in the future) \\
  \hline
  {\color{red}{19}}& {\color{red}{Eric Chu is criticized of breaking his promise that he would not run for the presidential election when he was the mayor of New Taipei city.}}& {\color{red}{诚信 (credibility)，人物 (figure)，政客 (politico)，政党 (party)，选举 (election)，说话 (speech)，利益 (benefit)，选票 (vote)，道德 (morality)，承诺 (promise)，结束 (end)，良心 (conscience)，面对 (confront with)，可耻 (shameless)，自私 (selfish)}}\\
  \hline
  20 & Commentators encourage Eric Chu to return to normal work after his failure in the election.& 市长 (mayor)，回去 (go back)，明天 (tomorrow)，回来 (go back)，记得 (remember)，上班 (go to work)，好好 (well)，回家 (go home)，市政府 (municipal government)，请假 (off work)，市民 (citizens)，明年 (next year)，乖乖 (quietly)，辛苦 (work hard)，伦伦 (a nickname of Eric Chu)\\
  \hline
  21 & Commentators support Taiwan independence.&中华民国 (Republic of China)，台湾 (Taiwan)，万岁 (long live)，国家 (country)，中国 (China)，国民党 (the KMT)，世界 (world)，独立 (independence)，政府 (government)，捍卫 (defend)，统一 (unite)，灭亡 (die out)，代表 (represent)，精神 (spirit)，认同 (identification)\\
  \hline
  22 & Commentators discuss Eric Chu's performance in the television debates.& 政见 (political views)，辩论 (debate)，主席 (chairman)，竞选 (campaign)，候选人 (candidate)，广告 (advertisement)，政策 (policy)，对手 (opponent)，电视 (television)，发表 (make a statement)，攻击 (attack)，网路 (Internet)，留言 (leave messages)，选民 (voter)，不错 (good)\\
  \hline
  23 & Commentators hope Eric Chu will bring a stable and peaceful life for Taiwan citizens.& 台湾 (Taiwan)，经济 (economic)，和平 (peaceful)，两岸 (mainland and Taiwan)，希望 (hope)，国家 (country)，安定 (stable)，百姓 (common people)，生活 (life)，发展 (develop)，执政 (take office)，未来 (in the future)，幸福 (happiness)，社会 (social)，稳定 (stable)\\
  \hline
\end{longtable}
\end{center}
\end{CJK}

\clearpage
\begin{CJK}{UTF8}{gkai}
\begin{center}
\begin{longtable}{|p{.1\textwidth}|p{.2\textwidth}|p{.7\textwidth}|}
\caption[Topics of overlapping users of James Soong]{Topics of overlapping users of James Soong}\label{tab:lc_soong} \\

\hline \multicolumn{1}{|c|}{\textbf{Topic Number}} & \multicolumn{1}{c|}{\textbf{Summary}} & \multicolumn{1}{c|}{\textbf{Keywords}} \\ \hline
\endfirsthead

\multicolumn{3}{c}%
{{\bfseries \tablename\ \thetable{} -- Topics of overlapping users who Like and commented on James Soong}} \\
\hline \multicolumn{1}{|c|}{\textbf{Topic Number}} &
\multicolumn{1}{c|}{\textbf{Summary}} &
\multicolumn{1}{c|}{\textbf{Keywords}} \\ \hline
\endhead

\hline \multicolumn{3}{|r|}{{Continued on next page}} \\ \hline
\endfoot

\hline \hline
\endlastfoot

\hline
  1 & Commentators praise James Soong for his diligence in government affairs.& 宋楚瑜 (James Soong)，总统 (president)，爱民 (love of effect)，勤政 (be assiduous in government affairs)，新年 (new year)，慈悲 (mercy)，稳健 (steady)，蔓延 (spread)，守护神 (patron saint)，徐欣莹 (Hsu Hsin-ying)，延续 (go on)，领导 (leader)，心肠 (heart)，组合 (group)，冲刺 (sprint)\\
  \hline
  2 & Commentators have confidence in James Soong.& 省长 (governor of a province)，人物 (figure)，民众 (the public)，政治 (politics)，时代 (times)，信心 (confidence)，真心 (sincerely)，成功 (success)，认同 (identification)，台南 (Tainan)，中小企业 (small- and medium-sized enterprises)，一半 (half)，宋镇迈 (Song Zhenmai)，曙光 (dawn)，废除 (abolish) \\
  \hline
  3 & Commentators hope that James Song will bring a better life for Taiwan.& 带领 (lead)，希望 (hope)，期待 (look forward)，公道 (fair)，边缘化 (marginalize)，政治家 (politico)，超越 (surpass)，理想 (ideal)，绑架 (kidnap)，成绩 (achievement)，当天 (that very day)，同心 (be of one heart)，感谢 (appreciate)，国父 (the Father of a nation)，胆识 (courage and insight) \\
  \hline
  4 & Commentators praise James Soong for his English proficiency.& 能力 (competence)，第一 (first)，总统 (president)，真心 (sincere)，回答 (answer)，说到做到 (suit the actions to the words)，翻译 (translate)，安心 (feel at ease)，摆脱 (get rid of)，主席 (chairman)，手势 (gesture)，严重 (serious)，厉害 (powerful)，鼓励 (encourage)，傻眼 (be dumbfounded) \\
  \hline
  5 & Commentators praise James Soong for his excellent performance in the television debates.& 宋楚瑜 (James Soong)，总统 (president)，徐欣莹 (Hsu Hsin-ying)，支持 (support)，现场 (on live)，辩论 (debate)，明日 (tomorrow)，亲爱 (dear)，今晚 (tonight)，甘苦 (sweetness and bitterness)，感人 (moving)，实况 (live)，喜乐 (happy)，报告 (report)，加油 (cheer up)\\
  \hline
  6 & Commentators show their firm support for James Soong.& 永远 (forever)，总统 (president)，心中 (in the heart)，唯一 (only)，我爱 (I love)，骄傲 (proud)，值得 (it deserves)，眼泪 (tear)，领袖 (leader)，珍惜 (cherish)，支持者 (supporter)，一家 (the whole family)，留下 (leave)，优质 (high quality)，好日子 (good days)\\
  \hline
  7 & Commentators appreciate the electioneering of James Soong's daughter.& 爸爸 (father)，选票 (vote)，下一代 (next generation)，心疼 (love dearly)，女儿 (daughter)，感动 (touched)，声音 (sound)，心情 (mood)，充满 (be filled with)，证明 (demonstrate)，空心菜 (hollow Tsai)，信心 (confidence)，当选 (be elected)，可爱 (lovely)，播放 (broadcast)\\
  \hline
  8 & Commentators accuse the KMT and the DPP of blocking James Soong's remarks.& 媒体 (media)，网路 (Internet)，国际 (international)，报导 (report)，新闻 (news)，骄傲 (proud)，电视 (television)，封锁 (be blocked)，格局 (setup)，高度 (lofty vision)，发声 (make a statement)，辩论会 (debate)，领导人 (leader)，直播 (broadcast)，伯伯 (uncle) \\
  \hline
  9 & Commentators praise James Soong for his competence and experience.&能力 (competence)，治国 (manage state affairs)，经验 (experience)，感谢 (appreciate)，环境 (environment)，担心 (worry)，感恩 (feel grateful)，市长 (mayor)，改善 (improve)，承担 (undertake)，家族 (the family)，定莹 (must win)，两岸 (mainland and Taiwan)，流泪 (burst into tears)，救国救民 (save the country and the people)\\
  \hline
  10 & Commentators cheer up James Soong.&加油 (cheer up)，找出路 (finding a way out for Taiwan)，宋北北 (a nickname of James Soong)，金门 (Kinmen)，子孙 (generation)，好友 (friends)，爱护 (cherish)，将心比心 (feel of others)，大头照 (headshot)，专程 (special trip)，承担 (undertake)，一路 (all the way)，本钱 (foundation)，太阳 (sun)，乡民 (villagers) \\
  \hline
  11 & Commentators praise Eric Chu for his hard work.&谢谢 (thanks)，福气 (good fortune)，付出 (efforts)，感谢 (appreciate)，做事 (work)，尽力 (try best)，领导人 (leader)，默默 (silently)，历史 (history)，恳请 (please)，简单 (simple)，抱歉 (sorry)，宋北北 (a nickname of James Soong)，优秀 (excellent)，老当益壮 (become vigorous with age)\\
  \hline
  12 & Commentators praise James Soong for his good quality.& 选贤 (select wisdom)，选民 (voter)，选举 (election)，民主 (democracy)，政党 (party)，投票 (vote)，明天 (tomorrow)，选票 (vote)，悲哀 (sad)，年龄 (age)，素养 (quality)，无愧于 (be worthy of)，选择 (select)，土地 (land)，救星 (savior)\\
  \hline
  13 & Commentators hope James Soong will provide a fair life for young people.& 希望 (hope)，执政 (take office)，选举 (election)，朋友 (friend)，年轻 (young)，带来 (bring)，坚定 (firm)，年轻人 (the young people)，选票 (vote)，日子 (life)，请问 (excuse me)，救救 (save)，说话 (speak)，公平 (fair)，将来 (in the future) \\
  \hline
  14 & Commentators encourage netizens to vote for James Soong.& 宋进总统府 (support James Soong to be elected)，宋莹 (James Soong and Hsu Hsin-ying)，当选 (be elected)，老天 (God)，改革 (reform)，见到 (see)，法院 (court)，恐龙法官 (the verdict of the social expectation)，边缘 (marginal)，逆转 (reverse)，拒绝 (refuse)，才会莹 (will win)，人山人海 (huge crowds of people)，骨力 (vigorous strokes)\\
  \hline
  15 & Commentators wish James Soong a happy new year.&快乐 (happy)，新年 (new year)，健康 (health)，平安 (safe)，祝福 (bless)，早安 (good morning)，领导人 (leader)，幸福 (happiness)，顺利 (everything goes well)，力挺 (fervently support)，愉快 (happy)，顺心 (satisfactory)，特地 (specially)，期望 (look forward)，热潮 (upsurge)\\
  \hline
  16 & Commentators hope James Soong will be elected and believe his victory would end the partisan battles of the KMT and the DPP.& 恶斗 (tough fight)，终结 (end)，总统府 (presidential palace)，团结 (unite)，颜色 (color)，经济 (economic)，选民 (voter)，未来 (in the future)，唯一 (only)，发展 (develop)，反省 (self-reflection)，抹黑 (disgrace)，人心 (the will of people)，身边 (those around you)，凝聚 (condensation)\\
  \hline
\end{longtable}
\end{center}
\end{CJK}

\clearpage
\begin{CJK}{UTF8}{gkai}
\begin{center}
\begin{longtable}{|p{.1\textwidth}|p{.2\textwidth}|p{.7\textwidth}|}
\caption[Topics of users who commented on James Soong]{Topics of users who commented on James Soong}\label{tab:c_soong} \\

\hline \multicolumn{1}{|c|}{\textbf{Topic Number}} & \multicolumn{1}{c|}{\textbf{Summary}} & \multicolumn{1}{c|}{\textbf{Keywords}} \\ \hline
\endfirsthead

\multicolumn{3}{c}%
{{\bfseries \tablename\ \thetable{} -- Topics of users who commented on James Soong}} \\
\hline \multicolumn{1}{|c|}{\textbf{Topic Number}} &
\multicolumn{1}{c|}{\textbf{Summary}} &
\multicolumn{1}{c|}{\textbf{Keywords}} \\ \hline
\endhead

\hline \multicolumn{3}{|r|}{{Continued on next page}} \\ \hline
\endfoot

\hline \hline
\endlastfoot
\hline
  1 & Commentators criticize Tsai Ing-wen for her empty policies.& 主席 (chairman)，总统 (president)，英文 (English)，空心菜 (hollow Tsai)，候选人 (candidate)，年轻人 (the young people)，辩论 (debate)，参选 (participate in the election)，不用 (don't have to)，青年 (the youth)，黑心 (evil mind)，投票权 (voting right)，讲话 (speech)，喜欢 (like)，实在 (sincerely)\\
  \hline
  2& Commentators hope James Soong will provide a hopeful life with dignity for Taiwan citizens.& 希望 (hope)，找回 (find back)，未来 (in the future)，找出路 (finding a way out for Taiwan)，带领 (lead)，尊严 (dignity)，困境 (dilemma)，骄傲 (proud)，人员 (staff)，下一代 (next generation)，历史 (history)，中华民国 (Republic of China)，创造 (create)，自信 (confidence)，光荣 (honor) \\
  \hline
  3& Commentators praise James Soong for his diligence in government affairs.& 爱民 (love of effect)，精神 (spirit)，蒋经国 (Jingguo Jiang)，省长 (governor of a province)，百姓 (the common people)，勤政 (be assiduous in government affairs)，感动 (touched)，值得 (it deserves)，理念 (idea)，理想 (ideal)，参选 (participate in the election)，找出路 (finding a way out for Taiwan)，记得 (remember)，国父 (the Father of a nation)，亲民爱民 (kind) \\
  \hline
  4 & Commentators cheer up James Soong.&加油 (cheer up)，胜利 (victory)，找出路 (finding a way out for Taiwan)，信心 (confidence)，爷爷 (grandpa)，成功 (success)，温暖 (warm)，胜利在望 (victory is in sight)，冲冲 (come on)，帅哥 (handsome man)，一同 (together)，加把劲 (step on it)，成绩 (achievement)，情势 (situation)，是非 (right and wrong) \\
  \hline
  5 & Commentators show their firm support for James Soong.& 支持 (support)，宋楚瑜 (James Soong)，唯一 (only)，总统 (president)，选择 (select)，全家 (the whole family)，好友 (friends)，坚定 (firmly)，贤能 (wisdom)，立场 (standpoint)，刘文雄 (Liu Wenxiong)，力挺 (fervently support)，当选 (be elected)，基隆 (Keelung)，强力 (strong)\\
  \hline
  6 & Commentators hope James Soong will change the current disappointing life in Taiwan.& 改变 (change)，现状 (status)，救救 (save)，机会 (opportunity)，觉醒 (awake)，失望 (disappointed)，小宋 (a nickname of James Soong)，沉沦 (sink into)，司法 (judicial)，完蛋 (be done)，醒醒 (awake)，王道 (king)，强力 (strong)\\
  \hline
  7 & Commentators hope James Soong will be elected and end the partisan battles of the KMT and the DPP.& 蓝绿 (the other two factions)，恶斗 (tough fight)，终结 (end)，团结 (unite)，选票 (vote)，改变 (change)，终止 (termination)，执政 (take office)，摆脱 (get rid of)，抛开 (throw off)，厌恶 (hate)，对立 (oppose)，停止 (stop)，拒绝 (refuse) \\
  \hline
  8 & Commentators accuse the KMT and the DPP of blocking James Soong's remarks.& 媒体 (media)，新闻 (news)，误报 (misinform)，电视 (television)，封锁 (be blocked)，不公 (unfair)，网路 (Internet)，公平 (fair)，选举 (election)，民主 (democracy)，封杀 (be banned)，边缘化 (marginalized)，蓝绿 (two factions)，办法 (methods)，曝光 (exposure)\\
  \hline
  9 & Commentators believe that James Soong will win the election.& 天佑 (bless)，领袖 (leader)，超越 (surpass)，高度 (lofty vision)，领导人 (leader)，关键 (key)，逆转 (reverse)，时刻 (the moment)，格局 (setup)，赢得 (win)，民心 (popular support)，勇敢 (brave)，选票 (vote)，突破 (break through)，再造 (reconstruction)\\
  \hline
  10 & Commentators encourage netizens to vote for James Soong.& 宋进总统府 (support the election of James Soong)，一定莹 (must win)，宋莹 (James Soong and Hsu Hsin-ying)，法院 (court)，立法 (legislation)，救星 (savior)，挺进 (advance)，才会莹 (will win)，瑜定莹 (must win)，东山再起 (stage a comeback)，打破 (breakthrough)，前途 (promise)，倾听 (listen attentively to)，台湾 (Taiwan)\\
  \hline
  11 & Commentators persuade Eric Chu's voters to vote for James Soong.& 宋进总统府 (support the election of James Soong)，幸福 (happiness)，全民 (the whole people)，感动 (touched)，找出路 (finding a way out for Taiwan)，弃朱保宋 (support James Soong and give up Eric Chu)，唯一 (only)，天佑 (bless)，全台 (Taiwan)，为国 (for the country)，欧吉桑 (uncle)，儿女 (children)，还给 (give back)，受苦 (suffer)，党籍 (party membership)\\
  \hline
  12& Commentators express their expectations for James Soong's policies on economics and education.& 政策 (policy)，政府 (government)，解决 (solve)，教育 (education)，经济 (economic)，办法 (methods)，改革 (reform)，未来 (in the future)，重点 (focal point)，青年 (the youth)，面对 (confront with)，重视 (pay attention to)，提出 (propose)，财团 (financial group)，整合 (integrate) \\
  \hline
  13& Commentators expect James Soong to propose polices in favor of young people.& 希望 (hope)，未来 (in the future)，年轻人 (young people)，下一代 (next generation)，年轻 (young)，国家 (nation)，期待 (look forward)，幸福 (happiness)，世代 (generation)，生活 (life)，子孙 (descendants)，孩子 (children)，机会 (opportunity)，日子 (life)，小孩 (children) \\
  \hline
  14 & Commentators are impressed by James Soong's campaign advertising.& 感动 (touched)，主席 (chairman)，政见 (political views)，感觉 (feel)，影片 (video)，声音 (sound)，现场 (on live)，广告 (advertisement)，发表 (make a statement)，电话 (telephone)，听到 (hear)，开心 (happy)，画面 (picture)，辩论 (debate)，可爱 (lovely)\\
  \hline
  15& Commentators remind Eric Chu to look after himself.&辛苦 (work hard)，身体 (health)，保重 (take care of)，伯伯 (uncle)，谢谢 (thanks)，希望 (hope)，永远 (forever)，宋北北 (a nickname of James Soong)，努力 (make efforts)，休息 (have a rest)，支持 (support)，健康 (healthy)，照顾 (take care of)，心疼 (love dearly)，敬爱 (beloved)\\
  \hline
\end{longtable}
\end{center}
\end{CJK}

\clearpage
\section{Event Study and Results}
\subsection{Event Detection and Event Study}
\begin{figure}
\centering
\subfigure[Tsai Ing-wen]{
\label{fig:ldtsai}
\includegraphics[width=\textwidth]{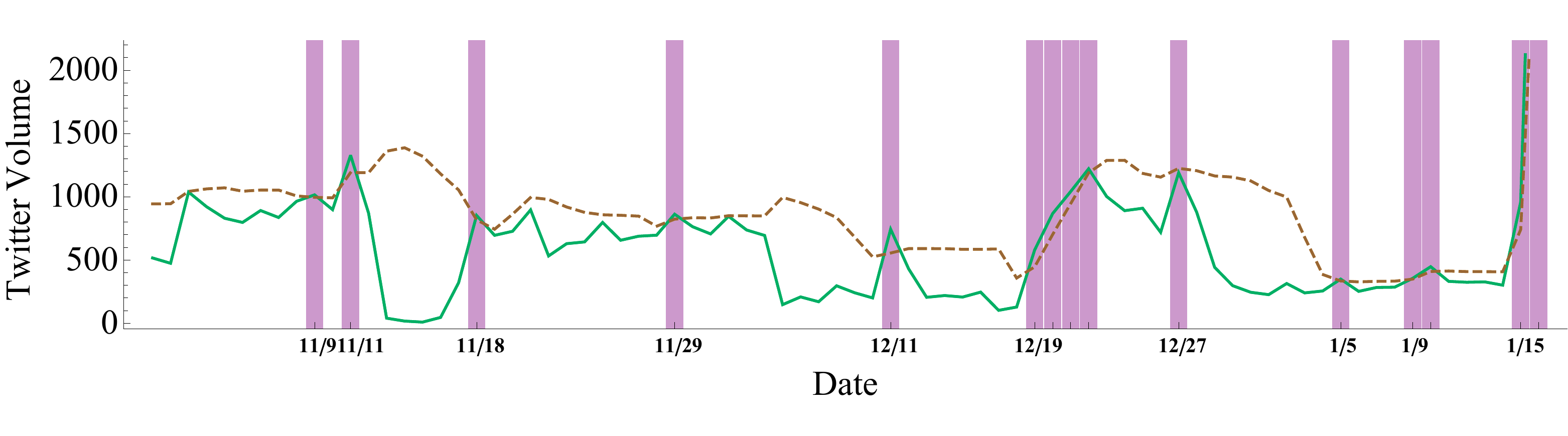}}
\subfigure[Eric Chu]{
\includegraphics[width=\textwidth]{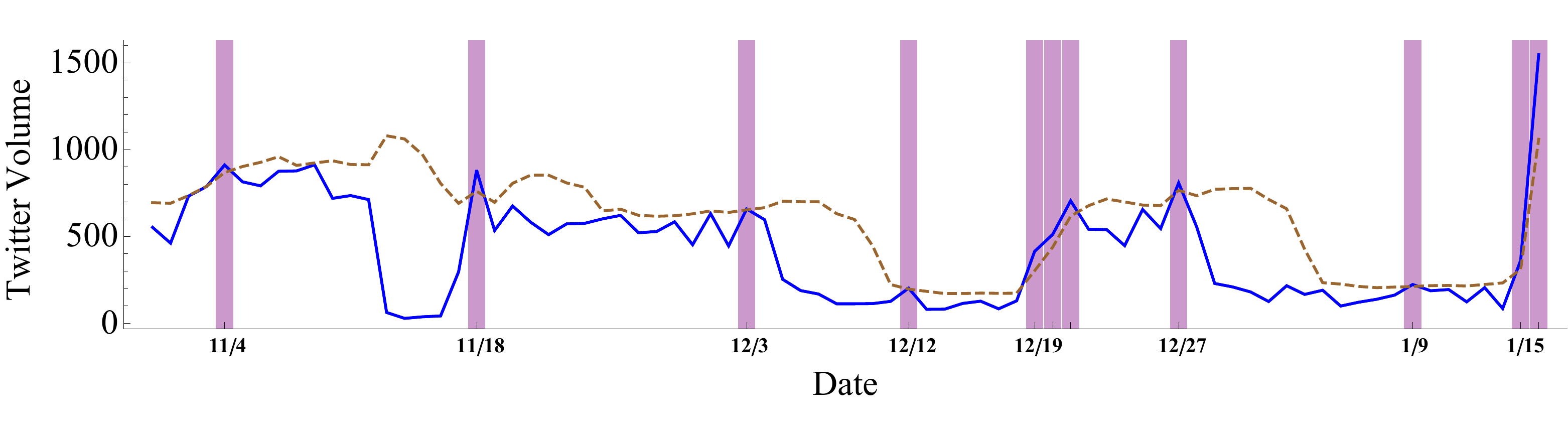}}
\subfigure[James Soong]{
\includegraphics[width=\textwidth]{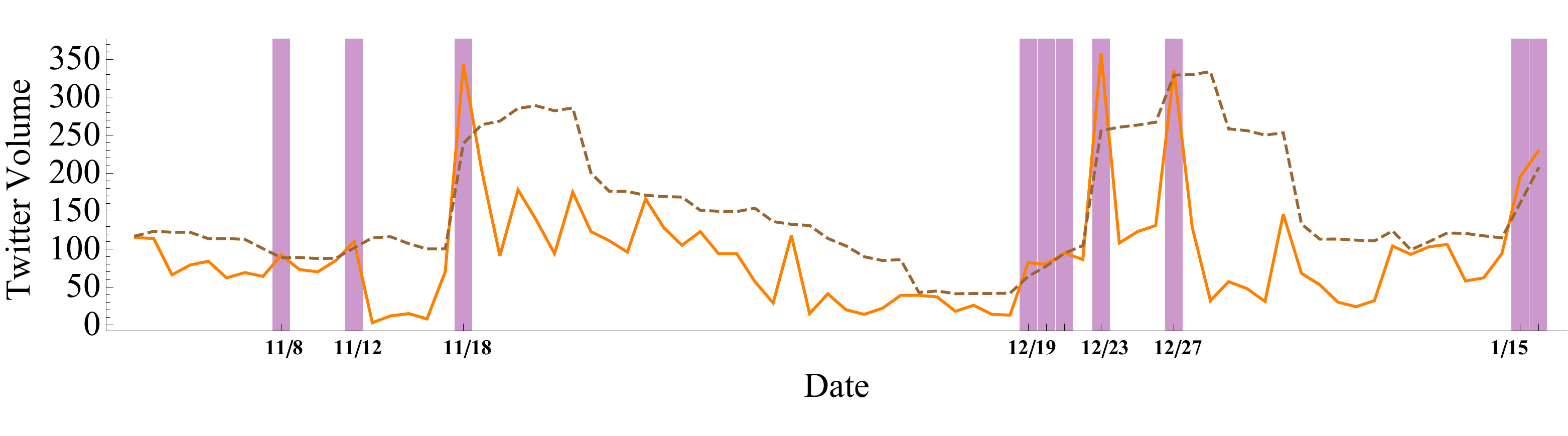}}
\caption{Bursty days. The solid lines in the figures are the daily Twitter volumes of the three candidates. The dashed lines represent the daily upper bounds of the confidence intervals. If the Twitter volume exceeds the upper bound, then the date is marked as a bursty day. The bursty days are represented by purple bars in the figures.}
\label{fig:burst}
\end{figure}

 We adopt an event study to measure the effects of campaign events on public opinion.
  First, we use daily Twitter volume to detect bursty days. Twitter is an online platform that aggregates information about candidates during the campaign. Therefore, the number of tweets related to candidates is a signal that is related to events. We select candidate keywords to retrieve tweets via Twitter API, as illustrated in Section~\ref{sub:data}, and then count the number of tweets $tw^c_k$ for each candidate $c$ on each day $k$. Applying moving average model, we calculate the confidence intervals for daily Twitter volume $u^c_{k+1}$ as follows:
 \begin{equation}
 u^c_{k+1}=\hat{n}+\frac{s}{\sqrt{m}}t_{\alpha/2}(m-1),
 \end{equation}
 where $\hat{n}$ is the average of $tw^c_k$ for the previous $m$ days and $s$ is the standard deviation. Based on a t-test with significance level $\alpha$, there exists an influential event if $tw^c_k$ exceeds $u^c_{k}$. Here, we set $m=7$ and $\alpha=0.05$. Daily Twitter volume and the confidence interval are plotted in FIG.~\ref{fig:burst}.

Second, we estimate the event timespan of each detected burst. The daily tweets for each candidate are first integrated into a single document, and the terms in the documents are weighted by the $tf-idf$ method. $tf-idf$ is a numerical statistic that is intended to reflect how important a word is to a document in a collection of corpora. The $tf-idf$ value increases proportionally with the number of times a word appears in a document but is often offset by the frequency of the word in the corpus, which helps to adjust for the fact that some words appear more frequently in general. $tf-idf$ is calculated as follows,
\begin{equation}
\begin{split}
&tf^c(t,d^c_k)=\frac{f_{t,d^c_k}}{\sum_t f_{t,d^c_k}},\\
&idf^c(t,D^c)=log \frac{N^c}{1+|d^c_k \in D^c:t \in d^c_k|},\\
&tf-idf^c(t,d^c_k,D^c)=tf^c(t,d^c_k)idf^c(t,D^c),\\
\end{split}
\end{equation}
where $f_{t,d^c_k}$ is the term count of $t$ in daily Twitter document $d^c_k$ of candidate $c$ at time $k$.
$D^c$ is the total Twitter document of candidate $c$, $N^c=|D^c|$.
$|d^c_k \in D^c:t \in d^c_k|$ is the number of documents in which the term $t$ appears.

We extract the top 30 words as keywords at the burst. Then, for each bursty day, we compare the keywords to those of each day from day $k-5$ to day $k+5$. If any keywords from the bursty do not appear in the time window, the event is eliminated. 
Otherwise, we take the first date that any of the keywords appear as the start date of the event and the date that any of keywords no longer appear as the end date. If the keywords on the bursty day also appear in the keywords for day $k-5$ (or day $k+5$), we continue to check the keywords of the 5 days before day $k-5$(or the 5 days after day $k+5$). The results are shown in TABLE~\ref{tab:event_tsai}, TABLE~\ref{tab:event_llchu}, and TABLE~\ref{tab:event_soong}.

In the three tables, the column 'timespan' represents the event window derived by keyword matching. The column 'event' provides a descriptive summary of events based on the keywords. The 'Checking date' column is the dates with which the keywords coincide with those of the bursty days. Dates with extended event timespans (that is dates for which the search continued to the five days before $k-5$ or the five days after $k+5$) are marked in purple. 
The overlapping keywords are listed in the last column. Bursty days detected based on Twitter volume are highlighted in red.

To propagandize their political views, the three candidates held two television debates on December 27, 2015 and January 2, 2016. However, we only obtain Twitter references to Tsai Ing-wen during the first debate. 
 To examine and compare the influence of the two debates on perceptions of all three candidates, we modify the debate timespan of Tsai Ing-wen and James Soong to be the period from December 27, 2015 to January 4, 2016, which coincides with that of Eric Chu.

After determining the events and their timespans, we calculate the impact of events on public opinion using vote state estimates derived via Kalman filter. We augment the state transition equation with dummy variables to capture the effects of events. We use a dummy variable to $D_{j,k}$ to indicate each detected event:
\begin{equation}
p^c_k = a+p^c_{k-1}+\sum^{J_c}_{j=1}\gamma_j D^c_{j,k}+\varepsilon,
\end{equation}
where $D^c_{j,k}=1$ during the event timespan of event $j$ of candidate $c$, and $D^c_{j,k}=0$ otherwise. $J_c$ is the number of detected events of candidate $c$. The coefficient $\gamma_j$ captures the event effect, which is an estimate of the average effect across event $j$. If the event has a significant effect on public opinion, then $\gamma_j$ will pass the t-test. It is worth mentioning that $p^c_k$ is equal to the posterior estimate of the state value $\hat{x}^{c}_{k|k}$.

The results are shown in TABLE~\ref{tab:ee_tsai}, TABLE~\ref{tab:ee_llchu}, and TABLE~\ref{tab:ee_soong}. The statistical results of $\gamma_i,i \in \{1,...,21\}$ correspond to effects of 21 events marked in $E_i,i \in \{1,...,21\}$.

\subsection{Introduction of Significant Influential Events}
By conducting event study, we find that several events related to cross-strait relations significantly influence public opinion, such as the Chou Tzu-yu flag incident and the Xi-Ma meeting. Because of the complex backgrounds of these events, we describe them in detail and cite a number of news reports from the mainstream media.

\emph{Chou Tzu-yu flag incident.} Chou Tzu-yu, a 16-year-old Taiwan singer and a member of the South Korean K-pop girl group Twice, attracted attention with her appearance on a South Korean show in which she introduced herself as Taiwanese and waved the "Republic of China" (ROC) flag alongside those of South Korea and Japan. Soon after the episode was broadcast, she was accused of being a "Pro-Taiwanese independence activist" by Taiwan-born China-based singer Huang An. This incident sparked a major controversy with mainland Chinese Internet users. The controversy further led to cancellations of Chou's scheduled performances on several television stations in mainland China. With the uproar over the issue, JYP entertainment, the talent agency and music production company representing Chou, released a video showing Chou reading an apology, the night before the election. She mentioned in part:

"There is only one China. The two shores are one. I feel proud being a Chinese. I, as a Chinese, have hurt the company and netizens' emotions due to my words and actions during overseas promotions. I feel very, very sorry and guilty."~\footnote[7]{New York Times report on the Chou Tzu-yu flag incident, \protect\url{http://www.nytimes.com/2016/01/17/world/asia/taiwan-china-singer-chou-tzu-yu.html?_r=0}}

Nevertheless, many Taiwanese saw her apology as humiliating and a sign of Taiwan's predicament that Chou had to apologize for expressing her Taiwan identity and showing the ROC flag. Facebook and other social media were flooded with angry comments on the eve and day of the election. "If someone forces you to forget something you shouldn't forget, you should revolt and fight...Those of you who haven't voted yet, use your vote to fight back. Today is D-Day." wrote one person who identified himself as Chung Nian-huang on Facebook~\footnote[8]{BBC News report on the Chou Tzu-yu flag incident, \protect\url{http://www.bbc.com/news/world-asia-35340530}}.
"In Taiwan, online commentators compared her apology to hostage videos released by the Islamic State, although it was probably more reminiscent of the sort of humiliating confessions that dissidents are increasingly forced to make on Chinese state television," notes \emph{the Washington Post}'s China bureau chief Simon Denyer~\footnote[9]{Washington Post report on the Chou Tzu-yu flag incident,
\protect\url{https://www.washingtonpost.com/news/worldviews/wp/2016/01/16/watch-teenage-pop-stars-humiliating-apology-to-china-for-waving-taiwan-flag/?tid=pm_world_pop_b}}.

This incident is believed to potentially contribute to pro-independence politicians. A number of media have reported on and analyzed this incident. "Any boost to the turnout likely helps Tsai and the DPP, particularly since this has gone viral among young people," said Clayton Dube of the University of Southern Cailifornia's U.S.-China Institute (Cited from Slatest~\footnote[10]{ Slatest report on the Chou Tzu-yu flag incident, \protect\url{http://www.slate.com/blogs/the_slatest/2016/01/16/pop_star_may_have_helped_pro_independence_party_win_taiwan_presidency.html}}). BBC News cited remarks of Xiao Xinhuang, a Taiwan sociologist, and reported that "Ms Tsai would have won even if the video hadn't been posted, but the incident may have contributed another one or two percentage points"~\footnotemark[7].
In addition, traditional pollsters also investigated the influence of the incident. TVBS announced on January 20th that "Polls show that about 500 thousand voters, accounting for 4\%, who previously did not want to go out to vote might show up to cast their ballots to Taiwan-centric candidates"~\footnote[11]{TVBS reports on the influences of the Chou Tzu-yu flag incident, \protect\url{http://news.ltn.com.tw/news/politics/breakingnews/1579332}}. Taiwan Think Tank also released a poll on January, which reported that 11.9\% of the respondents were influenced by the incident in the legislative election, and 11.4\% of the respondents were influenced in the presidential election~\footnote[12]{Taiwan Think Tank reports on the influences of the Chou Tzu-yu flag incident, \protect\url{http://www.cna.com.tw/news/aipl/201601210169-1.aspx}}.

\emph{Xi-Ma meeting.} On 7 November, 2015, Xi Jin-ping, the president of the People's Republic of China, and Ma Ying-Jeou, the president of the Republic of China, met in Singapore. The meeting was the first between the leaders of the two sides of the Taiwan Strait since the end of the Chinese Civil War in 1949. Both men spoke of the historic moment after 66 years of separation and the need to preserve the stable ties of recent years.

The meeting provoked diverse responses from parties, civil society and countries. Ma Ying-Jeou said that "this meeting with Xi is not aimed at boosting his personal legacy after he steps down in May 2016, nor is it intended to salvage the flagging ruling Kuomintang campaign in the runnup to the Jan.16 presidential election, but is designed entirely for the good of the next generation"~\footnote[13]{Focus Taiwan report on the Xi-Ma meeting,
\protect\url{http://focustaiwan.tw/news/acs/201511050016.aspx}}. But the presidential candidate and incumbent chairwoman of the DPP Tsai Ing-wen countered by stating that this meeting "is a manipulation of the January elections and labelling the decision-making process as opaque"~\footnote[14]{Taipei Times report on the Xi-Ma meeting,
\protect\url{http://www.taipeitimes.com/News/front/archives/2015/11/05/2003631718}}. The youth division of the Taiwan Solidarity Union and third-force party leaders, such as New Power Party Chairman Huang Kuo-chang, Social Democratic Party chairman Fan Yun, and Green Party Taiwan convener Lee Ken-Cheng, also opposed this meeting.
After the Xi-Ma meeting, Tsai said "she was disappointed. President Ma Ying-Jeou made no reference to preserving the island's democracy and freedom during his landmark meeting with China President Xi Jinping in Singapore". On the contrary, several Taiwan business organizations, such as the Chinese National Federation of Industries, Taipei Chamber of Commerce, and Allied Association for Science Park Industries, expressed support for the meeting. The meeting was perceived to help to improve Taiwan's economic prospects~\footnotemark[13]. The Ministry of Foreign Affairs of Singapore, a Conservative politician of the United Kingdom, and a spokesman of the U.S. State Department also expressed their supportive attitude.

Several agencies conducted polls to investigate public attitudes toward the Xi-Ma meeting. The results of the Cross Strait Policy Association, EBC, SET, TISR, and Taiwan Think Tank polls show that the meeting was supported by a majority.

\clearpage
\begin{CJK}{UTF8}{gkai}
\begin{longtable}
{|m{1cm}|m{4cm}|m{1cm}|m{9cm}|}
\caption[Events and Timespans: Tsai Ing-wen]{Events and Timespans: Tsai Ing-wen}\label{tab:event_tsai} \\

\hline \multicolumn{1}{|c|}{\textbf{Timespan}}& \multicolumn{1}{c|}{\textbf{Event}} & \multicolumn{1}{c|}{\textbf{Checking Date}}& \multicolumn{1}{c|}{\textbf{Keywords}} \\ \hline
\endfirsthead

\multicolumn{4}{c}%
{{\bfseries \tablename\ \thetable{} -- Events and Timespans:Tsai Ing-wen}} \\
\hline \multicolumn{1}{|c|}{\textbf{Timespan}} &
\multicolumn{1}{c|}{\textbf{Event}} &
\multicolumn{1}{c|}{\textbf{Checking Date}}&
\multicolumn{1}{c|}{\textbf{Keywords}} \\ \hline
\endhead

\hline \multicolumn{4}{|r|}{{Continued on next page}} \\ \hline
\endfoot

\hline \hline
\endlastfoot

\centering
\multirow{7}{1cm}{2015/11/4-2015/11/10} &\multirow{7}{4cm}{\textbf{E1: Xi-Ma Meeting}} &2015/11/4& 马习 (Ma Ying-Jeou and Xi Jin-ping)\\
\cline{3-4}
&&2015/11/5&马习 (Ma Ying-Jeou and Xi Jin-ping), 习马 (Xi Jin-ping and Ma Ying-Jeou), 香格里拉饭店 (Shangri-La Hotel Singapore, the meeting place)\\
\cline{3-4}
&&2015/11/6&马习 (Ma Ying-Jeou and Xi Jin-ping)\\
\cline{3-4}
&&2015/11/7&马习 (Ma Ying-Jeou and Xi Jin-ping), 失望 (disappoint)\\
\cline{3-4}
&&2015/11/8&马习 (Ma Ying-Jeou and Xi Jin-ping), 失望 (disappoint), 习马 (Xi Jin-ping and Ma Ying-Jeou)\\
\cline{3-4}
&&\textcolor[rgb]{1.00,0.00,0.00}{2015/11/9}& 论调 (view), 支持率 (vote share), 周玉蔻 (Clara Chou, a Taiwan journalist, she asked a number of questions related to cross-strait relations at the Xi-Ma meeting press conference), 失实 (inconsistent with the facts), 选择权 (the right to vote), 剥夺 (deprive), 群聚 (meet), 马习 (Ma Ying-Jeou and Xi Jin-ping), 失望 (disappoint), 应限 (should limit), 台湾队 (Taiwan), 习马 (Xi Jin-ping and Ma Ying-Jeou), 参访 (visit), 基隆 (Keelung, a place in Taiwan), 棒球 (baseball), 习马两人 (Xi Jin-ping and Ma Ying-Jeou), 香格里拉 (Shangri-La Hotel Singapore), 公表 (make a public statement), 论谈 (talk about), 马办(Ma Ying-Jeou's office), 声明 (statement), 诺贝尔和平奖 (the Nobel Peace Prize), 殊荣 (honor), 苏姬 (Suu Kyi), 翁山 (Aung San, Aung San Suu Kyi is a Burmese politician, diplomat, author, and a winner of the Nobel Peace Prize), 宪章 (charter), 驳斥(refute),  零八年 (the year of 2008), 习想(Xi Jin-ping intends to),  吓跑 (scare away)\\
\hline
&&2015/11/10&马习 (Ma Ying-Jeou and Xi Jin-ping), 习马 (Xi Jin-ping and Ma Ying-Jeou), 翁山苏姬 (Aung San Suu Kyi)\\
\hline
\end{longtable}
\label{tab:tsai_ekw}
\end{CJK}

\clearpage
\begin{CJK}{UTF8}{gkai}
\begin{longtable}
{|m{2cm}|m{4cm}|m{1cm}|m{9cm}|}
\caption[Events and Timespans: Eric Chu]{Events and Timespans: Eric Chu}\label{tab:event_llchu} \\

\hline \multicolumn{1}{|c|}{\textbf{Timespan}}& \multicolumn{1}{c|}{\textbf{Event}} & \multicolumn{1}{c|}{\textbf{Checking Date}}& \multicolumn{1}{c|}{\textbf{Keywords}} \\ \hline
\endfirsthead

\multicolumn{4}{c}%
{{\bfseries \tablename\ \thetable{} -- Events and Timespans: Eric Chu}} \\
\hline \multicolumn{1}{|c|}{\textbf{Timespan}} &
\multicolumn{1}{c|}{\textbf{Event}} &
\multicolumn{1}{c|}{\textbf{Checking Date}}&
\multicolumn{1}{c|}{\textbf{Keywords}} \\ \hline
\endhead

\hline \multicolumn{4}{|r|}{{Continued on next page}} \\ \hline
\endfoot

\hline \hline
\endlastfoot

\centering
\multirow{10}{2cm}{2015/11/4-2015/11/12} &\multirow{10}{4cm}{\textbf{E9: Xi-Ma Meeting}} &\textcolor[rgb]{1.00,0.00,0.00}{2015/11/4}& 马习 (Ma Ying-Jeou and Xi Jin-ping), 美国务院 (the State Department of the United States), 规格 (specifications), 台南 (Tainan), 谈过 (talk about), 跳楼 (jump off a building), 便当 (lunch), 习近平 (Xi Jin-ping), 言辞 (words), 有助于 (contribute to), 总会 (association), 历史 (history), 出任 (take up the post of), 太平 (peaceful), 义正 (justice), 中国队 (China), 危朱立 (Eric Chu), 美国队 (the United States), 张荣祥 (Zhang Rong-xinag), 习马 (Xi Jin-ping and Ma Ying-Jeou), 花莲 (Hualien, a place in Taiwan), 白宫 (White House), 台南市 (Tainan), 态势 (situation), 劳工 (labor), 对垒 (oppose each other), 永康 (Yongkang), 米 (rice), 互信 (mutual trust)\\
\cline{3-4}
&&2015/11/5&马习 (Ma Ying-Jeou and Xi Jin-ping), 白宫 (White House), 原住民 (natives)\\
\cline{3-4}
&&2015/11/6&马习 (Ma Ying-Jeou and Xi Jin-ping)\\
\cline{3-4}
&&2015/11/7&马习 (Ma Ying-Jeou and Xi Jin-ping), 习马 (Xi Jin-ping and Ma Ying-Jeou)\\
\cline{3-4}
&&\textcolor[rgb]{0.50,0.00,1.00}{2015/11/8}& 马习 (Ma Ying-Jeou and Xi Jin-ping), 一场 (one), 远大 (of far-reaching significance), 意义 (significance), 祸害 (disaster), 见到 (meeting), 神隐 (disappear), 闪躲 (evade), 有利于 (be propitious to), 徐巧芯 (Xu Qiaoxin), 习马 (Xi Jin-ping and Ma Ying-Jeou), 见面 (meeting), 周日 (Sunday), 害怕 (be afraid of), 不利 (disadvantage), 东宁 (Dongning), 谈论 (talk about), 正妹 (beauty), 忠于 (be loyal to), 介绍 (introduce), 海峡 (strait), 平面 (plane), 晚餐 (dinner), 地位 (status), 畅谈 (talk freely), 国教 (state religion), 新加坡 (Singapore), 黄国 (a name, this may be a mistake of word segmentation), 面前 (in the face of)\\
\cline{3-4}
&&2015/11/9&马习 (Ma Ying-Jeou and Xi Jin-ping), 地位 (status)\\
\cline{3-4}
&&2015/11/10&马习 (Ma Ying-Jeou and Xi Jin-ping), 习马 (Xi Jin-ping and Ma Ying-Jeou)\\
\cline{3-4}
&&2015/11/11&马习 (Ma Ying-Jeou and Xi Jin-ping), 畅谈 (talk freely)\\
\hline
\end{longtable}
\label{tab:tsai_ekw}
\end{CJK}

\clearpage
\begin{CJK}{UTF8}{gkai}
\begin{longtable}
{|m{2cm}|m{4cm}|m{2cm}|m{8cm}|}
\caption[Events and Timespans: James Soong]{Events and Timespans: James Soong}\label{tab:event_soong} \\

\hline \multicolumn{1}{|c|}{\textbf{Timespan}}& \multicolumn{1}{c|}{\textbf{Event}} & \multicolumn{1}{c|}{\textbf{Checking Date}}& \multicolumn{1}{c|}{\textbf{Keywords}} \\ \hline
\endfirsthead

\multicolumn{4}{c}%
{{\bfseries \tablename\ \thetable{} -- Events and Timespans: James Soong}} \\
\hline \multicolumn{1}{|c|}{\textbf{Timespan}} &
\multicolumn{1}{c|}{\textbf{Event}} &
\multicolumn{1}{c|}{\textbf{Checking Date}}&
\multicolumn{1}{c|}{\textbf{Keywords}} \\ \hline
\endhead

\hline \multicolumn{4}{|r|}{{Continued on next page}} \\ \hline
\endfoot

\hline \hline
\endlastfoot

\centering
\multirow{7}{2cm}{2015/11/4-2015/11/10} &\multirow{7}{4cm}{\textbf{E16: Xi-Ma Meeting}} &2015/11/4& 马习 (Ma Ying-Jeou and Xi Jin-ping), 肯定 (sure)\\
\cline{3-4}
&&2015/11/5&马习 (Ma Ying-Jeou and Xi Jin-ping)\\
\cline{3-4}
&&2015/11/6&马习 (Ma Ying-Jeou and Xi Jin-ping)\\
\cline{3-4}
&&2015/11/7&马习 (Ma Ying-Jeou and Xi Jin-ping)\\
\cline{3-4}
&&\textcolor[rgb]{1.00,0.00,0.00}{2015/11/8}& 马习 (Ma Ying-Jeou and Xi Jin-ping), 关圣 (Guan Yu), 帝君 (Guan Yu), 马忘 (James Soong criticized Ma Ying-Jeou), 不足 (insufficient), 检讨 (self-criticism), 尊敬 (respect), 部署 (deploy), 李显龙 (Lee Hsien Loong), 陆飞 (Lu Fei), 鸿泰 (Hongtai), 宫庙 (temple), 马记 (this may be a mistake of word segmentation), 听信 (believe), 小兵 (people under the leadership of somebody), 台中 (Taichung), 团队 (team), 主将 (chief commander), 有情 (interest), 宫庙参观 (visit temples), 邱毅 (Chiu Yi),不全 (incomplete), 肯定 (sure), 自称 (claim), 真相 (truth), 骨力 (vigorous strokes), 暴力 (violence), 留言 (comment), 边缘化 (marginalize), 欧吉 (this may be a mistake of word segmentation)\\
\cline{3-4}
&&2015/11/9&马习 (Ma Ying-Jeou and Xi Jin-ping), 检讨 (self-criticism)\\
\cline{3-4}
&&2015/11/10&马习 (Ma Ying-Jeou and Xi Jin-ping)\\
\hline
\multirow{2}{2cm}{2015/11/12-2015/11/13} &\multirow{2}{4cm}{\textbf{E17: James Song takes part in the commemoration meeting of Sun Yat-sen}} &\textcolor[rgb]{1.00,0.00,0.00}{2015/11/12}& 国父 (the father of a nation), 身手 (skill), 诞辰 (birthday), 帝君 (king), 黄义交 (Huang Yijiao), 化身 (embodiment), 有情 (interest), 关圣 (Guan Yu), 纪念 (in commemoration of), 部落格 (blog), 神隐 (disappear), 沦为 (become), 体力 (physical strength), 缅怀 (remember), 公开信 (open letters), 宋神掌 (a nickname of James Soong), 壹起 (one), 低调 (low-key), 骨力 (vigorous strokes), 欧吉 (this may be a mistake of word segmentation), 现身 (appear), 不想 (do not want to), 翻转 (reversal), 精神 (spirit), 官方 (official), 行程 (itinerary), 世界 (world), 台湾人 (Taiwanese), 先生 (Mr.), 马英九 (Ma Ying-Jeou)\\
\hline
\end{longtable}
\label{tab:tsai_ekw}
\end{CJK}

\clearpage
\begin{table}[htbp]
\caption{Effects of events detected in Tsai Ing-wen's Twitter on public opinion}
\begin{tabular}{ccccc}
\toprule
&Estimate&Std.Error&t value&Pr(>|t|)\\
\midrule
c&-0.0261644&0.0223770&-1.169&0.2464\\
$p^{tsai}_{k-1}$&1.0503217&0.0430209&24.414&< 2e-16 ***\\
$\gamma_1$&-0.0055497&0.0032195&-1.724&0.0894 *\\
$\gamma_2$&-0.0078498&0.0034143&-2.299&0.0246 **\\
$\gamma_3$&-0.0016263&0.0026600&-0.611&0.5430\\
$\gamma_4$&0.0007406&0.0038982&0.190&0.8499\\
$\gamma_5$&-0.0024433&0.0029135&-0.839&0.4047\\
$\gamma_6$&0.0011313&0.0018736&0.604&0.5480\\
$\gamma_7$&0.0007497&0.0020732&0.362&0.7188\\
$\gamma_8$&0.0365850&0.0056551&6.469&1.33e-08***\\
\hline
Residual standard error&\multicolumn{4}{c}{0.005508 on 67 degrees of freedom}\\
Multiple R-squared&\multicolumn{4}{c}{0.9529}\\
Adjusted R-squared&\multicolumn{4}{c}{0.9459}\\
F-statistic&135.6 on 10 and 67 DF&p-value&<2.2e-16\\
\bottomrule
\end{tabular}
\begin{tablenotes}
\item[1] ***, **, and * indicate that the coefficients are significantly different from zero at the level of 1\%, 5\%, and 10\%, respectively.
\end{tablenotes}
\label{tab:ee_tsai}
\end{table}

\clearpage
\begin{table}[htbp]
\caption{Effects of events detected in Eric Chu's Twitter on public opinion}
\begin{tabular}{ccccc}
\toprule
&Estimate&Std.Error&t value&Pr(>|t|)\\
\midrule
c&-0.0055981&0.0182615&-0.307&0.76012\\
$p^{llchu}_{k-1}$&1.0101605&0.0480926&21.004&< 2e-16 ***\\
$\gamma_9$&0.0058044&0.0019974&2.906&0.00494 ***\\
$\gamma_{10}$&0.0025127&0.0026600&0.945&0.34819\\
$\gamma_{11}$&0.0010210&0.0024453&0.418&0.67761\\
$\gamma_{12}$&0.0023037&0.0026457&0.871&0.38696\\
$\gamma_{13}$&-0.0002331&0.0021261&-0.110&0.91303\\
$\gamma_{14}$&0.0048592&0.0026373&1.842&0.06977 *\\
$\gamma_{15}$&-0.0261781&0.0051982&-5.036&3.73e-06***\\
\hline
Residual standard error&\multicolumn{4}{c}{0.005007 on 68 degrees of freedom}\\
Multiple R-squared&\multicolumn{4}{c}{0.9529}\\
Adjusted R-squared&\multicolumn{4}{c}{0.9466}\\
F-statistic&152.7 on 9 and 68 DF&p-value&<2.2e-16\\
\bottomrule
\end{tabular}
\begin{tablenotes}
\item[1] ***, **, and * indicate that the coefficients are significantly different from zero at the level of 1\%, 5\%, and 10\%, respectively.
\end{tablenotes}
\label{tab:ee_llchu}
\end{table}

\clearpage
\begin{table}[htbp]
\caption{Effects of events detected in James Soong's Twitter on public opinion}
\begin{tabular}{ccccc}
\toprule
&Estimate&Std.Error&t value&Pr(>|t|)\\
\midrule
c&-0.0008878&0.0012076&-0.735&0.465\\
$p^{llchu}_{k-1}$&1.0171937&0.0113960&89.259&< 2e-16 ***\\
$\gamma_{16}$&-0.0003667&0.0007516&-0.488&0.627\\
$\gamma_{17}$&-0.0006323&0.0010710&-0.590&0.557\\
$\gamma_{18}$&0.0009155&0.0007376&1.241&0.219\\
$\gamma_{19}$&-0.0005611&0.0010016&-0.560&0.577\\
$\gamma_{20}$&0.0003622&0.0006590&0.550&0.584\\
$\gamma_{21}$&-0.0054510&0.0012896&-4.227&7.13e-05***\\
\hline
Residual standard error&\multicolumn{4}{c}{0.001664 on 69 degrees of freedom}\\
Multiple R-squared&\multicolumn{4}{c}{0.9963}\\
Adjusted R-squared&\multicolumn{4}{c}{0.9959}\\
F-statistic&2331 on 8 and 69 DF&p-value&<2.2e-16\\
\bottomrule
\end{tabular}
\begin{tablenotes}
\item[1] ***, **, and * indicate that the coefficients are significantly different from zero at the level of 1\%, 5\%, and 10\%, respectively.
\end{tablenotes}
\label{tab:ee_soong}
\end{table}

\clearpage
\section*{References}
\bibliographystyle{naturemag}


\end{document}